\documentclass[final,leqno]{siamltex}
\usepackage{amsfonts}
\usepackage{color, xcolor, colortbl}
\usepackage{amsmath,amssymb}
\usepackage{float}
\usepackage{algorithm}
\usepackage{bm}
\newfloat{algorithm}{t}{lop}
\usepackage{algorithmic}
\usepackage{graphicx,epstopdf}
\usepackage[caption=false]{subfig}
\usepackage{appendix}
\usepackage{multirow}

\newcommand{\norm}[1]{\lVert#1\rVert}

\newcommand{\tup}{\bigtriangleup}
\newcommand{\tdown}{\bigtriangledown}

\newcommand{\Or}{\mathcal{O}}

\newcommand{\NN}{\mathbb{N}}
\newcommand{\RR}{\mathbb{R}}

\title{Low-rank representation of tensor network operators with long-range pairwise interactions}

\author{Lin Lin\thanks{Department of Mathematics, University of California, Berkeley, and Computational Research Division, Lawrence Berkeley National Laboratory, Berkeley, CA 94720. Email: \texttt{linlin@math.berkeley.edu}} \and
Yu Tong\thanks{Department of Mathematics, University of California,
Berkeley, CA 94720. Email: \texttt{yu\_tong@berkeley.edu}}}

\begin{document}
\maketitle

\begin{abstract}
Tensor network operators, such as the matrix product operator (MPO) and the
projected entangled-pair operator (PEPO), can provide efficient representation
of certain linear operators in high dimensional spaces. This paper focuses on
the efficient representation of tensor network operators with long-range pairwise
interactions such as the Coulomb interaction. For MPOs, we find that all
existing efficient methods exploit a peculiar ``upper-triangular low-rank''
(UTLR) property, i.e. the upper-triangular part of the matrix can be well
approximated by a low-rank matrix, while the matrix itself can be full-rank.
This allows us to convert the problem of finding the efficient MPO
representation into a matrix completion problem.  We develop a modified
incremental singular value decomposition method (ISVD) to solve this
ill-conditioned matrix completion problem. This algorithm yields equivalent MPO
representation to that developed in [Stoudenmire and White, Phys. Rev. Lett.
2017].  In order to efficiently treat more general tensor network operators, we
develop another strategy for compressing tensor network operators based on
hierarchical low-rank matrix formats, such as the hierarchical off-diagonal
low-rank (HODLR) format, and the $\mathcal{H}$-matrix format. Though the
pre-constant in the complexity is larger, the advantage of using the hierarchical low-rank matrix
format is that it is applicable to both MPOs and PEPOs. 
For the Coulomb interaction, the operator can be represented by a linear combination of $\Or(\log(N)\log(N/\epsilon))$ MPOs/PEPOs, each with a constant bond dimension, where $N$ is the system size and $\epsilon$ is the accuracy of the low-rank truncation. Neither the modified ISVD nor the hierarchical low-rank algorithm assumes that the long-range interaction takes a translation-invariant form.
\end{abstract}



\pagestyle{myheadings}
\thispagestyle{plain}

\section{Introduction}\label{sec:intro}

Tensor network states~\cite{White1992,Vidal2007}, particularly presented by the matrix product states (MPS)~\cite{White1992,OstlundRommer1995,Schollwock2011} and projected entangled-pair states (PEPS)~\cite{VerstraeteCirac2004,VerstraeteWolfPerezEtAl2006,Orus2014}, are among the most promising classes of variational methods for approximating high-dimensional functions in quantum physics. Besides their successes for treating strongly correlated quantum systems~\cite{YanHuseWhite2011,Norman2016}, the MPS (also known as the tensor train method (TT)~\cite{OseledetsTyrtyshnikov2009,Oseledets2011,LubichRohwedderSchneiderEtal2013}), have become useful in a wide range of applications~\cite{RakhubaOseledets2016,StoudenmireSchwab2016,HanWangFanEtAl2018}. A core component of tensor network algorithms is the efficient representation of linear operators.  In MPS the operator representation is called the matrix product operator (MPO), and in PEPS the projected entangled-pair operator (PEPO). These forms of representation are naturally designed for short-range interactions, such as interactions of nearest-neighbor type on a regular lattice. For long-range interactions, such as those naturally appearing from electronic structure calculations due to the Coulomb interaction, a straightforward representation would lead to a large MPO/PEPO rank, which can be prohibitively expensive.  Therefore efficient representation of tensor network operators with long-range interactions is crucial for the success of the methods.

In this paper, we focus on the tensor network operators with long-range
pairwise interactions, such as the Coulomb interaction.  A tensor network operator with pairwise interaction can be defined as
\begin{equation}
\hat{V}=\frac{1}{2}\sum_{1\le i\neq j\le N}\mathbf{V}(i,j)\hat{n}_i\hat{n}_j=\sum_{1\le i<j\le N} \mathbf{V}(i,j) \hat{n}_i\hat{n}_j.
\label{eqn:pairwise}
\end{equation}
Here $\mathbf{V}\in \RR^{N\times N}$ is a symmetric matrix, called the
coefficient matrix of $\hat{V}$, and $\hat{n}_i$ is called a number operator.  The precise definition of
$\hat{n}_{i}$, as well as the connection to the Coulomb interaction will
be discussed in Section~\ref{sec:prelim}.

There have been a number of proposals to treat MPOs with long-range
pairwise interactions. Using the exponential fitting techniques based on finite state machines (FSM)~\cite{crosswhite2008finite,pirvu2010matrix,frowis2010tensor,ChanKeselmanEtAl2016,crosswhite2008applying}, the Coulomb interaction and other translation-invariant long-range interactions on one-dimensional lattice systems can be efficiently represented. For the Coulomb interaction, the MPO rank can be bounded by $\log (N/\epsilon)$, where $N$ is the number of sites and $\epsilon$ is the target accuracy.  For one-dimensional systems without the translation-invariance property, recently a method based on a sequence of singular value decomposition (SVD) compression operations~\cite{ChanKeselmanEtAl2016,StoudenmireWhite2017} has been developed. 
This method is observed to yield a compact MPO representation for the Coulomb interaction with basis functions arising from electronic structure calculations that are not translation-invariant. On the other hand, this is a greedy method, and there is no \textit{a priori} upper bound for the MPO rank.
For two-dimensional and higher dimensional lattice systems, PEPOs can provide more efficient representation of linear operators than MPOs. To our knowledge, the only efficient PEPO representation for long-range interaction is given by the recently developed correlation function valued PEPO (CF-PEPO)~\cite{o2018efficient}.  CF-PEPO uses a fitting method, which maps the original lattice system with translation-invariant long-range interactions to an effective system defined on a superlattice but with short-range interactions. There is no \textit{a priori} upper bound for the PEPO rank. Neither is it clear whether the CF-PEPO method can be efficiently generalized to non-translation-invariant systems.

\noindent\textbf{Contribution:}
The contribution of this paper is two-fold. First, we find that the success of both the exponential fitting method and the SVD compression method rests on the assumption that $\mathbf{V}$ satisfies the ``upper-triangular low-rank'' (UTLR) property, i.e. the upper-triangular part of $\mathbf{V}$ can be extended into a matrix that is approximately low-rank, while $\mathbf{V}$ can be a full-rank matrix.  Therefore the problem can be viewed as a special type of matrix completion problem~\cite{candes2009exact}. However, this matrix completion problem is highly ill-conditioned, and standard matrix completion algorithms suffer from numerical stability problems~\cite{cai2010singular,hu2008collaborative,mnih2008probabilistic,brand2002incremental,balzano2013grouse}. Based on the work of~\cite{brand2002incremental}, we develop a modified incremental singular value decomposition method (ISVD) to simultaneously find the low-rank structure in a numerically stable fashion, and to construct the MPO representation.   
The algorithm yields an MPO representation equivalent to that in the SVD compression method~\cite{StoudenmireWhite2017}. The ISVD method is not restricted to translation-invariant systems, and numerical results indicate that the performance of ISVD is comparable to (in fact is marginally better than) that of the exponential fitting methods for translation-invariant systems.

Second, we propose a new way for representing long-range interactions based on the framework of hierarchical low-rank matrices. In particular, we focus on the hierarchical off-diagonal low-rank (HODLR) format~\cite{AmbikasaranDarve2013}, and the $\mathcal{H}$-matrix format~\cite{grasedyck2003construction,hackbusch1999sparse}. The main advantage of the hierarchical low-rank format is that it allows us to construct efficient representation for both MPOs and PEPOs.  For the Coulomb interaction, we can represent $\hat{V}$ using a linear combination of $\Or(\log(N)\log(N/\epsilon))$ MPOs/PEPOs, and the bond dimension of each MPO/PEPO is bounded by a constant.  From a computational perspective, such a format can be more preferable than a single MPO/PEPO operator with bond dimension bounded by $\Or(\log(N)\log(N/\epsilon))$.  Furthermore, the hierarchical low-rank format can also be applied to systems without the translation-invariance property.



\noindent\textbf{Notation:}

Throughout the paper all vectors and operators are defined on finite-dimensional spaces. We use hatted letters such as $\hat{A},\hat{V}$ to denote high-dimensional linear operators. Vectors, matrices and tensor cores are denoted by bold font letters such as $\mathbf{A},\mathbf{X},\mathbf{V}$. The tensor product of $\mathbf{A}$ and $\mathbf{B}$ is denoted by $\mathbf{A}\otimes \mathbf{B}$. We use the matlab notation such as $\mathbf{A}(\mathcal{I},\mathcal{J})$ to denote sub-matrices, where $\mathcal{I},\mathcal{J}$ are index sets. Given a certain global ordering of all indices, the notation $\mathcal{I}\prec\mathcal{J}$ means that for any $i\in \mathcal{I}$ and $j\in\mathcal{J}$, we have $i<j$.

\noindent\textbf{Organization:}
The rest of the paper is organized as follows. 
In Section \ref{sec:prelim} we briefly introduce the MPO and PEPO representation. In Section \ref{sec:long_range_mpo_1d} we introduce and use the UTLR structure to construct MPO representation of long-range interaction for 1D systems. In Section \ref{sec:long_range_hierarchical} we introduce the method to construct MPO/PEPO representation using the hierarchical low-rank matrix format. Numerical results are presented in Section \ref{sec:numerical}. We conclude our work in Section \ref{sec:conclusion}. The rules of finite state machine (FSM) for representing non-overlapping MPOs and PEPOs are given in the Appendix~\ref{sec:appendix_mpo} and~\ref{sec:appendix_pepo}, respectively.  

\section{Preliminaries}\label{sec:prelim}

We first briefly review the construction of MPS/MPO, as well as of PEPS/PEPO. For more detailed information we refer readers to~\cite{Schollwock2011,Orus2014}.  A single vector $\mathbf{X}\in \RR^{n_1\times \cdots \times n_d}$ can be interpreted as an order-$d$ tensor,  where the $d$-tuple $(n_1,\ldots,n_d)$ is called the size of the tensor. We assume that the index follows a lexicographic ordering (a.k.a. row-major ordering). Each entry can be accessed using multi-indices $\mathbf{X}(i_1,\ldots,i_d),1\le i_\alpha \le n_\alpha, \alpha=1,\ldots,d$. A linear operator $\hat{A}\in \RR^{(m_1\times \cdots \times m_d)\times (n_1\times \cdots \times n_d)}$ is an order $2d$-tensor, with each entry denoted by 
$$
\hat{A}(i_1,\ldots,i_d;j_1,\ldots,j_d).
$$ 
The application of $\hat{A}$ to $\mathbf{X}$ gives a tensor $\mathbf{Y}\in \RR^{m_1\times \cdots\times m_d}$ as
$$
\mathbf{Y}(i_1,\ldots,i_d)=\sum_{j_1=1}^{n_1}\cdots\sum_{j_d=1}^{n_d}\hat{A}(i_1,\ldots,i_d;j_1,\ldots,j_d) \mathbf{X}(j_1,\ldots,j_d),
$$
which can be written in short-hand notation as $\mathbf{Y}=\hat{A}\mathbf{X}$. 

In~\eqref{eqn:pairwise}, we used $i,j$ as the site indices and $d=N$.  The sites can be organized into lattices of one, two and three dimensions. For Coulomb interaction in 1D $\mathbf{V}(i,j)=1/|i-j|$ for $i\ne j$, and we set $\mathbf{V}(i,i)=0$.  In higher dimensions we use the notation $\mathbf{i},\mathbf{j}$ to represent a lattice site.  For example, on an $N\times N$ 2D lattice we have $\mathbf{i}=(i_x,i_y)$ for $1\leq i_x,i_y\leq N$. For a pairwise interaction in the form of (\ref{eqn:pairwise}), the coefficient matrix has as its entries $\mathbf{V}(i_x+Ni_y,j_x+Nj_y)$, using a row-major order, and this is written more compactly as $\mathbf{V}(\mathbf{i},\mathbf{j})$ hereafter. The Coulomb interaction takes the form
$\mathbf{V}(\mathbf{i},\mathbf{j}) = \frac{1}{\|\mathbf{i}-\mathbf{j}\|}$, where $\|\cdot\|$ is the Euclidean distance.  Using this notation, the pairwise interaction is called 
translation-invariant if $\mathbf{V}(\mathbf{i},\mathbf{j})$ is a function of $\mathbf{i}-\mathbf{j}$ (in the presence of periodic boundary conditions, we may simply redefine $\mathbf{i}-\mathbf{j}$).
In Eq.~\eqref{eqn:pairwise}, we may assume for simplicity that $n_1=\cdots=n_d=n$. 
$\hat{n}_i$ acts only on a single-site $i$ and should be interpreted as 
$\mathbf{I}^{\otimes (i-1)}\otimes \mathbf{n}\otimes  \mathbf{I}^{\otimes
(n-i+1)}$, where $\mathbf{I}$ is the identity matrix of size $n$ and
$\mathbf{n}$ is an $n\times n$ matrix. $\hat{n}_i$ can be understood as a spin
operator for quantum spin systems, or a number operator for quantum fermionic
systems, though the precise form of $\mathbf{n}$ is not relevant for the purpose of this paper.


\subsection{Matrix product operators}\label{sec:MPO}

A vector $\mathbf{X}$ is represented by an MPS/TT format if each entry can be
represented as as a matrix product

\begin{equation}
\mathbf{X}(i_1,\ldots,i_d)=\mathbf{U}_1(i_1)\cdots \mathbf{U}_d(i_d),
\label{eqn:MPS}
\end{equation}
where each $\mathbf{U}_\alpha(i_\alpha)$ is a matrix of size $(r_{\alpha-1}, r_{\alpha})$. Since $\mathbf{X}(i_1,\ldots,i_d)$ is a scalar, by definition $r_0=r_d=1$. Each $\mathbf{U}_\alpha$, called a core tensor, can be viewed as a $3$-tensor of size $(r_{\alpha-1},n_\alpha,r_{\alpha})$, and the matrix $\mathbf{U}_\alpha(i_\alpha)$ is called the $i_\alpha$-th slice of $\mathbf{U}_\alpha$. Since the first and third indices of $\mathbf{U}_\alpha$ are to be contracted out via matrix-multiplication operations, they are called the internal indices, while the second index is called the external index. The $(d+1)$-tuple $(r_0,r_1,\ldots,r_d)$ is called the bond dimension (a.k.a. the MPO/TT rank). Sometimes the bond dimension also refers to a single scalar $\max_{1\le \alpha\le d} r_\alpha$.  

The MPO format of a linear operator $\hat{A}$ is
\begin{equation}
\hat{A}(i_1,\ldots,i_d;j_1,\ldots,j_d)=\mathbf{A}_{1}\left(i_{1}, j_{1}\right) \mathbf{A}_{2}\left(i_{2}, j_{2}\right) \cdots \mathbf{A}_{d}\left(i_{d}, j_{d}\right).
\label{eqn:MPO}
\end{equation}
Each $\mathbf{A}_\alpha$ is also called a core tensor, and is of size $(s_{\alpha-1},m_\alpha,n_\alpha,s_{\alpha})$.  The first and the fourth indices are called the internal indices, and the second and third indices the external indices. The matrix $\mathbf{A}_\alpha(i_1,j_1)\in\RR^{m_\alpha\times n_\alpha}$ is called a slice. Again $s_0=s_d=1$. The $(d+1)$-tuple $(s_0,\ldots,s_d)$, or sometimes simply $\max_{1\le \alpha\le d} s_\alpha$, is called the bond dimension. 

Consider the application of the linear operator $\hat{A}$ on $\mathbf{X}$, denoted by $\mathbf{Y}=\hat{A}\mathbf{X}$, then the vector $\mathbf{Y}$ has an MPS representation as
$$
\mathbf{Y}(i_1,\ldots,i_d)=\mathbf{V}_1(i_1)\cdots \mathbf{V}_d(i_d).
$$
Using the MPO representation, we may readily find that 
$$
\mathbf{V}_\alpha(i_\alpha)=\sum_{j_\alpha=1}^{n_\alpha} \mathbf{A}_{\alpha}(i_\alpha,j_\alpha)\otimes \mathbf{U}(j_\alpha).
$$
The bond dimension of $\mathbf{Y}$ is $(s_0 r_0,s_1r_1,\ldots,s_dr_d)$. 

The simplest example of the MPO is an operator in the tensor product form, i.e.
$$
\hat{A}=\mathbf{a}_1\otimes\cdots\otimes \mathbf{a}_n,
$$
where $\mathbf{a}_\alpha\in\RR^{n_{\alpha}\times n_{\alpha}}$ is a matrix, i.e.  $\mathbf{a}_\alpha(i_\alpha,j_\alpha)$ is a scalar. In the component form
$$
\hat{A}(i_1,\ldots,i_d;j_1,\ldots,j_d)=\mathbf{a}_{1}\left(i_{1}, j_{1}\right) \mathbf{a}_{2}\left(i_{2}, j_{2}\right) \cdots \mathbf{a}_{d}\left(i_{d}, j_{d}\right).
$$
Clearly the MPO bond dimension is $1$.

The second example is an operator with nearest-neighbor interaction, e.g.
$$
\hat{A}=\sum_{\alpha=1}^{d-1}(\mathbf{I}_1\otimes\cdots\otimes \mathbf{I}_{\alpha-1})\otimes \mathbf{a}_\alpha \otimes \mathbf{a}_{\alpha+1}\otimes(\mathbf{I}_{\alpha+2}\otimes\cdots\otimes \mathbf{I}_d).
$$
where $\mathbf{I}_\alpha$ is the identity matrix of size $n_\alpha$. In the component form
\begin{equation}
  \begin{split}
    &\hat{A}(i_1,\ldots,i_d;j_1,\ldots,j_d)\\
    =&\sum_{\alpha=1}^{d-1} (\delta_{i_1,j_1}\cdots\delta_{i_{\alpha-1},j_{\alpha-1}}) \mathbf{a}_{\alpha}\left(i_{\alpha}, j_{\alpha}\right) \mathbf{a}_{\alpha+1}\left(i_{\alpha+1}, j_{\alpha+1}\right)(\delta_{i_{\alpha+2},j_{\alpha+2}}\cdots\delta_{i_{d},j_{d}}) .
  \end{split}
\label{eqn:MPONNcomponent}
\end{equation}
$\hat{A}$ can be viewed as the linear combination of $(d-1)$ MPOs in the tensor product form, and if so the resulting MPO rank would be $(d-1)$. However, note that we may define
$$
\mathbf{A}_1(i_1,j_1)=\begin{pmatrix}
0 & \mathbf{a}_1(i_1,j_1) & \delta_{i_1,j_1}
\end{pmatrix}, \quad \mathbf{A}_d(i_d,j_d)=\begin{pmatrix}
\delta_{i_d,j_d}\\ \mathbf{a}_{d}(i_d,j_d) \\ 0
\end{pmatrix},
$$
and
$$
\mathbf{A}_\alpha(i_\alpha,j_\alpha)=\begin{pmatrix}
\delta_{i_{\alpha},j_{\alpha}}& 0 & 0\\
\mathbf{a}_{\alpha}(i_\alpha,j_\alpha) & 0 & 0\\
0 & \mathbf{a}_{\alpha}(i_\alpha,j_\alpha) & \delta_{i_{\alpha},j_{\alpha}}
\end{pmatrix},\quad 2\le\alpha\le d-1.
$$
Then it can be readily verified that the MPO of the form $\eqref{eqn:MPO}$ that agrees with the component form $\eqref{eqn:MPONNcomponent}$. Hence the MPO rank is independent of $d$ and is always $3$.  

When the context is clear, we may also identify 
 $\hat{a}_\alpha$ with $(\mathbf{I}_1\otimes\cdots\otimes \mathbf{I}_{\alpha-1})\otimes \mathbf{a}_\alpha \otimes(\mathbf{I}_{\alpha+1}\otimes\cdots\otimes \mathbf{I}_d)$. Then 
$$
\hat{A}=\sum_{\alpha=1}^{d-1}\hat{a}_\alpha \hat{a}_{\alpha+1}.
$$
In the MPO form, we may also omit the $(i,j)$ indices and write
$$
\mathbf{A}_1=\begin{pmatrix}
0 & \mathbf{a}_1 & \mathbf{I}_1
\end{pmatrix}, \quad \mathbf{A}_d=\begin{pmatrix}
\mathbf{I}_d\\ \mathbf{a}_{d} \\ 0
\end{pmatrix}, \quad \mathbf{A}_\alpha=\begin{pmatrix}
\mathbf{I}_\alpha & 0 & 0\\
\mathbf{a}_{\alpha} & 0 & 0\\
0 & \mathbf{a}_{\alpha} & \mathbf{I}_\alpha
\end{pmatrix},\quad 2\le\alpha\le d-1.
$$

The third example is an operator with a special form of long-range interaction:
$$
\hat{A}=\sum_{1\le \alpha<\beta\le d}
e^{\lambda(\alpha-\beta)}\hat{a}_\alpha \hat{a}_{\beta}.
\label{eq:exp_interaction}
$$
We assume $\lambda>0$, and omit the component form for simplicity. The length of the interaction is characterized by $1/\lambda$.  A straightforward term-by-term representation would lead to an MPO of rank $\Or(d^2)$. Nonetheless, using the fact that
$$
e^{-\lambda n}=\underbrace{e^{-\lambda} e^{-\lambda}\cdots e^{-\lambda}}_{n~\text{times}}, \quad n\in \NN_+,
$$
$\hat{A}$ can also be written as an MPO with rank-$3$ as
$$
\mathbf{A}_1=\begin{pmatrix}
0 & \mathbf{a}_1 & \mathbf{I}_1
\end{pmatrix}, \quad \mathbf{A}_d=\begin{pmatrix}
\mathbf{I}_d\\ e^{-\lambda} \mathbf{a}_{d} \\ 0
\end{pmatrix}, \quad \mathbf{A}_\alpha=\begin{pmatrix}
\mathbf{I}_\alpha & 0 & 0\\
e^{-\lambda}\mathbf{a}_{\alpha} & e^{-\lambda}\mathbf{I}_\alpha & 0\\
0 & \mathbf{a}_{\alpha} & \mathbf{I}_\alpha
\end{pmatrix},\quad 2\le\alpha\le d-1.
$$
In physics literature, this is a special case of the finite state machine (FSM)~\cite{crosswhite2008applying}. For more complex settings, the FSM rule often specifies the row/column indices of each core tensor as input/output signals. We will use the language of input/output signals only in the Appendices~\ref{sec:appendix_mpo} and~\ref{sec:appendix_pepo}.

\begin{figure}[ht]
    \centering
    \subfloat[A $4$-tensor]{\includegraphics[width=0.4\textwidth]{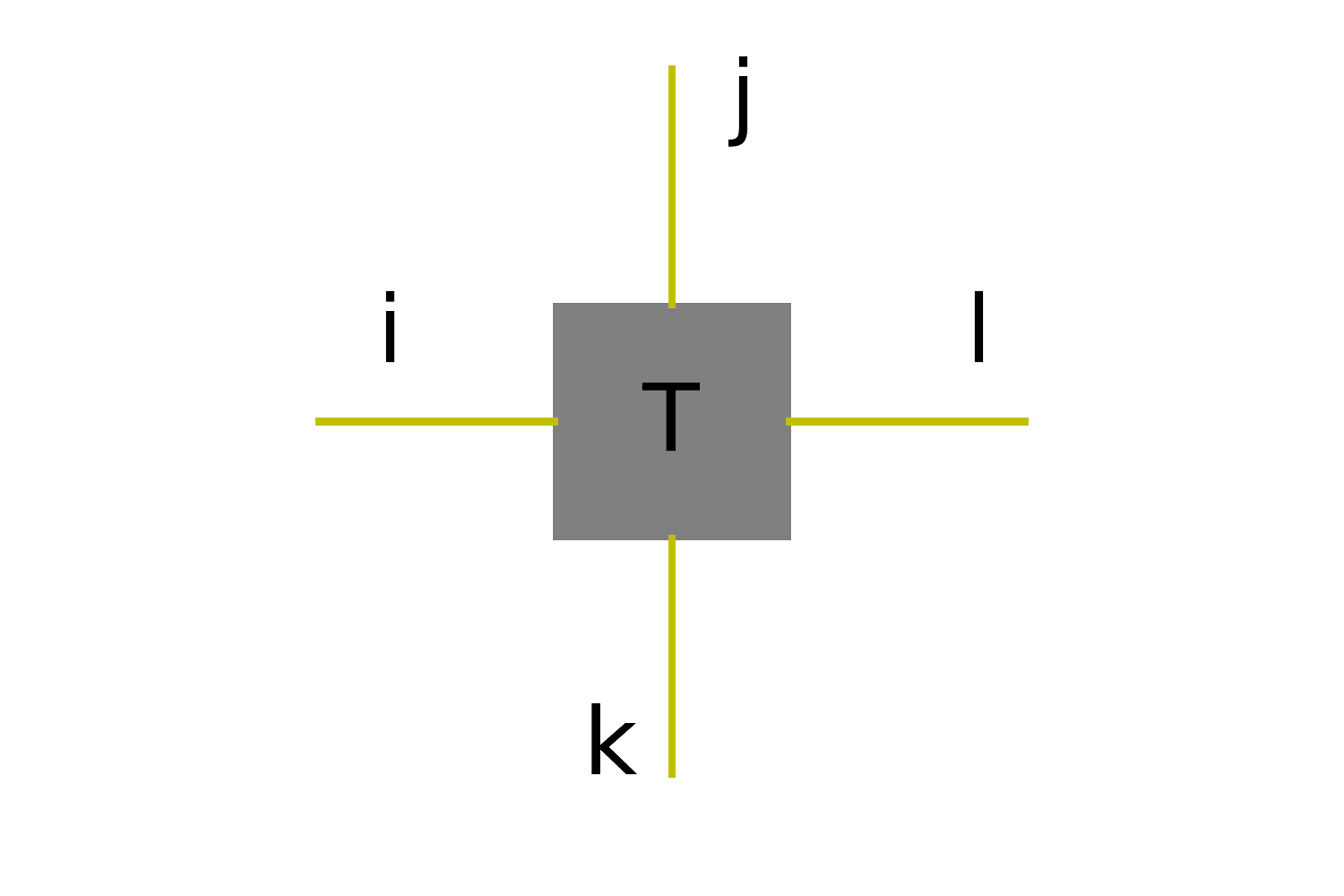}}
    \subfloat[Tensor contraction between a $4$-tensor and a $3$-tensor]{\includegraphics[width=0.4\textwidth]{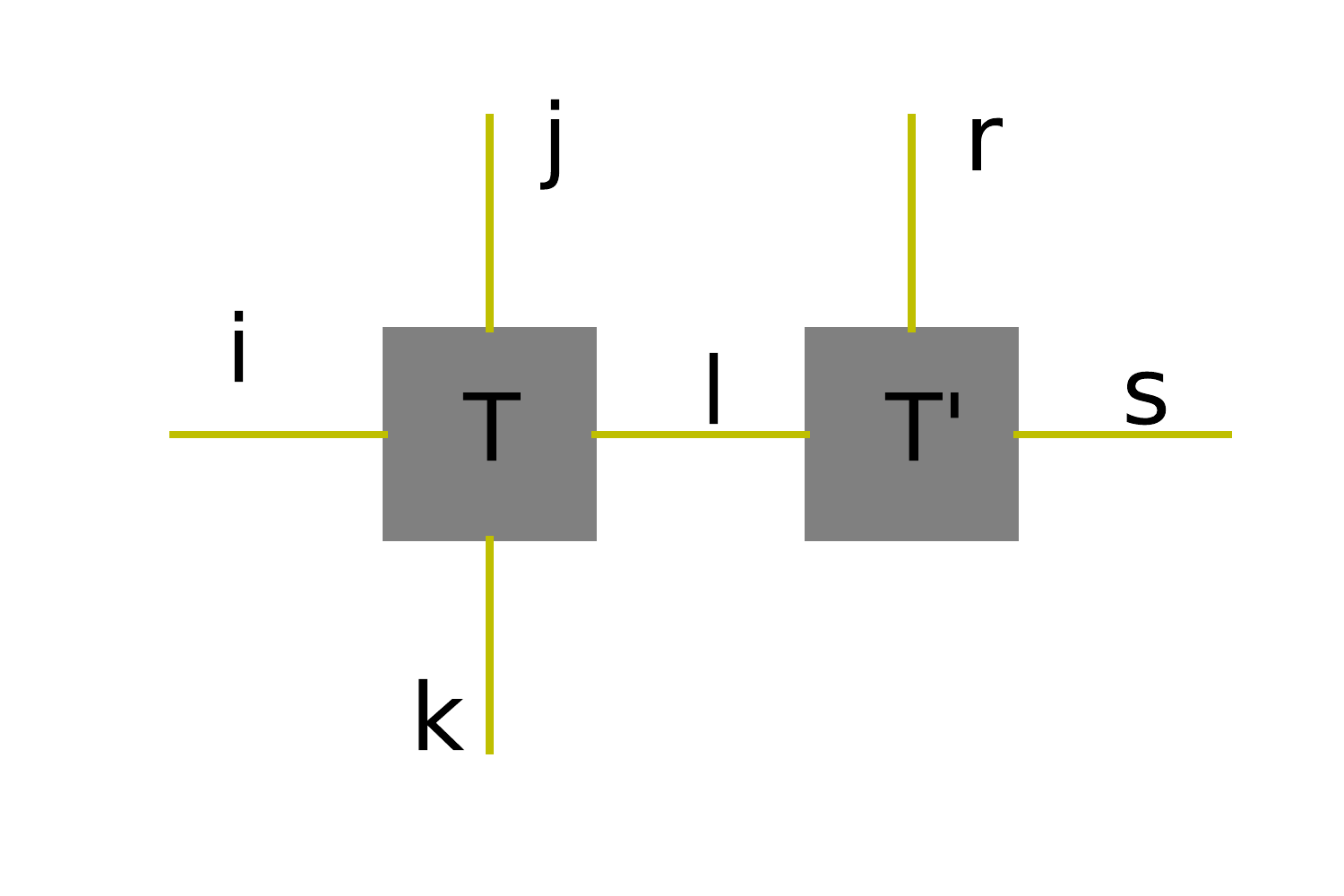}}
    \caption{(a) Graphical representation of a $4$-tensor $\mathbf{T}=(T_{ijkl})$. (b) Graphical representation of the tensor contraction between tensors $\mathbf{T}$ and $\mathbf{T'}$: $\sum_{l}T_{ijkl}T'_{lrs}$.}
    \label{fig:graph_tensor}
\end{figure}

\begin{figure}[ht]
    \centering
    \subfloat[MPS]{\includegraphics[width=0.45\textwidth]{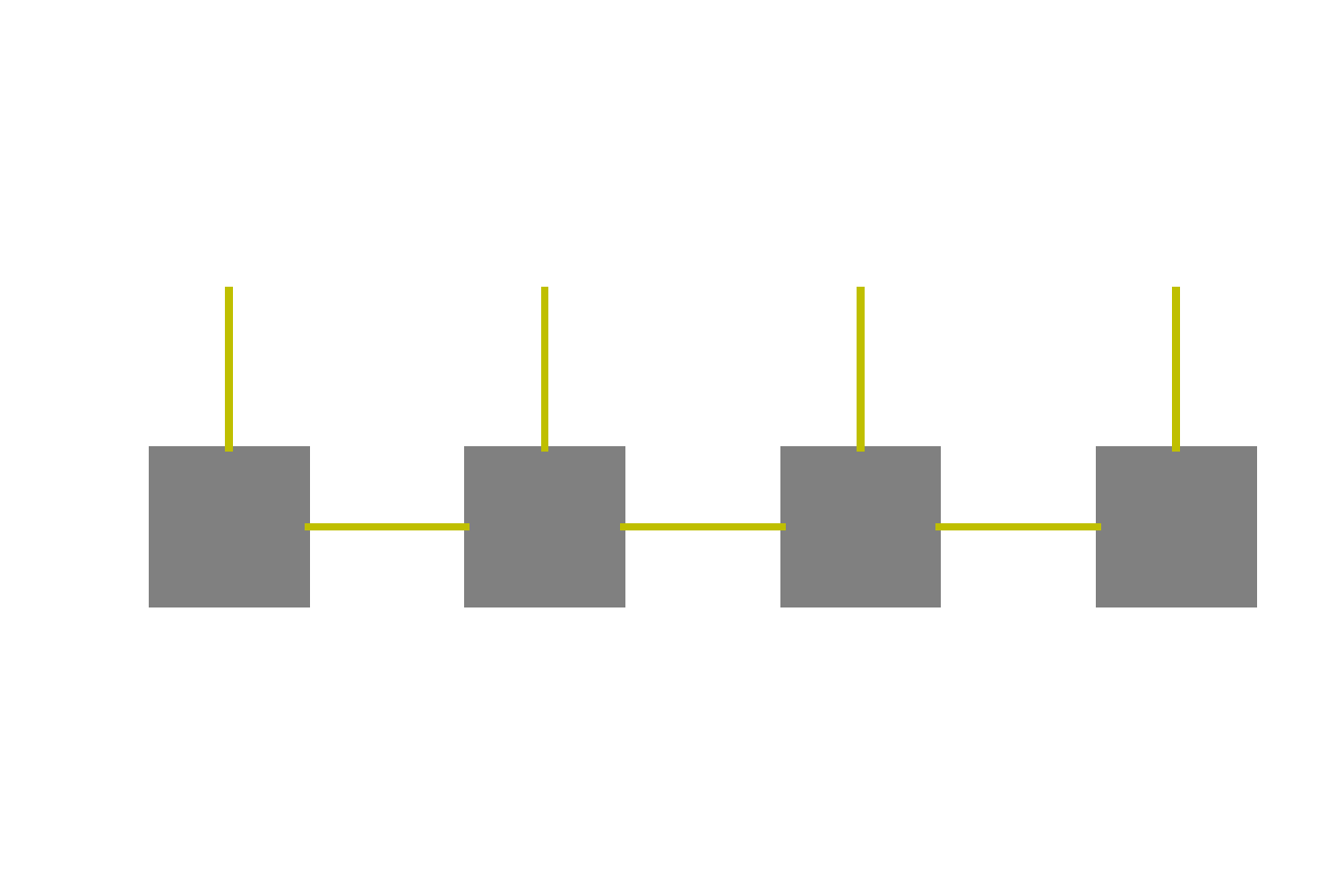}}
    \subfloat[MPO]{\includegraphics[width=0.45\textwidth]{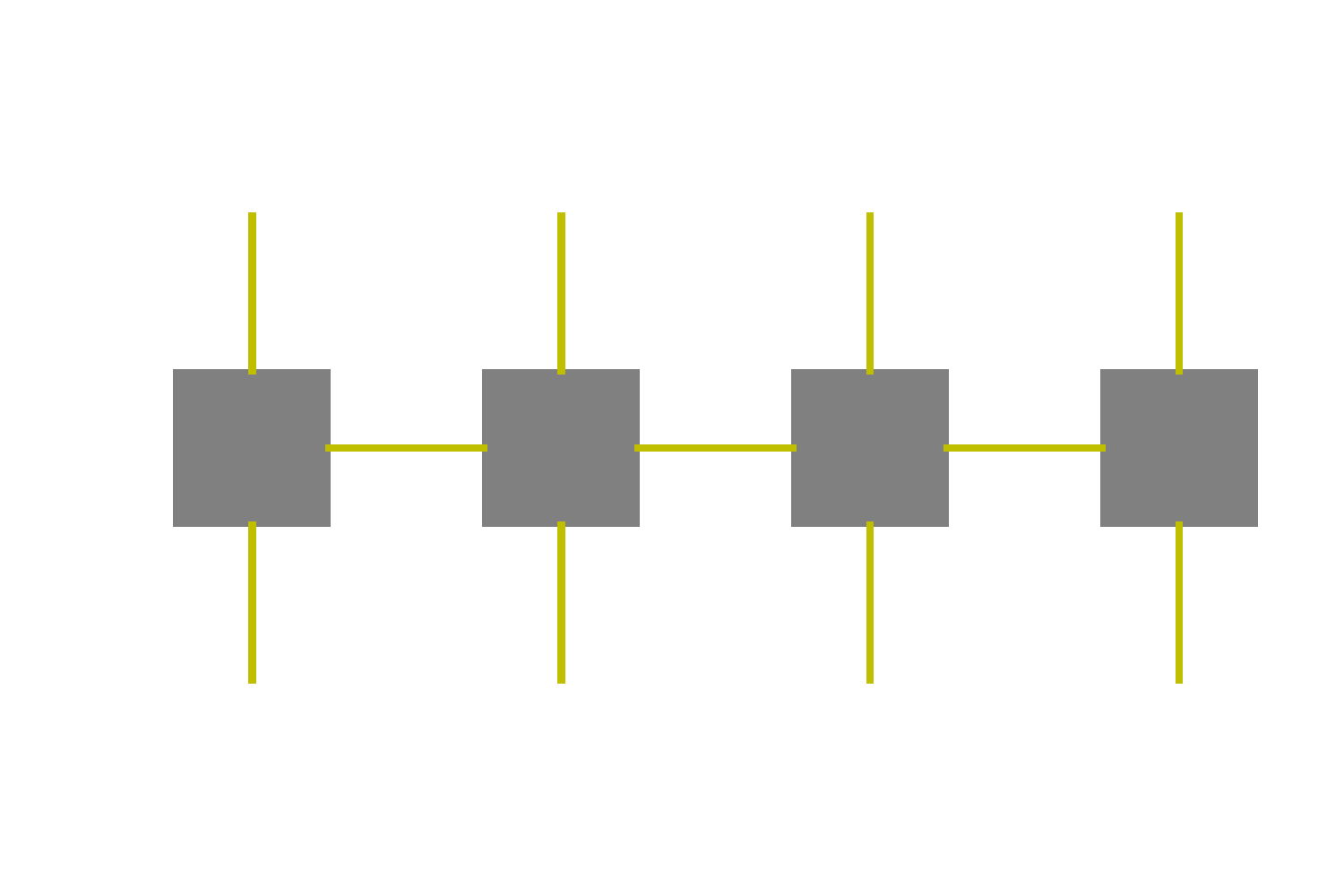}}
    \caption{Graphical representation of an MPS and an MPO.}
    \label{fig:mps_mpo}
\end{figure}

In order to proceed with the discussion of PEPS/PEPO, it is no longer productive to keep using the component form. Instead, the graphical representation of tensors is preferred.
Fig.~\ref{fig:graph_tensor} shows an example of the graphical representation of tensors and tensor contraction operations. A tensor is visualized as a vertex and each index (internal and external) is represented as an edge. When two vertices are linked by an edge, the corresponding index is contracted out. A more detailed introduction can be found in \cite{biamonte2017tensor}. Using this representation, MPS and MPO can be represented using graphs in Fig.~\ref{fig:mps_mpo} (a) and (b) respectively.

\subsection{Projected entangled-pair operators}\label{sec:PEPO}

\begin{figure}[ht]
    \centering
    \subfloat[PEPS]{\includegraphics[width=0.45\textwidth]{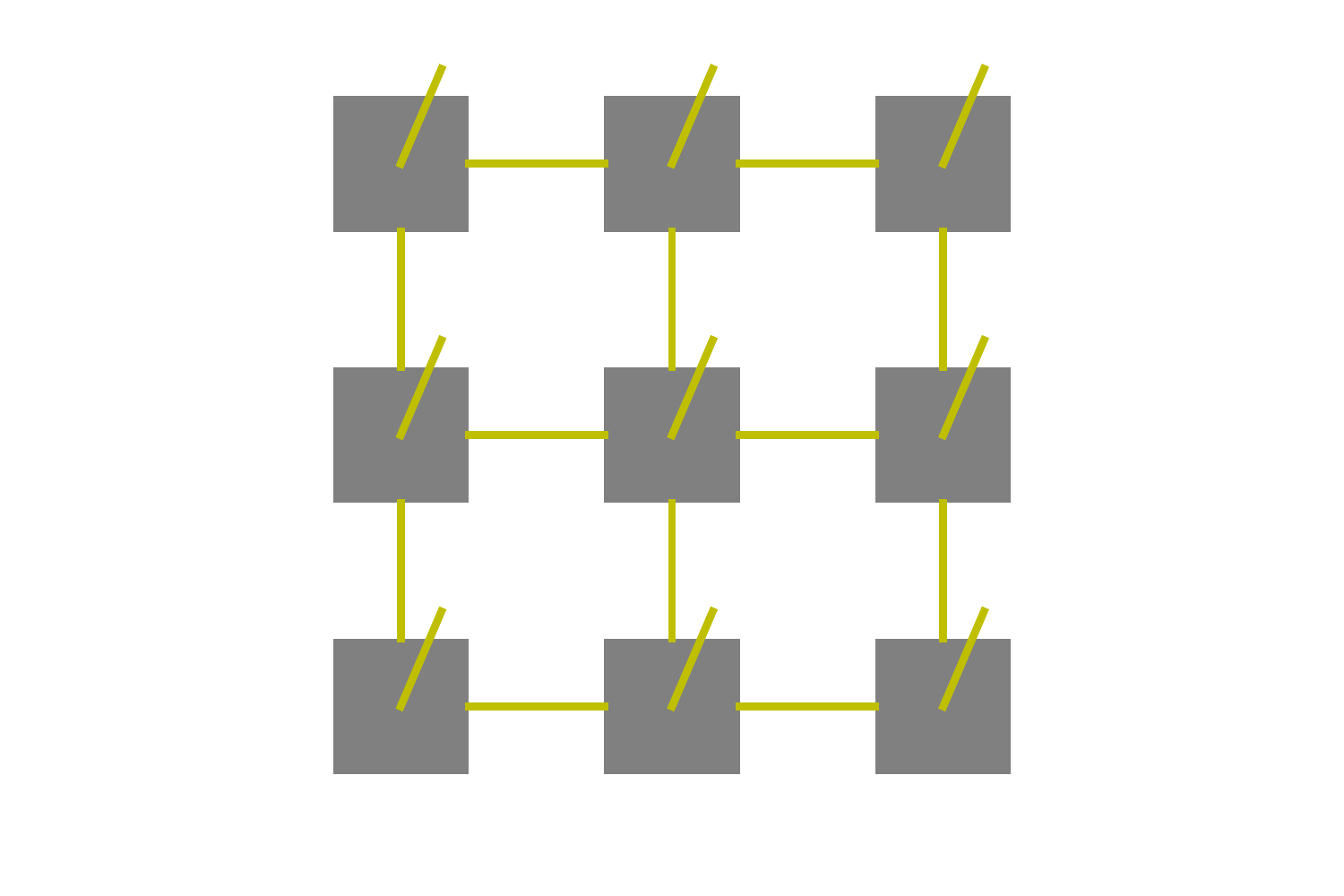}}
    \subfloat[PEPO]{\includegraphics[width=0.45\textwidth]{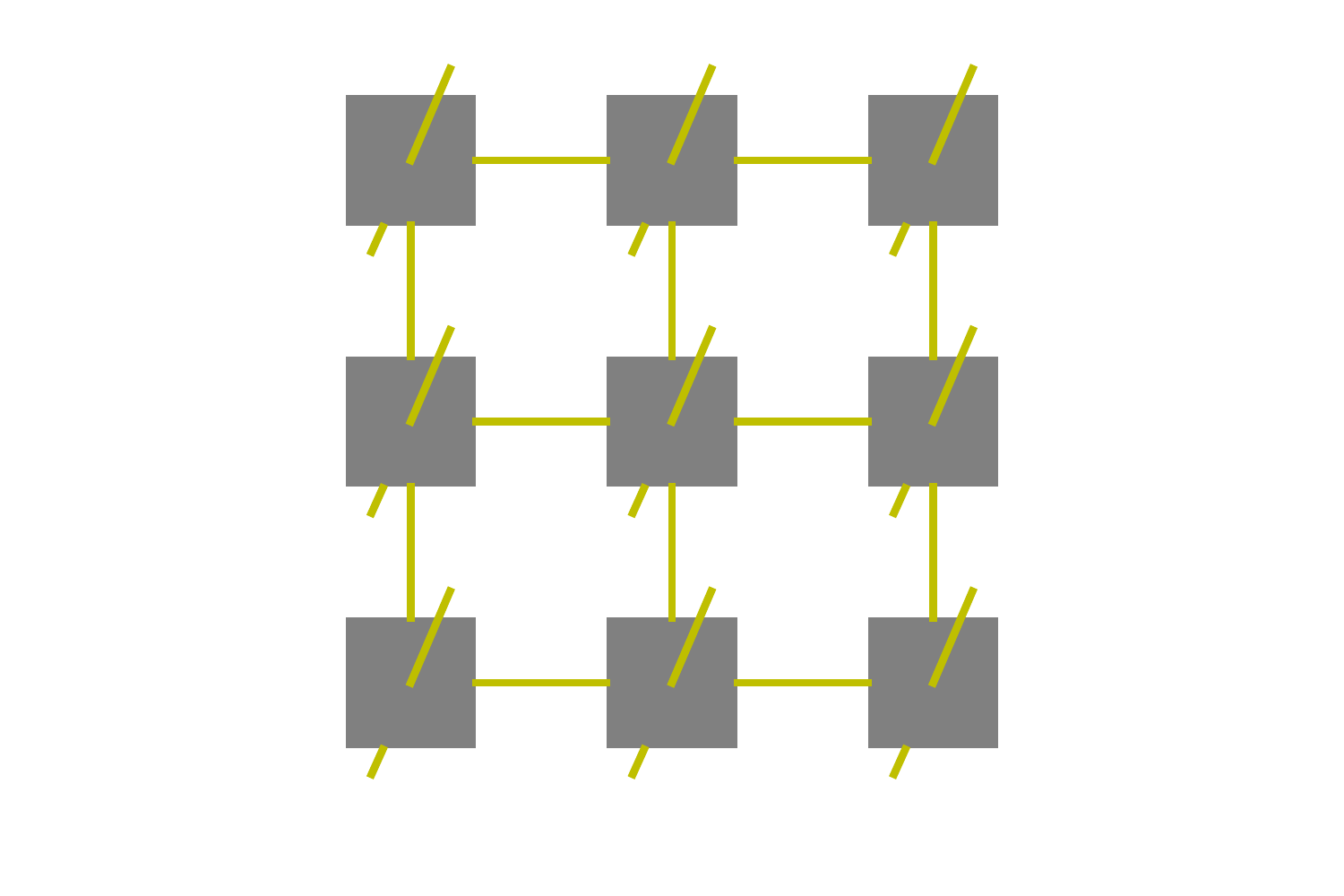}}
    \caption{Graphical representation of a PEPS and a PEPO corresponding to a two-dimensional lattice.}
    \label{fig:peps_pepo}
\end{figure}

The MPS and MPO impose an intrinsically one-dimensional structure on the tensor following the index $1,\ldots,d$. This is a very efficient representation for certain problems defined on a one-dimensional lattice. For problems defined on a two-dimensional lattice and beyond, the PEPS and PEPO often provide a more efficient representation of vectors and linear operators, respectively.  The PEPS/PEPO can be most conveniently represented using the graphical representation (see Fig.~\ref{fig:peps_pepo} (a) and (b)).

In the two-dimensional case, each core tensor in the PEPS/PEPO format has up to 5/6 indices, respectively. Similar to what we have done for MPS/MPO, the (up to 4) contracted indices are called internal indices, while remaining ones are called external indices.

\section{Long-range interaction MPO in 1D systems} \label{sec:long_range_mpo_1d}

In this section we focus on a 1D system consisting of $N$ sites.  The simplest example of constructing a long-range interaction MPO in 1D system is the exponential fitting method \cite{pirvu2010matrix,crosswhite2008finite,frowis2010tensor}. We briefly describe this method used for the Coulomb interaction. First we approximate the inverse-distance by the sum of exponentials
\begin{equation}
\frac{1}{r}\approx \sum_{k=1}^M a_k e^{-\lambda_k r}, \quad 1\leq r\leq N.
\label{eq:exp_fitting}
\end{equation}
Then the tensor corresponding to the Coulomb interaction can be approximated by
\[
\hat{V}\approx \sum_{k=1}^M a_k \sum_{i<j}e^{-\lambda_k (j-i)}\hat{n}_i\hat{n}_j,
\]
where $a_k,\lambda_k\geq 0$. This can be done analytically through the quadrature of an integral representation~\cite{braess2009efficient}, or numerically through a least squares procedure. Following the discussion of the finite state machine in Section~\ref{sec:MPO}, the operator on the right-hand side can be represented by an MPO of bond dimension $M+2$, the MPO tensor of which at site $i$ has the form 
\begin{equation}
\left(
\begin{array}{ccc}
\mathbf{I} & 0 & 0 \\
\mathbf{b}\otimes\mathbf{n}_i & e^{-\mathbf{\Lambda}}\otimes \mathbf{I} & 0 \\
0 & \mathbf{a}^{T}\otimes \mathbf{n}_i & \mathbf{I}
\end{array}
\right),
\label{eq:exp_mpo}
\end{equation}
where $\mathbf{a}=(a_1,\cdots,a_M)^T$, $\mathbf{b}=(e^{-\lambda_1},\cdots,e^{-\lambda_M})^T$, and $\mathbf{\Lambda} = \diag(\lambda_1,\cdots,\lambda_M)$. This is true for $2\leq i\leq N-1$. For the tensors at the beginning and end of the MPO we only need to use the last row and first column of the above matrix respectively. 
Then in order to achieve a target accuracy $\epsilon$ for a system of size $N$, we only need $M=\mathcal{O}(\log(N/\epsilon))$ \cite{beylkin2002numerical,braess2005approximation}.

For the Coulomb interaction, there are two main methods for performing the exponential fitting procedure. 
One way is to represent $1/r$ as an integral, e.g.
\begin{equation*} 
  \frac{1}{r}=\int^{\infty}_0 e^{-\lambda r}\mathrm{d}\lambda,
\end{equation*}
and then approximate the right hand side with a quadrature scheme with $M$ points, such as the Gauss quadrature. This procedure results in pointwise error of $\epsilon$ in a fixed size finite interval $[a,b]$ where $0<a<b<\infty$ with $M$ upper bounded by $\mathcal{O}(\log(1/\epsilon))$ terms~\cite{beylkin2005approximation,braess2009efficient,braess2005approximation}. Another way is to 
solve a nonlinear least squares problem to find the fitting parameters. However, this optimization problem can have many stationary points, making finding the global minimum difficult. We can only start from different reasonable initial guesses to increase the chance of obtaining the global minimum. 


The exponential fitting method is efficient when $\mathbf{V}$ is strictly decaying and translation-invariant.  When either of these two conditions is not satisfied, this method needs some modification. There are methods to deal with the former by introducing complex exponents \cite{pirvu2010matrix}. The generalization to non-translation-invariant systems can be more difficult.

In this work we propose a new perspective based on matrix completion, which
naturally generalizes the exponential fitting method to situations where
$\mathbf{V}$ is neither strictly decaying nor translation-invariant. We define
a rank $M$ matrix $\tilde{\mathbf{V}}$ by
\begin{equation}
\tilde{\mathbf{V}}(i,j) = \sum_{k=1}^M a_k e^{-\lambda_k (j-i)} = \sum_{k=1}^M a_k e^{-\lambda_k j}e^{\lambda_k i}.
\label{eq:exp_decomp}
\end{equation}
The success of the exponential fitting method rests on the fact that 
$\tilde{\mathbf{V}}$ can agree very well with $\mathbf{V}$ in the
upper-triangular part, but can differ with $\mathbf{V}$ \textit{arbitrarily} on the
diagonal part and the lower-triangular part. Indeed, the lower-triangular part of
$\tilde{\mathbf{V}}$ grows exponentially with respect to $(i-j)$, while 
that of $\mathbf{V}$ decays with respect to $(i-j)$ due to the symmetry of 
$\mathbf{V}$. Despite the difference, the tensor
$\sum_{i<j}\tilde{\mathbf{V}}(i,j)\hat{n}_i\hat{n}_j$ very well approximates
$\sum_{i<j}\mathbf{V}(i,j)\hat{n}_i\hat{n}_j$. 

More generally, we
shall demonstrate that it is possible to construct an efficient MPO
representation of $\hat{V}$, if the matrix $\mathbf{V}$ satisfies the following
upper triangular low rank (UTLR) property.
\begin{definition}[Upper-triangular low-rank matrix] A symmetric matrix
  $\mathbf{A}\in \RR^{N\times N}$ is an upper-triangular low-rank (UTLR)
  matrix, if for any $\epsilon>0$, there exists $r\lesssim \log(N/\epsilon)$, and a
  rank-$r$ matrix $\tilde{\mathbf{A}}$, such that for any index sets
  $\mathcal{I}\prec \mathcal{J}$,
  \[
  \norm{\mathbf{A}(\mathcal{I},\mathcal{J})-\tilde{\mathbf{A}}(\mathcal{I},\mathcal{J})}_{2} \le \epsilon
  \norm{\mathbf{A}(\mathcal{I},\mathcal{J})}_{2}.
  \]
  \label{def:UTLR}
\end{definition}

For a UTLR matrix $\mathbf{V}$, we may find its rank-$M$ approximation by solving the following optimization problem
\begin{equation}
\begin{aligned}
\min_{\tilde{\mathbf{V}}} \quad & \|P_\Omega (\tilde{\mathbf{V}})- P_\Omega (\mathbf{V})\|_{F}^2 \\
\mathrm{s.t.} \quad & \mathrm{rank}(\tilde{\mathbf{V}})\leq M,
\end{aligned}
\label{eq:matrix_completion}
\end{equation}
where $\Omega$ denotes the set of indices corresponding to the upper-triangular part of the matrix, and $P_\Omega$ is an restriction operation that returns only the elements in the set $\Omega$.  Eq.~\eqref{eq:matrix_completion} is a matrix completion problem~\cite{candes2009exact}. 

Once the low-rank solution $\tilde{\mathbf{V}}=\mathbf{L}\mathbf{R}^T$ is obtained, where $\mathbf{L}$ and $\mathbf{R}$ are both $N\times M$ matrices, we let $\mathbf{l}_i$, $\mathbf{r}_i$ denote the transpose of the $i$-th row of $\mathbf{L}$ and $\mathbf{R}$ respectively. Then $\tilde{\mathbf{V}}(i,j)=\mathbf{l}_i^T \mathbf{r}_j$. Therefore we can construct an MPO with the following tensor at site $i$ ($2\le i\le N-1$):
\begin{equation}
\left(
\begin{array}{ccc}
\mathbf{I} & 0 & 0 \\
\mathbf{r}_i \otimes \mathbf{n}_i & \mathbf{I}\otimes \mathbf{I} & 0 \\
0 & \mathbf{l}_i^T\otimes \mathbf{n}_i & \mathbf{I}
\end{array}
\right).
\label{eq:mc_mpo}
\end{equation}
Again the core tensor for $i=1$ and $i=N$ are given by the last row and the first column of Eq.~\eqref{eq:mc_mpo}. 
In order to recover the MPO construction in the exponential fitting method \eqref{eq:exp_mpo}, we may define
\[
\mathbf{l}_i = e^{-(N-i)\mathbf{\Lambda}}\mathbf{a}, \quad \mathbf{r}_i = e^{(N-i+1)\mathbf{\Lambda}}\mathbf{b}.
\]
The extra term $e^{N\mathbf{\Lambda}}$ is introduced to keep the notation consistent with a more general case discussed later.

\subsection{An online matrix completion algorithm}

The matrix completion problem (\ref{eq:mc_mpo})  has several features:
\begin{enumerate}
\item The sparsity pattern $\Omega$ is always fixed to be all the upper-triangular entries.
\item We are actually not concerned about restoring the missing data. The missing data only serves to support a low-rank structure, and they do not necessarily need to be explicitly computed. 
\item The matrix completion problem is ill-conditioned, i.e. the ratio between the largest singular value and the smallest nonzero singular value scales exponentially with respect to the system size. Therefore explicit computation of the missing entries can easily lead to an overflow error. 
\end{enumerate}
The last feature requires some explanation. Consider the matrix restored by the exponential fitting procedure as in (\ref{eq:exp_decomp}). The lower left corner of the matrix is $\tilde{\mathbf{V}}(N,1)=\sum_{k}a_k e^{\lambda_k(N-1)}$. Therefore this entry, and nearby entries, grow exponentially with respect to the system size. This also results in an exponential growth of the largest singular value.

For the matrix completion problem, both convex optimization-based approaches~\cite{cai2010singular} and alternating least squares (ALS)-based approaches~\cite{hu2008collaborative, mnih2008probabilistic} require some regularization term. In our case, the regularization term will become dominant even if the regularization parameter is chosen to be very small. Online matrix completion algorithms, which applies SVD compression operations incrementally~\cite{brand2002incremental,balzano2013grouse}, also suffer from the exponential growth of matrix entries in the low-rank approximation. More importantly, the largest singular value, which grows exponentially with system size, also makes the matrix completion problem difficult to solve.

We introduce a modified incremental SVD (ISVD) algorithm with missing data, which avoids the computation of the SVD of the whole matrix and therefore improves the numerical stability. This method is modified from the ISVD algorithm with missing data introduced in \cite{balzano2013grouse, brand2002incremental}. The main modification is that we add a QR factorization step which makes the method stable even for extremely ill-conditioned matrices. We also find that the resulting MPO representation is equivalent to that in \cite{StoudenmireWhite2017}. For clarity, we first introduce how to apply the original ISVD with missing data on the present matrix completion problem in this section and Section~\ref{sec:mc_construct_mpo}. In Section~\ref{sec:practical_alg} we introduce the modified algorithm.





The algorithm proceeds in a row-by-row fashion. Define $\Omega_{p}:=\{p+1,\ldots,N\}$ the set of column indices 
corresponding to the sparsity pattern $\Omega$ in the $p$-th row, and correspondingly $\Omega^c_{p}:=\{1,\ldots, p\}$. Suppose we have already completed the first $p$ rows, and the completed
matrix is $\tilde{\mathbf{V}}_{p}$. This matrix is rank-$M$ and has an SVD
$\tilde{\mathbf{V}}_{p}=\mathbf{U}_{p}\mathbf{S}_{p}\mathbf{W}_{p}^{T}$. Now we want to complete the
first $p+1$ rows denoted by the matrix $\mathbf{\bar{V}}_{p+1}$. Note that 
the first $p$ rows of $\mathbf{\bar{V}}_{p+1}$ are just $\tilde{\mathbf{V}}_p$, and on the $(p+1)$-th row we have   $\mathbf{\bar{V}}(p{+}1,\Omega_{p+1})=\mathbf{V}(p{+}1,\Omega_{p+1})$. Our goal is to choose the unknown entries $\mathbf{\bar{V}}(p{+}1,\Omega_{p+1}^{c})$, 
so that $\mathbf{\bar{V}}_{p+1}$ can also be approximated by a rank-$M$ matrix.
Now assume $\mathbf{\bar{V}}(p{+}1,:)=\mathbf{x}^{T}\mathbf{W}_{p}^{T}+\mathbf{r}^{T}$, where $\mathbf{r}^{T}\mathbf{W}_{p}=0$. If $\mathbf{r}=0$, then $\mathbf{\bar{V}}$ is already
a rank-$M$ matrix. Otherwise
\begin{equation}
  \begin{split}
    \mathbf{\bar{V}}_{p+1} & =\left(\begin{array}{c}
      \tilde{\mathbf{V}}_{p}\\
      \mathbf{\bar{V}}(p{+}1,:)
    \end{array}\right)=\left(\begin{array}{c}
      \mathbf{U}_{p}\mathbf{S}_{p}\mathbf{W}_{p}^{T}\\
      \mathbf{x}^{T}\mathbf{W}_{p}^{T}+\mathbf{r}^{T}
    \end{array}\right)\\
    & =\left(\begin{array}{cc}
      \mathbf{U}_{p}\mathbf{S}_{p} & 0\\
      \mathbf{x}^{T} & \|\mathbf{r}\|
    \end{array}\right)\left(\begin{array}{c}
      \mathbf{W}_{p}^{T}\\
      \mathbf{r}^{T}/\|\mathbf{r}\|
    \end{array}\right)\\
    & =\left(\begin{array}{cc}
      \mathbf{U}_{p} & 0\\
      0 & 1
    \end{array}\right)\left(\begin{array}{cc}
      \mathbf{S}_{p} & 0\\
      \mathbf{x}^{T} & \|\mathbf{r}\|
    \end{array}\right)\left(\begin{array}{c}
      \mathbf{W}_{p}^{T}\\
      \mathbf{r}^{T}/\|\mathbf{r}\|
    \end{array}\right).
  \end{split}
  \label{eqn:Vbar}
\end{equation}
Note that on the third line of Eq.~\eqref{eqn:Vbar}, both the first and the third matrices are orthogonal matrices. Hence the smallest singular value of $\mathbf{\bar{V}}_{p+1}$ is upper bounded by $\|\mathbf{r}\|$. In order to approximate $\mathbf{\bar{V}}_{p+1}$ by a rank-$M$ matrix, we want $\|\mathbf{r}\|$ to be as small as possible. This corresponds to the following optimization problem:
\[
\min_{\mathbf{\bar{V}}(p{+}1,\Omega_{p+1}^{c})}\min_{\mathbf{x}}\|\mathbf{\bar{V}}(p{+}1,:)-\mathbf{x}^{T}\mathbf{W}_{p}^{T}\|^{2}.
\]
It is easy to see that the solution is 
\begin{align*}
& \mathbf{x} =\mathbf{W}_{p}(:,\Omega_{p+1})^{+}\mathbf{\bar{V}}(p{+}1,\Omega_{p+1})^{T}\\
& \mathbf{\bar{V}}(p{+}1,\Omega_{p+1}^{c}) =\mathbf{x}^{T}\mathbf{W}_{p}(:,\Omega_{p+1}^{c}),
\end{align*}
where $\mathbf{A}^{+}$ denotes the Moore\textendash Penrose inverse of $\mathbf{A}$.
Then we obtain the truncated SVD 
\begin{equation}
\left(\begin{array}{cc}
\mathbf{S}_{p} & 0\\
\mathbf{x}^{T} & \|\mathbf{r}\|
\end{array}\right)\approx \mathbf{X}_{p+1}\mathbf{S}_{p+1}\mathbf{Y}_{p+1}^{T}
\label{eq:svd_in_isvd}
\end{equation}
where $\mathbf{S}_{p+1}$ is an $M\times M$ matrix, $\mathbf{X}_{p+1},\mathbf{Y}_{p+1}$ are
column orthogonal matrices of size $(M+1)\times M$. Thus we may approximate $\mathbf{\bar{V}}_{p+1}$
by 
\begin{align*}
\tilde{\mathbf{V}}_{p+1} & =\left[\left(\begin{array}{cc}
\mathbf{U}_{p} & 0\\
0 & 1
\end{array}\right)\mathbf{X}_{p+1}\right]\mathbf{S}_{p+1}\left[\mathbf{Y}_{p+1}^{T}\left(\begin{array}{c}
\mathbf{W}_{p}^{T}\\
\mathbf{r}^{T}/\|\mathbf{r}\|
\end{array}\right)\right].
\end{align*}
Thus letting 
\[
\mathbf{U}_{p+1}=\left(\begin{array}{cc}
\mathbf{U}_{p} & 0\\
0 & 1
\end{array}\right)\mathbf{X}_{p+1},\quad \mathbf{W}_{p+1}=\left(\begin{array}{cc}
\mathbf{W}_{p} & \mathbf{r}/\|\mathbf{r}\|\end{array}\right)\mathbf{Y}_{p+1},
\]
we can proceed to the next row. The $\mathbf{U}_{N-1}$, $\mathbf{S}_{N-1}$ and $\mathbf{W}_{N-1}$
we get in the end gives the SVD of the completed matrix (noting that $\Omega_{N}=\emptyset$).

\subsection{Constructing the MPO}
\label{sec:mc_construct_mpo}

We now apply the above procedure to the coefficient matrix $\mathbf{V}$.  Let $\mathbf{l}_i = \mathbf{U}_{N-1}(i,:)^T$ and $\mathbf{r}_i = \mathbf{S}_{N-1}\mathbf{W}_{N-1}^T(:,i)$, and use (\ref{eq:mc_mpo}), we get an MPO that represents the interaction $\hat{V}$.  However, this procedure is not numerically stable. Let us consider the Coulomb interaction approximated with exponential fitting method. Then the ratio $\|\mathbf{l}_{N-1}\|/\|\mathbf{l}_1\|$ grows exponentially with respect to the system size, and so is $\|\mathbf{r}_2\|/\|\mathbf{r}_{N}\|$.

To solve this problem, we will make use of the interior part of the MPO tensor. Write 
\[
\mathbf{X}_{p}=\left(\begin{array}{c}
\tilde{\mathbf{X}}_{p}\\
\mathbf{a}_{p}^{T}
\end{array}\right),
\]
then we have
\begin{align*}
\mathbf{U}_{p} & =\left(\begin{array}{cc}
\mathbf{U}_{p-1} & 0\\
0 & 1
\end{array}\right)\left(\begin{array}{c}
\tilde{\mathbf{X}}_{p}\\
\mathbf{a}_{p}^{T}
\end{array}\right)=\left(\begin{array}{c}
\mathbf{U}_{p-1}\tilde{\mathbf{X}}_{p}\\
\mathbf{a}_{p}^{T}
\end{array}\right)\\
 & =\left(\begin{array}{c}
\mathbf{U}_{p-2}\tilde{\mathbf{X}}_{p-1}\tilde{\mathbf{X}}_{p}\\
\mathbf{a}_{p-1}^{T}\tilde{\mathbf{X}}_{p}\\
\mathbf{a}_{p}^{T}
\end{array}\right)=\cdots
\end{align*}
Thus
\[
\mathbf{l}_i^T = \mathbf{U}_{N-1}(i,:) = \mathbf{a}_i^T\tilde{\mathbf{X}}_{i+1}\tilde{\mathbf{X}}_{i+2}\cdots\tilde{\mathbf{X}}_{N-1}
\]
where $\mathbf{a}_{1}=\mathbf{U}_{1}=(1)$. 
Therefore we have
\[
\tilde{\mathbf{V}}(i,j)=\mathbf{l}_i^T\mathbf{r}_j=\mathbf{a}_i^T\tilde{\mathbf{X}}_{i+1}\tilde{\mathbf{X}}_{i+2}\cdots\tilde{\mathbf{X}}_{N-1} \mathbf{r}_j.
\]
Now define 
\begin{equation}
\mathbf{b}_{j}=\tilde{\mathbf{X}}_{j}\tilde{\mathbf{X}}_{j+1}\cdots\tilde{\mathbf{X}}_{N-1}\mathbf{r}_j,
\label{eqn:bChoice1}
\end{equation}
we then have
\begin{equation}
\tilde{\mathbf{V}}(i,j)=\mathbf{a}_{i}^{T}\tilde{\mathbf{X}}_{i+1}\tilde{\mathbf{X}}_{i+2}\cdots\tilde{\mathbf{X}}_{j-1}\mathbf{b}_{j}.
\label{eq:isvd_approx}
\end{equation}
Therefore we can construct an MPO whose tensors are of the form
\begin{equation}
\left(\begin{array}{ccc}
\mathbf{I} & 0 & 0\\
\mathbf{b}_{i}\otimes\mathbf{n}_{i} & \tilde{\mathbf{X}}_{i}\otimes\mathbf{I} & 0\\
0 & \mathbf{a}_{i}^{T}\otimes\mathbf{n}_{i} & \mathbf{I}
\end{array}\right)\label{eq:mpo_tensor_mc}
\end{equation}
to represent $\sum_{i<j}\tilde{\mathbf{V}}(i,j)\hat{n}_{i}\hat{n}_{j}$. This is analogous to (\ref{eq:exp_mpo}) in the exponential fitting. The algorithm is summarized in Algorithm \ref{alg:mpo_mc}. Due to the fact that 
$\Omega_{p+1}\subset\Omega_p$, as a practical algorithm we do not need to keep track of $\mathbf{\bar{V}}_p$ or $\tilde{\mathbf{V}}_p$, but only $\mathbf{W}_p(:,\Omega_p)$. 

\begin{algorithm}[H]
\caption{Constructing the MPO representation via matrix completion}
\label{alg:mpo_mc}
\begin{algorithmic}[1]
\REQUIRE{$\mathbf{V}\in\mathbb{R}^{N\times N},M$}
\STATE{$\mathbf{S}_1 \leftarrow (\|\mathbf{V}(1,2{:}N)\|)$} 
\STATE{$\mathbf{W}_1 \leftarrow (\mathbf{V}(1,2{:}N)^T/\|\mathbf{V}(1,2{:}N)\|)\in\mathbb{R}^{(N-1)\times 1}$}
\STATE{$\mathbf{a}_1\leftarrow (1)$}
\STATE{$\mathbf{b}_2 \leftarrow (\mathbf{V}(1,2)) $}
\FOR{$p=1{:}N-2$}

\STATE{$\mathbf{x}\leftarrow\mathrm{argmin}_{\mathbf{x}}\|\mathbf{W}_p(2{:}N{-}p,:)\mathbf{x}-\mathbf{V}(p{+}1,p{+}2{:}N)^T\|$}
\STATE{$\mathbf{r} \leftarrow \mathbf{V}(p{+}1,p{+}2{:}N)^T -  \mathbf{W}_p(2{:}N{-}p,:)\mathbf{x}$}
\IF{$\|r\|>\mathrm{tol}$}
\STATE{$\mathbf{X}_{p+1},\mathbf{S}_{p+1},\mathbf{Y}_{p+1}\leftarrow\mathrm{SVD\ of\ }
\left( \begin{array}{cc}
\mathbf{S}_p & 0 \\
\mathbf{x}^T & \|\mathbf{r}\|
\end{array} \right)$} \COMMENT{keep $M$ singular values}

\STATE{$\mathbf{W}_{p+1}\leftarrow\left(\begin{array}{cc}
\mathbf{W}_p(2{:}N{-}p,:) & \mathbf{r}/\|\mathbf{r}\|
\end{array}\right)\mathbf{Y}_{p+1}$} 

\ELSE
\STATE{$\mathbf{X}_{p+1},\mathbf{S}_{p+1},\mathbf{Y}_{p+1}\leftarrow\mathrm{SVD\ of\ }
\left( \begin{array}{c}
\mathbf{S}_p  \\
\mathbf{x}^T
\end{array} \right)$} \COMMENT{keep $M$ singular values}

\STATE{$\mathbf{W}_{p+1}\leftarrow 
\mathbf{W}_p(2{:}N{-}p,:)
 \mathbf{Y}_{p+1}$}

\ENDIF

\STATE{$\mathbf{b}_{p+2}\leftarrow \mathbf{S}_{p+1}\mathbf{W}_{p+1}^T(:,1)$}
\IF{$p\neq N-1$}
\STATE{$\mathbf{a}_{p+1}\leftarrow\mathbf{X}_{M+1}(p{+}1,:)^T$, $\tilde{\mathbf{X}}_{p+1}\leftarrow\mathbf{X}_{p+1}(1{:}M,:)$}
\ENDIF

\ENDFOR
\ENSURE{$\mathbf{a}_i$, $\mathbf{b}_j$, $\tilde{\mathbf{X}}_k$ for $1 \leq i \leq N-1$, $2\leq j\leq N$, $2\leq k\leq N-1 $, and MPO according to (\ref{eq:mpo_tensor_mc}).}
\end{algorithmic}
\end{algorithm}

Note that Eq.~\eqref{eqn:bChoice1} is not the only possible choice for $\mathbf{b}_j$. If the matrix $\mathbf{V}$ is exactly of rank $M$, then 
\[
\mathbf{b}_j = \tilde{\mathbf{X}}_j\mathbf{S}_j\mathbf{W}_j^T(:,j)=\mathbf{S}_{j-1}\mathbf{W}_{j-1}^T(:,j).
\]
This relation becomes approximately correct for the case when $\mathbf{V}$ is numerically low-rank. Therefore we may set 
\begin{equation}
  \mathbf{b}_j=\mathbf{S}_{j-1}\mathbf{W}_{j-1}^T(:,j).
  \label{eqn:bChoice2}
\end{equation}

We argue that Eq.~\eqref{eqn:bChoice2} is in fact a better choice. Just like we can obtain $\mathbf{a}_i$ without computing $\mathbf{l}_i$ that grows exponentially with the system size, we want to do the same with $\mathbf{b}_j$ so that it does not rely on computing $\mathbf{r}_j$ either. And this is achieved by our new choice for the value of $\mathbf{b}_j$. Also, the new $\mathbf{b}_j$ can be expected to give a more accurate approximation than the original one, because for an element of $\mathbf{V}$ on the $p$-th column, the error of the approximation using this $\mathbf{b}_j$ only comes from the first $p$ SVDs, and does not depend on any future steps.

Furthermore, using the fact that $\Omega_{p+1}\subset\Omega_p$, we do not need to explicitly fill out the missing data either, thereby avoiding computing and storing the lower triangular part of the matrix $\mathbf{\tilde{V}}_p$. We can avoid computing the missing part $\mathbf{\bar{V}}(p{+}1,\Omega_{p+1}^{c})$ for each row as well.  The pseudocode for the resulting algorithm is presented in Algorithm \ref{alg:mpo_mc}. Note that there is some abuse of notation involved in the algorithm, as we use $\mathbf{W}_p$ to denote the last $N-p$ rows of the $\mathbf{W}_p$ matrix we described previously.

\subsection{A modified practical algorithm}
\label{sec:practical_alg}

\begin{algorithm}
\caption{A modified ISVD method for robust construction of the MPO representation}
\label{alg:mpo_mc_modified}
\begin{algorithmic}[1]
\REQUIRE{$\mathbf{V}\in\mathbb{R}^{N\times N},M$}
\STATE{$\mathbf{S}_1 \leftarrow (\|\mathbf{V}(1,2{:}N)\|)$}
\STATE{$\mathbf{W}_1 \leftarrow (\mathbf{V}(1,2{:}N)^T/\|\mathbf{V}(1,2{:}N)\|)\in\mathbb{R}^{(N-1)\times 1}$}
\STATE{$\mathbf{a}_1\leftarrow (1)$}
\STATE{$\mathbf{b}_2 \leftarrow (\mathbf{V}(1,2)) $}
\FOR{$p=1{:}N-2$}

\STATE{$\mathbf{Q}, \mathbf{R} \leftarrow$ QR factorization of $\mathbf{W}_p(2{:}N{-}p,:)$}
\STATE{$\mathbf{x}\leftarrow\mathrm{argmin}_{\mathbf{x}}\|\mathbf{W}_p(2{:}N{-}p,:)\mathbf{x}-\mathbf{V}(p{+}1,p{+}2{:}N)^T\|$}
\STATE{$\mathbf{r} \leftarrow \mathbf{V}(p{+}1,p{+}2{:}N)^T -  \mathbf{W}_p(2{:}N{-}p,:)\mathbf{x}$}
\IF{$\|r\|>\mathrm{tol}$}
\STATE{$\mathbf{X}_{p+1},\mathbf{S}_{p+1},\mathbf{Y}_{p+1}\leftarrow\mathrm{SVD\ of\ }
\left( \begin{array}{cc}
\mathbf{S}_p \mathbf{R}^T & 0 \\
\mathbf{x}^T \mathbf{R}^T & \|\mathbf{r}\|
\end{array} \right)$} \COMMENT{keep $M$ singular values}

\STATE{$\mathbf{W}_{p+1}\leftarrow\left(\begin{array}{cc}
\mathbf{Q} & \mathbf{r}/\|\mathbf{r}\|
\end{array}\right)\mathbf{Y}_{p+1}$}

\ELSE
\STATE{$\mathbf{X}_{p+1},\mathbf{S}_{p+1},\mathbf{Y}_{p+1}\leftarrow\mathrm{SVD\ of\ }
\left( \begin{array}{c}
\mathbf{S}_p \mathbf{R}^T  \\
\mathbf{x}^T \mathbf{R}^T
\end{array} \right)$} \COMMENT{keep $M$ singular values}

\STATE{$\mathbf{W}_{p+1}\leftarrow 
\mathbf{Q}
 \mathbf{Y}_{p+1}$}

\ENDIF
\STATE{$\mathbf{b}_{p+2}\leftarrow \mathbf{S}_{p+1}\mathbf{W}_{p+1}^T(:,1)$}
\IF{$p\neq N-1$}
\STATE{$\mathbf{a}_{p+1}\leftarrow\mathbf{X}_{M+1}(p{+}1,:)^T$, $\tilde{\mathbf{X}}_{p+1}\leftarrow\mathbf{X}_{p+1}(1{:}M,:)$}
\ENDIF

\ENDFOR
\ENSURE{$\mathbf{a}_i$, $\mathbf{b}_j$, $\tilde{\mathbf{X}}_k$ for $1 \leq i \leq N-1$, $2\leq j\leq N$, $2\leq k\leq N-1 $, and MPO according to (\ref{eq:mpo_tensor_mc}).}
\end{algorithmic}
\end{algorithm}

Despite that Algorithm~\ref{alg:mpo_mc} computes $\mathbf{a},\mathbf{b},\tilde{\mathbf{X}}$ directly, in practice it can still become unstable when $N$ becomes large. This is because the singular values of $\tilde{\mathbf{V}}$ grows exponentially with respect to $N$.  Then in Algorithm~\ref{alg:mpo_mc},  the SVD step (\ref{eq:svd_in_isvd}) also increasingly ill-conditioned as $p$ increases.
\begin{figure}[H]
    \centering
    \includegraphics[width=0.35\textwidth]{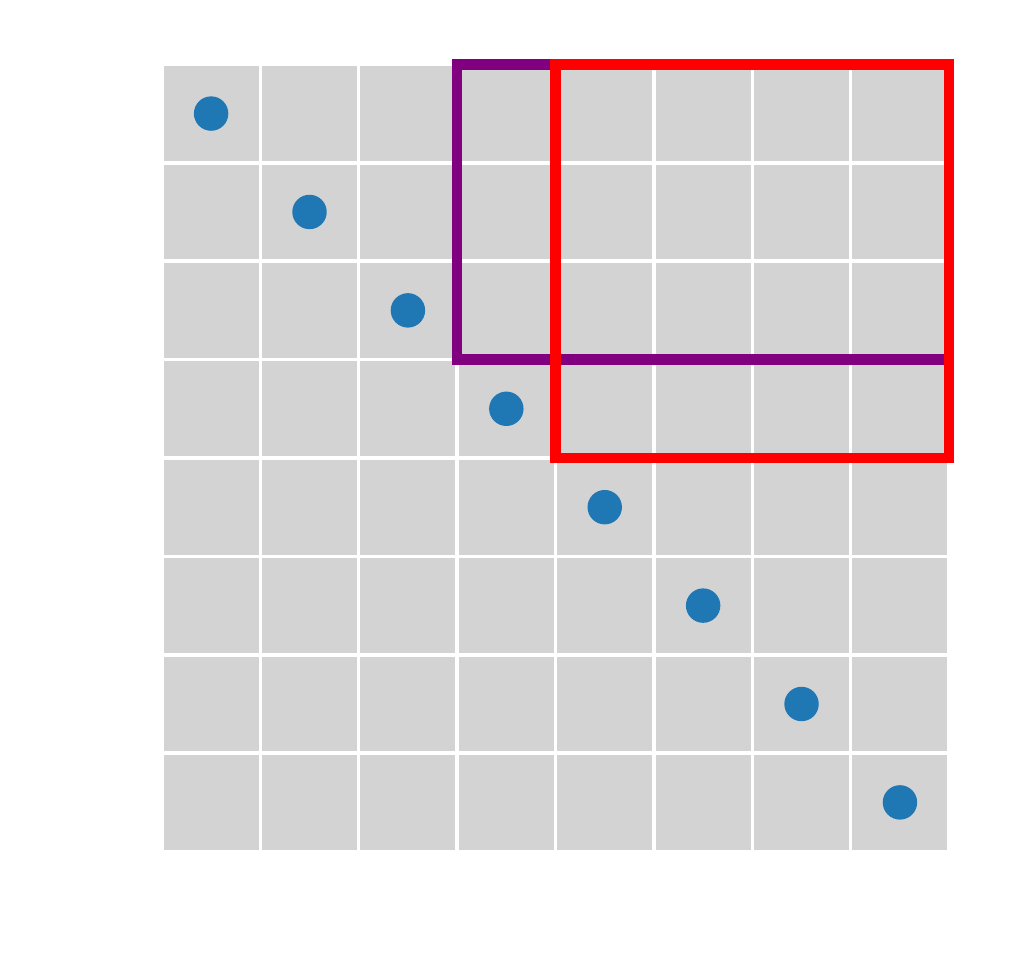}
    \caption{The purple and red rectangles are matrix blocks $\mathbf{V}(1{:}p,\Omega_p)$ with $p=3$ and 4, for which we are going to compute approximate truncated SVDs in Algorithm \ref{alg:mpo_mc_modified}. Diagonal entries are marked with blue dots.}
    \label{fig:isvd_blocks}
\end{figure}

We now further modify the ISVD algorithm as follows. Instead of trying to approximate the SVD of $\mathbf{V}(1{:}p,:)$ for each $p$ as is done in the previous sections, we compute the approximate SVD for each off-diagonal block $\mathbf{V}(1{:}p,\Omega_p)=\mathbf{V}(1{:}p,p{+}1{:}N)$, which avoids the exponential growth of singular values. This can be implemented by introducing only an extra QR factorization step.

Denoting the approximate SVD obtained for $\mathbf{V}(1{:}p,p{+}1{:}N)$ as $\mathbf{U}_p\mathbf{S}_p\mathbf{W}_p^T$ (note this $\mathbf{W}_p$ is different from the one defined previously), we perform a QR factorization on $\mathbf{W}_p(2{:}N{-}p,:)=\mathbf{QR}$. Then we have
\begin{equation*}
    \left(
    \begin{array}{c}
         \mathbf{U}_p\mathbf{S}_p[\mathbf{W}_p(2{:}N{-}p,:)]^T  \\
         \mathbf{V}(p{+}1,p{+}2{:}N) 
    \end{array}
    \right)
    =\left(\begin{array}{cc}
    \mathbf{U}_{p} & 0\\
    0 & 1
    \end{array}\right)\left(\begin{array}{cc}
    \mathbf{S}_{p}\mathbf{R}^T & 0\\
    \mathbf{x}^{T}\mathbf{R}^T & \|\mathbf{r}\|
    \end{array}\right)\left(\begin{array}{c}
    \mathbf{Q}^T\\
    \mathbf{r}^{T}/\|\mathbf{r}\|
    \end{array}\right)
\end{equation*}
Then we perform SVD on the matrix in the middle of the right-hand side, and use this to obtain $\mathbf{U}_{p+1}$, $\mathbf{S}_{p+1}$, $\mathbf{W}_{p+1}$. The pseudocode for the algorithm is described in Algorithm \ref{alg:mpo_mc_modified}. The extra QR factorization step is in Line 6 in Alg.~\ref{alg:mpo_mc_modified}  We remark that the MPO representation obtained from Algorithm~\ref{alg:mpo_mc_modified} is equivalent to that obtained in~\cite{StoudenmireWhite2017}. 


Below we demonstrate that one can reconstruct the UTLR low-rank factors from the $\mathbf{a}_i$, $\tilde{\mathbf{X}}_i$, $\mathbf{b}_i$ obtained from Algorithm \ref{alg:mpo_mc_modified}.
First, note that not all $\tilde{\mathbf{X}}_i$'s are square matrices, though all $\tilde{\mathbf{X}}_i$'s are full-rank matrices.
Each $\tilde{\mathbf{X}}_i$ for $i=2,3,\cdots,M$ is an $(i-1) \times i$ matrix, and each $\tilde{\mathbf{X}}_i$ for $i=N-M+1,N-M+2,\cdots,N-1$ is an $(N-i+1)\times (N-i)$ matrix. Each $\tilde{\mathbf{X}}_i$'s in the middle is an $M\times M$ matrix. Therefore the first $M-1$ $\tilde{\mathbf{X}}_i$'s have right inverses, the last $M-1$ $\tilde{\mathbf{X}}_i$'s have left inverses, and every $\tilde{\mathbf{X}}_i$ in the middle has an inverse. We denote these left or right inverses as $\tilde{\mathbf{X}}_i^{+}$ 

Now we define 
\begin{align*}
    \mathbf{l}_i^T &= \mathbf{a}_i^T \tilde{\mathbf{X}}_{i+1}\tilde{\mathbf{X}}_{i+2}\cdots \tilde{\mathbf{X}}_{N-M},\quad i=1,2,\cdots,N-M \\
    \mathbf{l}_i^T &= \mathbf{a}_i^T \tilde{\mathbf{X}}_{i}^{+}\tilde{\mathbf{X}}_{i-1}^{+}\cdots \tilde{\mathbf{X}}_{N-M+1}^{+}\quad i=N-M+1,N-M+2,\cdots,N-1 \\
    \mathbf{r}_j &= \tilde{\mathbf{X}}^{+}_{N-M} \tilde{\mathbf{X}}^{+}_{N-M-1} \cdots \tilde{\mathbf{X}}^{+}_{j} \mathbf{b}_j \quad j=2,3,\cdots,N-M \\
    \mathbf{r}_j &= \tilde{\mathbf{X}}_{N-M+1}\tilde{\mathbf{X}}_{N-M+2}\cdots \tilde{\mathbf{X}}_{j-1} \mathbf{b}_j \quad j=N-M+1,N-M+2,\cdots,N
\end{align*}
and it can be checked that $\mathbf{l}_i^T \mathbf{r}_j=\mathbf{a}_i^T \tilde{\mathbf{X}}_{i+1}\cdots\tilde{\mathbf{X}}_{j-1}\mathbf{b}_j$ for any $1\leq i<j\leq N$. Therefore we have a upper-triangular low-rank approximation for the matrix $\mathbf{V}$.

\section{Representing long-range interaction via the hierarchical low-rank format} \label{sec:long_range_hierarchical}

In this section we present an alternative method to construct tensor network operators with long-range pairwise interactions using the hierarchical low-rank format. The advantage of this method is that it can be naturally generalized to higher dimensional tensors represented by PEPOs. Compared to the CF-PEPO approach~\cite{o2018efficient} which relies on a fitting procedure and therefore is difficult to establish the \textit{a priori} error bound, we demonstrate that our algorithm yields a sum of $\Or(\log(N)\log(N/\epsilon))$ PEPOs whose bond dimension is bounded by a constant for the Coulomb interaction.

We study the case when $\mathbf{V}$ can be represented as a hierarchical off-diagonal low-rank (HODLR) matrix and a $\mathcal{H}$-matrix respectively. Furthermore, the hierarchical low-rank representation is not restricted to translation-invariant operators.


\subsection{Rank-one MPOs and PEPOs}
\label{sec:rank_one_mpos}

We first introduce some terminologies with respect to the MPOs and PEPOs we are going to use in this section. There are mainly two types of operators we are going to use extensively, which can be efficiently represented by MPOs and PEPOs. Consider a rank-one operator in 1D
\[
\sum_{i<j}\hat{a}_i\hat{b}_j,
\label{eq:rank_one}
\]
where $\hat{a}_i$ and $\hat{b}_j$ are local operators defined on sites $i$ and $j$, and as a result they commute with each other. 
This operator can be represented by an MPO with bond dimension 3, since we can treat it as a special case of the interaction in (\ref{eq:exp_interaction}) in which $\lambda=0$. 

\begin{figure}
    \centering
    \includegraphics[width=0.6\textwidth]{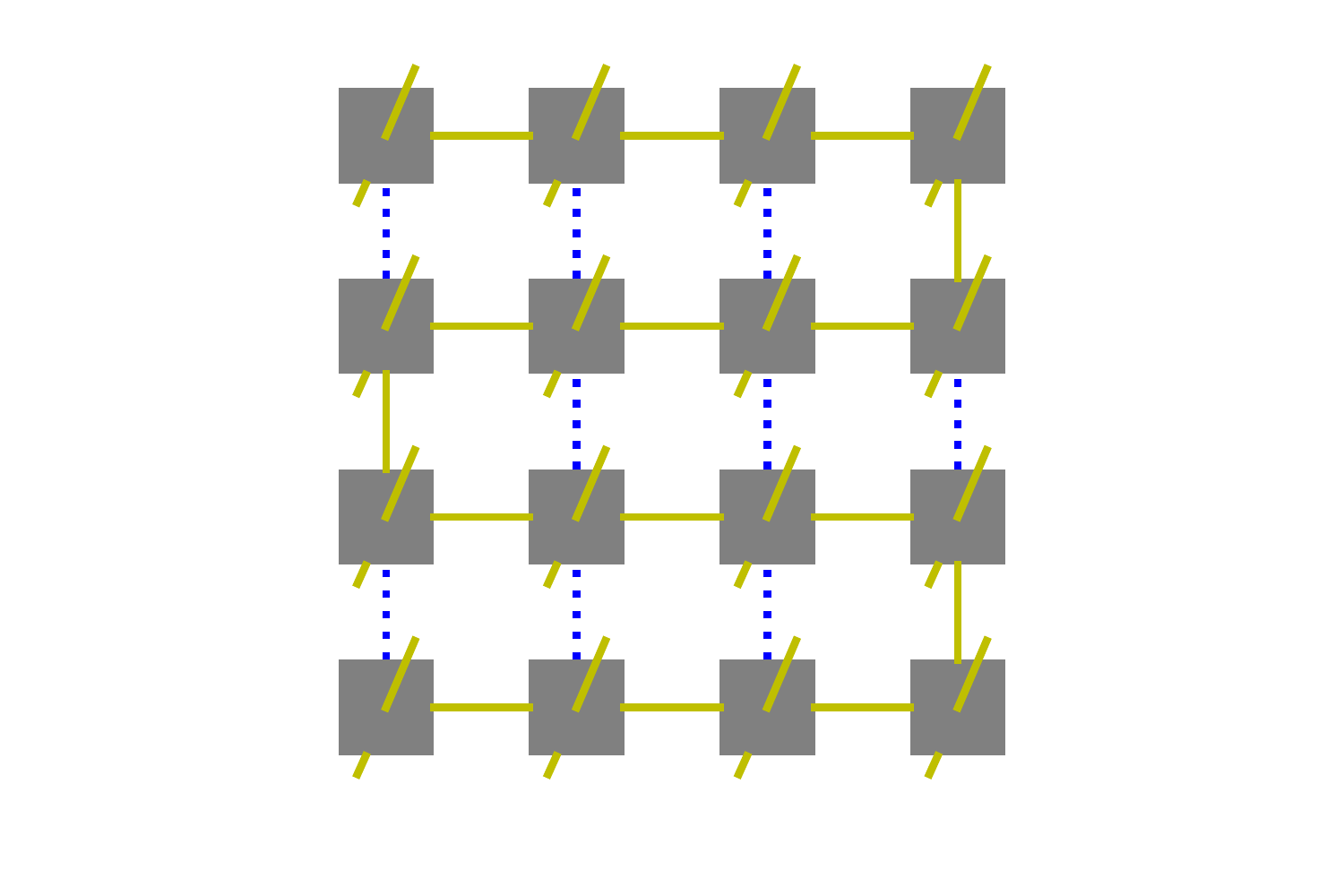}
    \caption{A PEPO constructed from the snake-shaped MPO by adding bonds with bond dimension 1 (blue dotted lines).}
    \label{fig:snake_shaped_pepo}
\end{figure}

The notion of rank-one operators can be easily generalized to a 2D system, for operators in the form
\[
\sum_{\mathbf{i}\prec \mathbf{j}}\hat{a}_{\mathbf{i}}\hat{b}_{\mathbf{j}},
\label{eq:rank_one_2d}
\]
where $\prec$ is some order assigned to the system. This can be represented by a snake-shaped MPO \cite{o2018efficient} with bond dimension 3. The coefficient matrix is a UTLR matrix of rank 1.
We add some bonds with bond dimension 1 to make it a PEPO. The bond dimension of the PEPO is still 3. A graphical illustration of the PEPO is provided in Fig.~\ref{fig:snake_shaped_pepo}.

Sometimes the operator acts trivially everywhere except for a given region (described by an index set $\mathcal{I}$) in the system, we may order the sites in such a way that all sites in $\mathcal{I}$ are labeled before the rest of the sites. Then the operator is $\hat{O}=\hat{O}_\mathcal{I}\otimes \hat{I}_{\mathcal{I}^c}$. In such cases we call the set $\mathcal{I}$ the \textit{support} of operator $\hat{O}$.  Now suppose $\hat{O}_1,\hat{O}_2,\cdots,\hat{O}_m$ are $m$ rank-one operators in a 1D system. Integer intervals $\mathcal{I}_1,\mathcal{I}_2,\cdots,\mathcal{I}_m$ are the supports of each operator respectively. If $\mathcal{I}_i\cap \mathcal{I}_j=\emptyset$, then the operators $\hat{O}_{i},\hat{O}_j$ are \textit{non-overlapping}. The sum of these operators can be expressed as an MPO with bond dimension 5. We give the MPO rules in Appendix~\ref{sec:appendix_mpo}. 

This notion of two operators being non-overlapping can be easily extended to 2D systems. Instead of intervals the supports are replaced by boxes. The sum of these operators can be expressed as a PEPO with bond dimension 5. We will also give the PEPO rules in Appendix~\ref{sec:appendix_pepo}. As mentioned before, in 2D we need to define an order $\prec$ for the rank-one operators. However when we consider a sum of non-overlapping rank-one operators, the order only needs to be defined locally within each support, i.e. we do not need to impose a global ordering valid for all rank-one operators.


\subsection{Hierarchical low-rank matrix}

In this section we introduce the hierarchical low-rank matrix, and its use to construct MPO and PEPO representation of tensor network operators with long-range interactions. Consider the 1D system first. We denote by
$\mathcal{I}_0=\{0,1,\cdots,N-1\}$ the set of all sites in the system. Note that we use zero-based indexing
for sites in the system from now on.  We divide the domain
hierarchically. At the first level, we divide the system into two equal
parts $\mathcal{I}_{1;0}=\{0,1,\cdots,N/2-1\}$ and
$\mathcal{I}_{1;1}=\{N/2,N/2+1,\cdots,N-1\}$. At the second level there
will be four such intervals, and third level eight intervals, and so on.
In general we define
$\mathcal{I}_{\ell;\alpha}=\{\alpha 2^{N-\ell},\alpha 2^{N-\ell}+1,\cdots,(\alpha+1)2^{N-\ell}-1\}$, where $0\le \alpha \le 2^{\ell}-1$. 
There are in total $L=\log_2(N)$ levels.
Each interval has a neighbor list $\mathrm{NL}(\mathcal{I})$ containing all the other intervals on the same level that are neighbors of $\mathcal{I}$.

The intervals defined in this way give rise to a tree structure. When an interval $\mathcal{I}_{\ell;\alpha}\subset\mathcal{I}_{\ell-1;\beta}$, then we say $\mathcal{I}_{\ell;\alpha}$ is a child of $\mathcal{I}_{\ell-1;\beta}$, or equivalently $\mathcal{I}_{\ell-1;\beta}$ is a parent of $\mathcal{I}_{\ell;\alpha}$. When two intervals are children of the same parent we call them siblings.

A key component of the definition of the hierarchical low-rank matrix is the \textit{interaction list}, which we denote by $\mathrm{IL}(\mathcal{I})$ for each interval $\mathcal{I}$. If $\mathbf{A}$ is a hierarchical low-rank matrix, and $\mathcal{J}\in \mathrm{IL}(\mathcal{I})$, then $\mathbf{A}(\mathcal{I},\mathcal{J})$ can be approximated by a low-rank matrix.  The two hierarchical low-rank matrix formats considered in this paper, the HODLR and the $\mathcal{H}$-matrix formats, differ only in the choice of interaction lists. Given a set of interaction lists, we are ready to define the hierarchical low-rank matrix:

\begin{definition}[Hierarchical low-rank matrix] A symmetric matrix $\mathbf{A}\in
  \RR^{N\times N}$ is a hierarchical low-rank matrix, if for any index set $\mathcal{I}$ and any index
  set $\mathcal{J}\in\mathrm{IL}(\mathcal{I})$, there exists a rank
  $r\lesssim \log(N/\epsilon)$ matrix $\tilde{\mathbf{A}}_{\mathcal{I},\mathcal{J}}$ such that
  \[
  \norm{\mathbf{A}(\mathcal{I},\mathcal{J})-\tilde{\mathbf{A}}_{\mathcal{I},\mathcal{J}}}_{2} \le \epsilon
  \norm{\tilde{\mathbf{A}}(\mathcal{I},\mathcal{J})}_2.
  \]
  \label{def:Hierarchical}
\end{definition}

From Definition~\ref{def:Hierarchical}, we immediately recognize
that a UTLR matrix must be a hierarchical low-rank matrix. To see this,
we simply pick $\tilde{\mathbf{A}}_{\mathcal{I},\mathcal{J}}$ to be the
$(\mathcal{I},\mathcal{J})$-th matrix block of $\tilde{\mathbf{A}}$ given
by the UTLR format. The reverse statement is not true, since the matrix
blocks $\tilde{\mathbf{A}}_{\mathcal{I},\mathcal{J}}$ in a hierarchical
low-rank matrix may not be related at all to each other.

The concept above can be naturally extends to two-dimensional and higher dimensional systems. Instead of intervals, the interaction lists are defined using boxes. 
We will first use hierarchical low-rank structure in 1D to build MPOs, and then
discuss its generalization to 2D for constructing PEPOs.

\subsection{HODLR format}
\label{sec:hodlr}

\begin{figure}[t]
\centering
\subfloat[HODLR]{\includegraphics[width=0.3\textwidth]{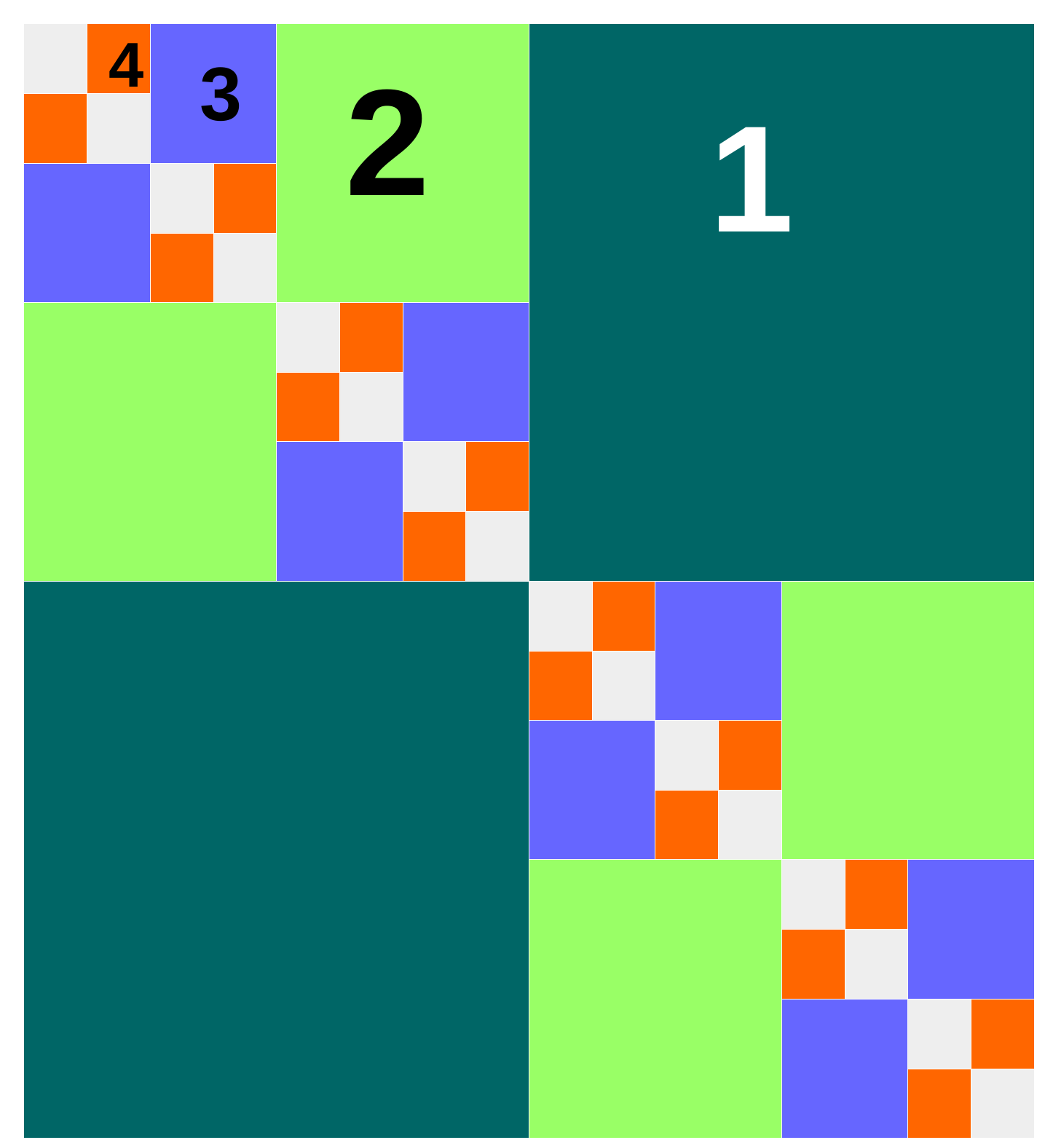}}
\subfloat[$\mathcal{H}$-matrix]{\includegraphics[width=0.3\textwidth]{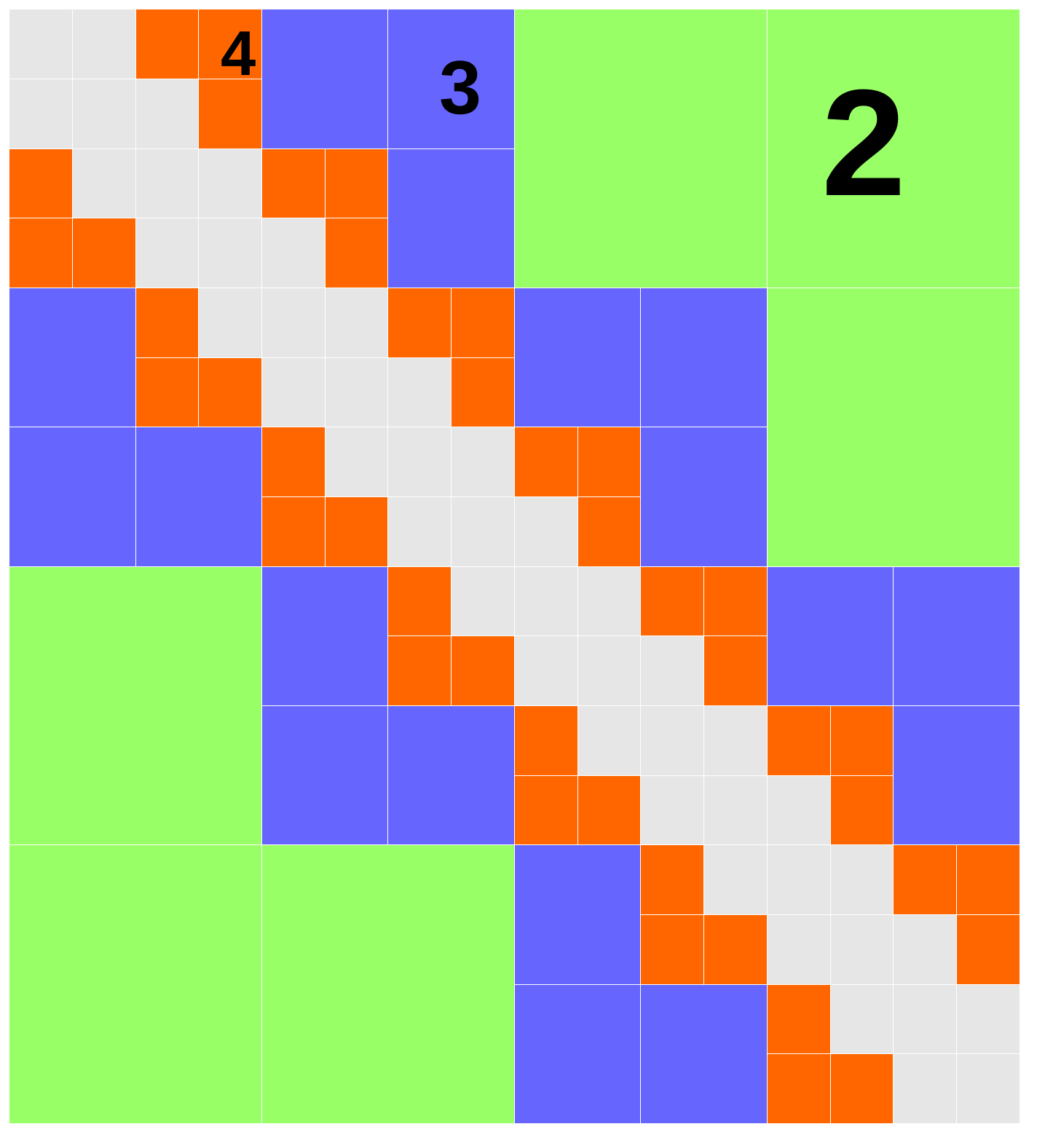}}
\subfloat[Type-0 and type-1 interactions]{\includegraphics[width=0.3\textwidth]{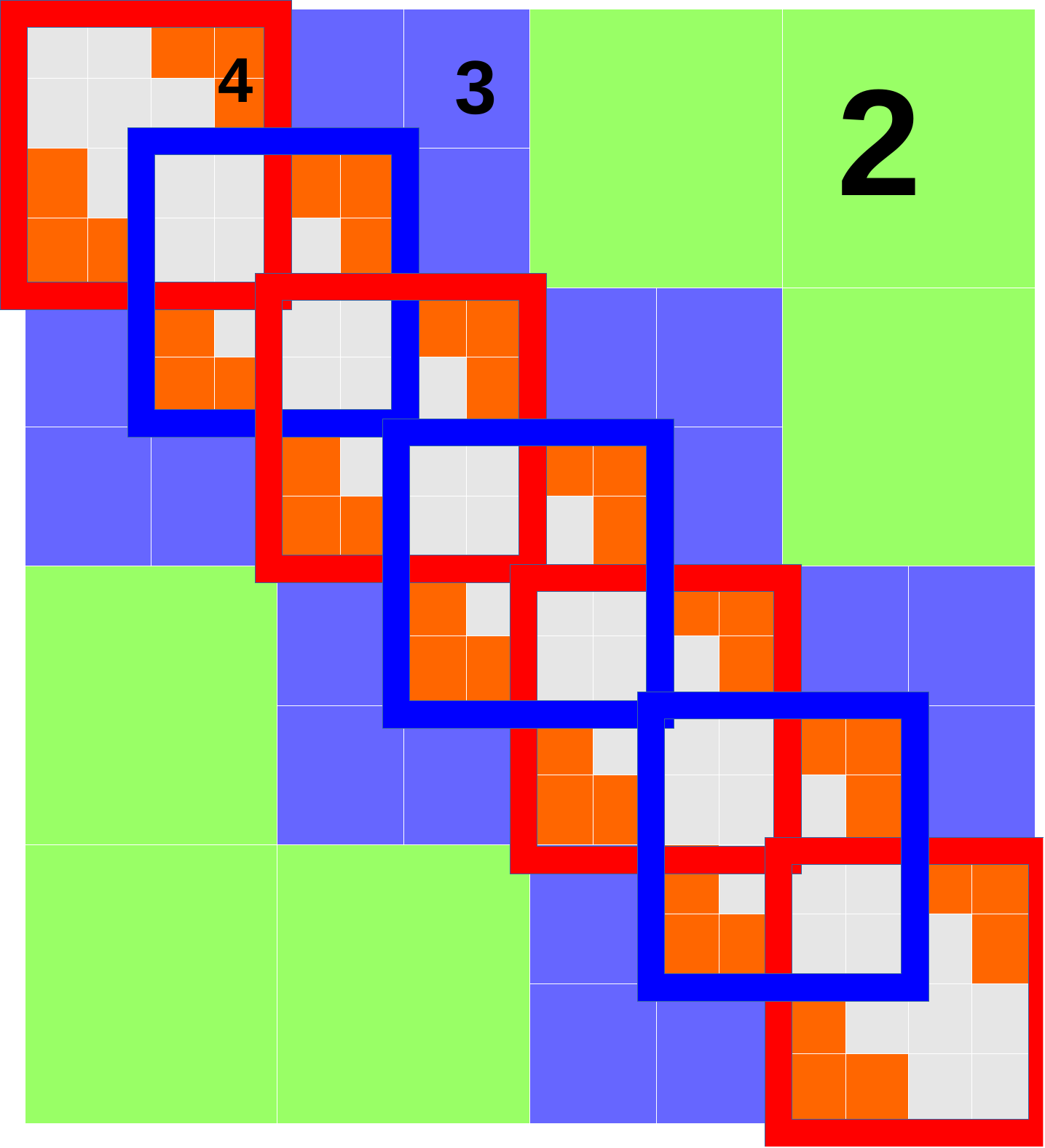}}
\caption{A matrix satisfying the HODLR format (a) and the $\mathcal{H}$-matrix format (b). (c) shows the two types of interactions at level $4$ for $\mathcal{H}$-matrix format. In red frames are type-0 interactions and in blue frames type-1 interactions. In (a)(b)(c) The numbers show the level of each submatrix.}
\label{fig:hmatrix}
\end{figure}

A matrix $\mathbf{A}$ of the HODLR format can be constructed as follows. For an interval $\mathcal{I}$ on level $\ell>1$, we define the interaction list $\mathrm{IL}(\mathcal{I})$ to be the set of all $\mathcal{J}$ such that $\mathcal{J}$ is a sibling of $\mathcal{I}$. For level 1 the two intervals have each other as the only member in their respective interaction list. Each matrix block $\mathbf{A}(\mathcal{I},\mathcal{J})$ for $\mathcal{J}\in\mathrm{IL}(\mathcal{I})$ then has a low-rank structure and therefore admits a low-rank approximation. Using this interaction list we can divide the whole matrix into non-overlapping blocks each with a low-rank structure. The division is shown in Fig.~\ref{fig:hmatrix} (a).

Now we construct the MPOs from this format. At level $\ell$, define
\begin{equation}
\hat{V}^{(\ell)}_{\alpha\beta} = \sum_{i\in\mathcal{I}_{\ell;\alpha}}\sum_{j\in\mathcal{I}_{\ell;\beta}}\mathbf{V}(i,j)\hat{n}_i\hat{n}_j,
\label{eq:interval_pair}
\end{equation}
for $\mathcal{I}_{\ell;\beta}\in\mathrm{IL}(\mathcal{I}_{\ell;\alpha})$. The sum of all the interactions at level $\ell$ can be written as
\[
\hat{V}^{(\ell)}=\sum_{\alpha=0}^{2^{\ell-1}}\hat{V}_{2\alpha,2\alpha+1}^{(\ell)}.
\]
Note only interactions allowed in the interaction list is covered on this level, which is an important fact that enables us to construct MPOs with a small bond dimension. Then we can decompose the tensor as
\[
\hat{V} = \sum_{\ell=1}^{L}\hat{V}^{(\ell)}.
\]
This sum covers all the interactions in $\hat{V}$.



Because of the hierarchical low-rank structure, we can yield some approximate low-rank decomposition $\mathbf{V}(\mathcal{I}_{\ell;2\alpha},\mathcal{I}_{\ell;2\alpha+1})\approx \mathbf{A}\mathbf{B}^T$. 
Then
\[
\sum_{i\in \mathcal{I}_{\ell;2\alpha}}\sum_{j\in \mathcal{I}_{\ell;2\alpha+1}}\mathbf{V}(i,j)\hat{n}_{i}\hat{n}_{j}\approx 
\sum_{r=1}^{M}  \sum_{i\in \mathcal{I}_{\ell;2\alpha}}\sum_{j\in \mathcal{I}_{\ell;2\alpha+1}}(\mathbf{A}(i,r)\hat{n}_{i})(\mathbf{B}(j,r)\hat{n}_{j}).
\]
Note that on the right-hand side each $r$ indexes a rank-one operator. The support of each rank-one operator is $\mathcal{I}_{\ell;2\alpha}\cup \mathcal{I}_{\ell;2\alpha+1}=\mathcal{I}_{\ell-1;\alpha}$.

Therefore, for each $\alpha=0,1,\cdots,2^{\ell-1}-1$
, we have $M$ rank-one operators. For each $\alpha$, we pick one from these $M$ operators correspond to a given index $r$, and sum them up to get,
\[
\hat{V}^{(\ell)}_{r} = \sum_{\alpha}\sum_{i\in \mathcal{I}_{\ell;2\alpha}}\sum_{j\in \mathcal{I}_{\ell;2\alpha+1}}(\mathbf{A}(i,r)\hat{n}_{i})(\mathbf{B}(j,r)\hat{n}_{j}),
\]
and then we have
\[
\hat{V}^{(\ell)} = \sum_{r=1}^{M}\hat{V}^{(\ell)}_{r}.
\]
Hence all the interactions at level $\ell$ can be collected into $M$ operators $\hat{V}^{(\ell)}_{r}$, each one is 
a sum of non-overlapping rank-one operators. From this procedure, we have collected all interactions at level $\ell$ into a sum of $M$ MPOs each with bond dimension 5. $M$ can be chosen to be $\Or(\log(N/\epsilon))$ for the coefficient matrix to have a 2-norm error of $\epsilon$. The system contains $L$ levels so in total we need $ML=\mathcal{O}(\log(N)\log(N/\epsilon))$ MPOs, each with a bounded rank. 

\subsection{$\mathcal{H}$-matrix format}

In the HODLR matrix, a matrix block
$\mathbf{A}(\mathcal{I},\mathcal{J})$ is low-rank as long as
$\mathcal{I},\mathcal{J}$ are non-overlapping and are at the same level.
This is also called the weak-admissibility
condition~\cite{HackbuschKhoromskijKriemann2004}. The $\mathcal{H}$-matrix
is a more general class of hierarchical low-rank matrices, which
requires the strong-admissibility condition, i.e.
$\mathcal{I},\mathcal{J}$ are not only non-overlapping, but their
distance should be comparable to their sizes, i.e.
$\mathrm{dist}(\mathcal{I},\mathcal{J})\gtrsim
\max\{|\mathcal{I}|,|\mathcal{J}|\}$.

In the $\mathcal{H}$-matrix format, we define the interaction list $\mathrm{IL}(\mathcal{I})$ of  an interval
$\mathcal{I}$ in the following way: an interval $\mathcal{J}\in\mathrm{\mathcal{I}}$ if and only if the parent of 
$\mathcal{J}$ and the parent of $\mathcal{I}$ are neighbors, and $\mathcal{J}$ is not a neighbor of $\mathcal{I}$. Therefore in 1D it is guaranteed that $\mathrm{dist}(\mathcal{I},\mathcal{J})\geq
\max\{|\mathcal{I}|,|\mathcal{J}|\}$ if $\mathcal{J}\in\mathrm{IL}(\mathcal{I})$.
A hierarchical low-rank matrix with interaction list
defined in this way is called an $\mathcal{H}$-matrix. A graphical illustration is shown in Fig.~\ref{fig:hmatrix} (b).

We first consider a 1D system. The system is divided into intervals hierarchically the same way as in Section~\ref{sec:hodlr}. Sill we divide the interactions into different levels
\begin{equation}
\hat{V} = \sum_{\ell=2}^{L}\hat{V}^{(\ell)} + \hat{D},
\label{eq:hierarchical_operator}
\end{equation}
where $\hat{D}$ contains only nearest-neighbor interactions. Note that the lowest level is $2$ in this format, as the interaction list at level $1$ is empty. Based on the interaction list of the $\mathcal{H}$-matrix, $\hat{V}^{(\ell)}$ is defined as 
\[
\hat{V}^{(\ell)} = \sum_{\alpha=0}^{2^{\ell-1}-2}\left(\hat{V}^{(\ell)}_{2\alpha,2\alpha+2}+\hat{V}^{(\ell)}_{2\alpha,2\alpha+3}+\hat{V}^{(\ell)}_{2\alpha+1,2\alpha+3}\right),
\]
where $\hat{V}^{(\ell)}_{\alpha\beta}$ is defined in Eq. (\ref{eq:interval_pair}). One can see how this corresponds to the $\mathcal{H}$-matrix interaction list from Fig.~\ref{fig:hmatrix} (b).

Now we want to efficiently represent the operator $\hat{V}^{(\ell)}$ as linear combination of MPOs. Similar to  the previous section we first write
\[
\hat{V}^{(\ell)}_{\alpha\beta} \approx \sum_{r=1}^{M}\hat{V}^{(\ell)}_{\alpha\beta,r},\quad \hat{V}^{(\ell)}_{\alpha\beta,r}=\sum_{i\in \mathcal{I}_{\ell;\alpha}}\sum_{j\in \mathcal{I}_{\ell;\beta}}(\mathbf{A}(i,r)\hat{n}_{i})(\mathbf{B}(j,r)\hat{n}_{j}),
\label{eq:rank_one_hmatrix}
\]
where $\hat{V}^{(\ell)}_{\alpha\beta,r}$ is a rank-one operator defined in (\ref{eq:rank_one}). The low-rank decomposition is done by performing truncated SVD and cutting off singular values smaller than $\epsilon'\|\mathbf{V}(\mathcal{I}_{\ell;\alpha},\mathcal{I}_{\ell;\beta})\|_2$ for some $\epsilon'>0$. 
Then
\begin{equation*}
\hat{V}^{(\ell)}=\sum_{r=1}^{M} \sum_{\alpha=0}^{2^{\ell-1}-2}\left(\hat{V}^{(\ell)}_{2\alpha,2\alpha+2,r}+\hat{V}^{(\ell)}_{2\alpha,2\alpha+3,r}+\hat{V}^{(\ell)}_{2\alpha+1,2\alpha+3,r}\right).
\end{equation*}
However, unlike in the HODLR case, these operators can be overlapping for different $\alpha,\beta$. For example $\hat{V}^{(\ell)}_{2\alpha,2\alpha+2,r}$ and $\hat{V}^{(\ell)}_{2\alpha,2\alpha+3,r}$ are overlapping. Also $\hat{V}^{(\ell)}_{2\alpha,2\alpha+2,r}$ and $\hat{V}^{(\ell)}_{2\alpha+2,2\alpha+4,r}$ are overlapping as well. Therefore we need to further separate the terms as
\begin{align}
\hat{V}^{(\ell)}&=\sum_{r=1}^{M} \left(  \sum_{\alpha=0}^{2^{\ell-2}-1}\hat{V}^{(\ell)}_{4\alpha,4\alpha+2,r}+ \sum_{\alpha=0}^{2^{\ell-2}-1}\hat{V}^{(\ell)}_{4\alpha,4\alpha+3,r}+ \sum_{\alpha=0}^{2^{\ell-2}-1}\hat{V}^{(\ell)}_{4\alpha+1,4\alpha+3,r}\right) \label{eq:type0}\\
&+ \sum_{r=1}^{M}\left(  \sum_{\alpha=0}^{2^{\ell-2}-2}\hat{V}^{(\ell)}_{4\alpha+2,4\alpha+4,r}+ \sum_{\alpha=0}^{2^{\ell-2}-2}\hat{V}^{(\ell)}_{4\alpha+2,4\alpha+5,r}+ \sum_{\alpha=0}^{2^{\ell-2}-2}\hat{V}^{(\ell)}_{4\alpha+3,4\alpha+5,r} \right) \label{eq:type1}
\end{align}
For each of the six terms on the right-hand side we construct $M$ MPOs. Each term represents a sum of non-overlapping rank-one operators, and therefore has bond dimension 5. Operators on the first line we call type-0 and on the second line we call type-1. They can be visualized on the coefficient matrix, as shown in Fig.~\ref{fig:hmatrix}.

From the discussion we construct $6M$ MPOs each with bond dimension 5. Since the system contains $L$ levels in total we have $6LM$ MPOs. To ensure the 2-norm error of the coefficient matrix to be smaller than $\epsilon$, we need to choose $\epsilon'=C\epsilon/N$ for some constant $C$. Then we have $M=\Or(\log(1/\epsilon'))=\Or(\log(N/\epsilon))$. The total number of MPOs is $\mathcal{O}(\log(N)\log(N/\epsilon))$.

In practical calculation, we may further reduce the constant prefactor from $6$ to $4$ without increasing $M$. This is because the strong admissibility condition allows one to concatenate certain blocks together only with a mild increase of the numerical rank~\cite{HackbuschKhoromskijKriemann2004}. For example, matrices $\mathbf{V}(\mathcal{I}_{\ell;4\alpha},\mathcal{I}_{\ell;4\alpha+2})$ and $\mathbf{V}(\mathcal{I}_{\ell;4\alpha},\mathcal{I}_{\ell;4\alpha+3})$ can be concatenated into a single matrix $\mathbf{V}(\mathcal{I}_{\ell;4\alpha},\mathcal{I}_{\ell;4\alpha+2}\cup \mathcal{I}_{\ell;4\alpha+3})$. This allows us to combine some of the operators and thereby reduce the prefactor. The reduction of the prefactor becomes more significant in 2D and higher-dimensional cases.

\subsection{Higher dimensional cases}
\label{sec:higher_dims}

\begin{figure}[ht!]
    \centering
    \subfloat[Allowed interactions]{\includegraphics[width=0.3\textwidth]{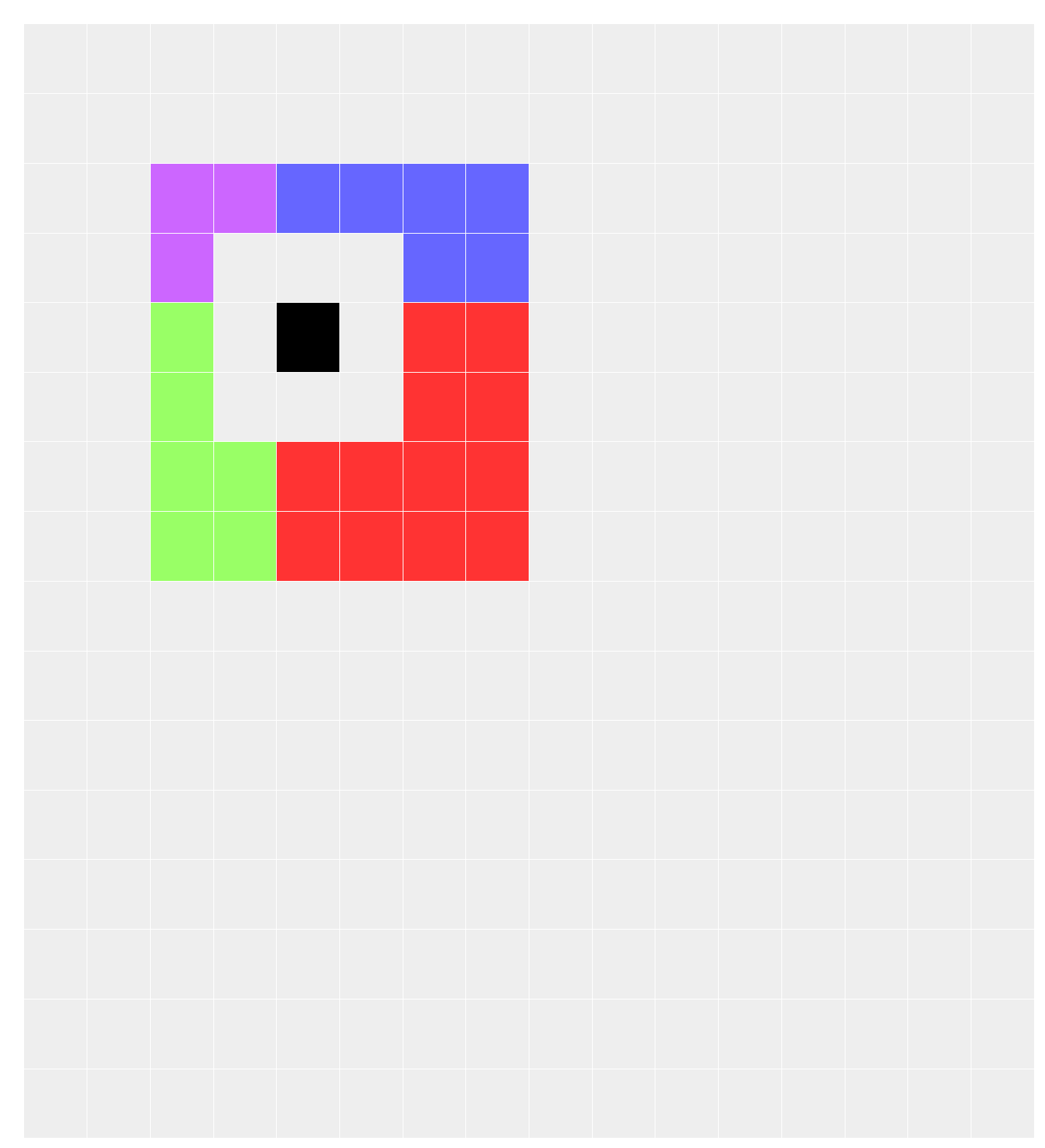}}
    \subfloat[$\hat{V}^{(l,p,q)}_{(\alpha_x,\alpha_y)}$ operators]{\includegraphics[width=0.3\textwidth]{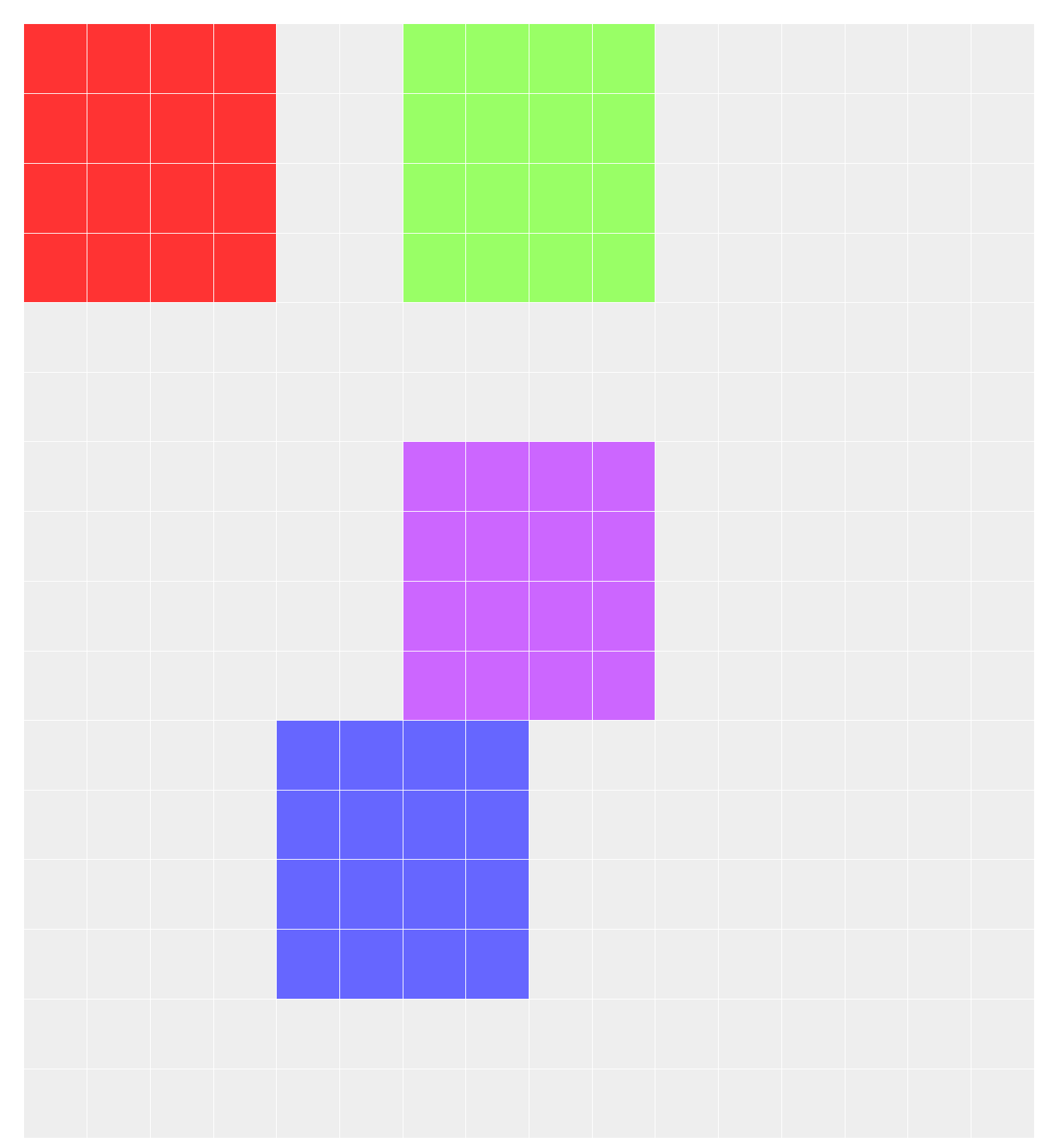}}
    \subfloat[Assignment rules]{\includegraphics[width=0.3\textwidth]{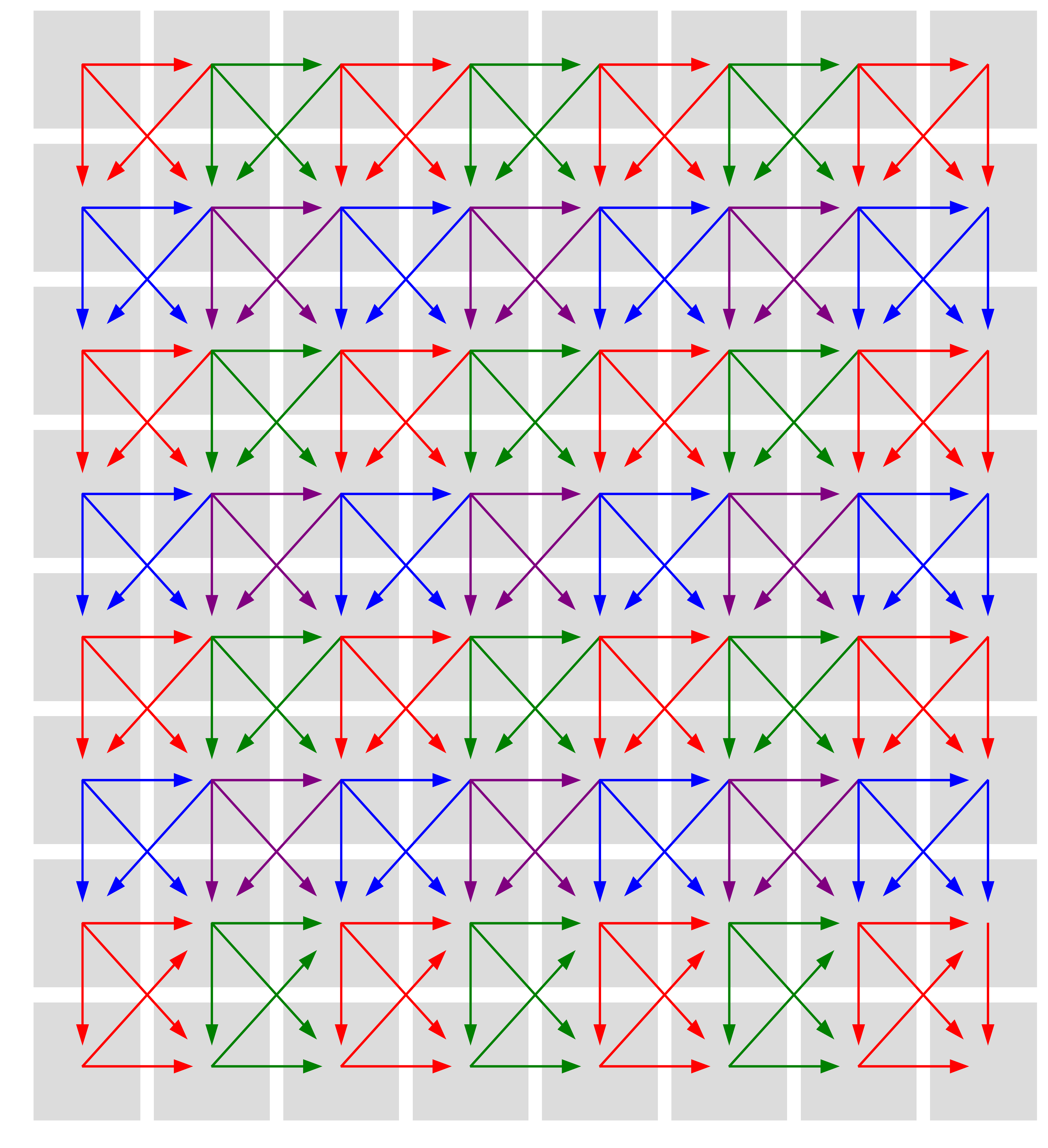}}
    \caption{An $N\times N$ 2D system. Different Colors represent the $(p,q)$ values of the relevant $\hat{V}^{(l,p,q)}_{(\alpha_x,\alpha_y)}$ operators. Red: (0,0); Green: (1,0); Blue: (0,1); Purple: (1,1). (a) A box at level 4 is colored
    black. Other colored boxes are those that are allowed to interact with it at level 4. (b) The regions colored red, green, purple and blue correspond respectively to $\hat{V}^{(4,0,0)}_{(0,0)}$, $\hat{V}^{(4,1,0)}_{(1,0)}$, $\hat{V}^{(4,1,1)}_{(1,1)}$, and $\hat{V}^{(4,0,1)}_{(1,2)}$. (c) An arrow here means we include all the interactions of the form $\mathbf{V}(\mathbf{i},\mathbf{j})\hat{n}_{\mathbf{i}}\hat{n}_{\mathbf{j}}$, where $\mathbf{i}$ is in the level-3 box at the origin of the arrow, $\mathbf{j}$ is in the level-3 box at the end of the arrow, and $\mathbf{i}$ and $\mathbf{j}$ are not neighbors, in the corresponding operator $\hat{V}^{(4,p,q)}_{(\alpha_x,\alpha_y)}$. Only interactions represented by arrows with the same color can be included in the same PEPO.}  
    \label{fig:hmatrix_2d}
\end{figure}

When the coefficient matrix $\mathbf{V}$ satisfies the  hierarchical low-rank
format, the tensor $\hat{V}$ can be efficiently represented by PEPOs
when the lattice dimension is larger than $1$. We consider the 2D case
below, and the 3D case follows similarly.  We consider an $N\times N:=2^L\times 2^L$ lattice, with each
site indexed by $\mathbf{i}=(\alpha_x,\alpha_y)\in \{1,2,\cdots,N\}^2$. 

In the 2D case, the intervals are replaced by boxes.  We define a box on level
$\ell$, indexed by $\bm{\alpha}=(\alpha_x,\alpha_y)$, by
\begin{align*}
\mathcal{I}_{\ell;\bm{\alpha}} &=\{\alpha_x 2^{L-\ell},\alpha_x 2^{L-\ell}+1,\cdots, (\alpha_x+1)2^{L-\ell}-1\} \\
&\times \{\alpha_y 2^{L-\ell},\alpha_y 2^{L-\ell}+1,\cdots, (\alpha_y+1)2^{L-\ell}-1\},
\end{align*}
and interaction between two level-$\ell$ boxes $\bm{\alpha}$, $\bm{\beta}$ is denoted as $\hat{V}^{(\ell)}_{\bm{\alpha}\bm{\beta}}$, similar to (\ref{eq:interval_pair}).

Similar to $\mathcal{H}$-matrices in 1D, the interaction list $\mathrm{IL}(\mathcal{I})$  is the set of all boxes $\mathcal{J}$ such that
\begin{enumerate}
    \item The parent of $\mathcal{J}$ is a neighbor of the parent of $\mathcal{I}$,
    \item $\mathcal{I}$ and $\mathcal{J}$ are not neighbors.
\end{enumerate}
Fig.~\ref{fig:hmatrix_2d} (a) shows an example of the allowed interactions at level $4$. All the interactions at level $\ell$ are collected into an operator $\hat{V}^{(\ell)}$, and we have the same decomposition as in (\ref{eq:hierarchical_operator}).

In the previous section we have divided the interactions at level $\ell$ into type-0 and type-1, and further into six terms in (\ref{eq:type0}) and (\ref{eq:type1}) respectively. Here we divide all interactions into four types: type-$(p,q)$ for $p,q\in \{0,1\}$. And we write
\[
\hat{V}^{(\ell)} =\sum_{pq} \sum_{\alpha_x=0}^{2^{\ell-2}-1-p}\sum_{\alpha_y=0}^{2^{\ell-2}-1-q}\hat{V}^{(l,p,q)}_{(\alpha_x,\alpha_y)},
\]
where $\hat{V}^{(l,p,q)}_{(\alpha_x,\alpha_y)}$ covers most, but not all  interactions at level $\ell$ (to avoid double counting) between boxes $\mathcal{I}_{\ell;\bm{\gamma}}$ for $\bm{\gamma}=(4\alpha_x+2p+\beta_x,4\alpha_y+2q+\beta_y)$, $\beta_x,\beta_y\in\{0,1,2,3\}$. The boxes can be divided into a $4\times 4$ array. Some examples are given in Fig.~\ref{fig:hmatrix_2d} (b). In general, $\hat{V}^{(l,p,q)}_{(\alpha_x,\alpha_y)}$
with the same $p,q$ but different $\alpha_x,\alpha_y$ are non-overlapping, and therefore can be summed up without increasing the bond dimension beyond 5. A graphical description of the rules is given in Fig.~\ref{fig:hmatrix_2d} (c).


Using the simplest scheme each $\hat{V}^{(l,p,q)}_{(\alpha_x,\alpha_y)}$ contains at most 78 interactions between pairs of boxes at level $\ell$. The number 78 is obtained as follows: there are $16\times 15/2=120$ pairwise interactions between boxes in $\hat{V}^{(l,p,q)}_{(\alpha_x,\alpha_y)}$, as can be seen in Fig.~\ref{fig:hmatrix_2d} (b). 42 of these pairwise interactions are between adjacent boxes (sharing an edge or vertex), and therefore in total there are 78 permissible ones. Note that there will be double counting if we include all 78 interactions in each $\hat{V}^{(l,p,q)}_{(\alpha_x,\alpha_y)}$, so in fact this maximum is only attained at the lower-right corner according to the assignment rule in Fig.~\ref{fig:hmatrix_2d} (c).

However, using the fact that for $\mathcal{I}_{\ell;\bm{\beta}_1},\mathcal{I}_{\ell;\bm{\beta}_2}\in\mathrm{IL}(\mathcal{I}_{\ell;\bm{\alpha}})$, we may directly compress the interactions between $\mathcal{I}_{\ell;\bm{\alpha}}$ and $\mathcal{I}_{\ell;\bm{\beta}_1}\cup \mathcal{I}_{\ell;\bm{\beta}_2}$. This is the same as what we have done for the 1D case in the previous section. Using this method, all pairwise interactions involving one box in the $4\times 4$ box array can be merged into one. Any interaction included has to involve one of the 12 boxes that are not among the 4 boxes at the lower-right corner. The result is that we can reduce the number of interactions to 12 for $p=q=0$ and 8 for all other cases.

Each of the 12 or 8 interactions, which we index by $k$, is then approximated by the linear combination of rank-one operators. Therefore
\[
\hat{V}^{(l,p,q)}_{(\alpha_x,\alpha_y)} = \sum_{k=1}^{K_{pq}}\sum_{r=1}^{M}\hat{V}^{(l,p,q)}_{(\alpha_x,\alpha_y),kr},
\]
where $K_{pq}=12$ for $p=q=0$ and 8 otherwise, and each $\hat{V}^{(l,p,q)}_{(\alpha_x,\alpha_y),kr}$ is a rank-one operator. We simply add zero operators when there is not enough $k$. It is clear that this is not the most efficient method, but it gives us the desired scaling in the end. The rank-one operators are obtained using a low-rank decomposition similar to (\ref{eq:rank_one_hmatrix}), which is in turn obtained from a truncated SVD. 

Then we can define
\[
\hat{V}^{(l,p,q)}_{kr} = \sum_{\alpha_x=0}^{2^{\ell-2}-1-p}\sum_{\alpha_y=0}^{2^{\ell-2}-1-q}\hat{V}^{(l,p,q)}_{(\alpha_x,\alpha_y),kr}.
\]
Everything we have done above is to make this a sum of non-overlapping rank-one operator introduced in Section \ref{sec:rank_one_mpos}. It can be represented by a PEPO with bond dimension 5.

Therefore all the interactions at level $\ell$ can be written as
\[
\hat{V}^{(\ell)}=\sum_{p=0}^1\sum_{q=0}^1\sum_{k=1}^{K_{pq}}\sum_{r=1}^M\hat{V}^{(l,p,q)}_{kr}.
\]
This is a linear combination of $36M$ PEPOs with bond dimension 5.

The total number of PEPOs needed for the system will be $36ML$. 
To ensure the coefficient matrix 2-norm error to be below $\epsilon$ we need $M=\mathcal{O}(\log(N/\epsilon))$ 
and $L=\log(N)$. The number of PEPOs needed will scale like $\mathcal{O}(\log(N/\epsilon)\log(N))$. We remark that this is in fact a rather loose upper bound. Our numerical experiments indicate that for small-to-medium sized systems (e.g. $N=128$), the actual number of PEPOs needed is usually much smaller.

\section{Error analysis for the Coulomb matrix}
\label{sec:error_analysis}

%


The modified ISVD algorithm and the hierarchical low-rank representation can be applied to general pairwise interactions with or without the translation-invariance property. To illustrate the efficiency of the compressed representation, in this section we briefly list some known results for the Coulomb interaction.

For the Coulomb interaction, the error bound for the exponential fitting method using the quadrature scheme is given in~\cite{beylkin2002numerical,braess2005approximation}.  In particular,  $0<\delta<r<D$ and any $\epsilon>0$, there exist $M=\mathcal{O}\{\log [D /(\delta \varepsilon)]\}$ numbers $\alpha_{l}, \lambda_{l}>0$  such that
\begin{equation}
  \left|\frac{1}{r}-\sum_{l=1}^{M} \alpha_{l} \exp \left(-\lambda_{l} r\right)\right|<\epsilon.
  \label{}
\end{equation}
A more general approximation can be obtained from classical results in the fast multipole method (e.g.  \cite[Theorem 3.2]{greengard1997new}). Consider the two-dimensional case for example, we may obtain the following approximate low-rank decomposition of the off-diagonal part of the Coulomb kernel: for two regions $\mathcal{A}\subset \mathbb{R}^2$ and $\mathcal{B}\subset \mathbb{R}^2$, such that $\mathcal{A}\subset B(\mathbf{r}_0,a)$ and $\mathcal{B}\subset B(\mathbf{r}_0,r)^c$, $a<r$, then for any $p\geq 0$, there exists $a_k\in C^{\infty}(\mathcal{A})$ and $b_k\in C^{\infty}(\mathcal{B})$, $1\leq k\leq 2p+1=:M$ such that for any $\mathbf{r}\in\mathcal{A}$, $\mathbf{r}'\in\mathcal{B}$,
\begin{equation}
  \left|\frac{1}{\norm{\mathbf{r}-\mathbf{r'}}}-\sum_{k=1}^{M} a_k(\mathbf{r})b_k(\mathbf{r}') \right|
    \leq \frac{1}{r-a}\left(\frac{a}{r}\right)^{p+1}.
\end{equation}
This immediately implies that in order to control the pointwise error up to $\epsilon$, the rank $M$ can be bounded by $\log (1/\epsilon)$. In order to control the 2-norm of each low-rank block, $M$ can be bounded by $\log(N/\epsilon)$.  Note that the separation between the sets $\mathcal{A},\mathcal{B}$ is crucial for the error estimate, which corresponds to the strong admissibility condition. For 2D and higher dimensional Coulomb interaction, the separation is indeed crucial to bound the PEPO rank. Hence we only consider the $\mathcal{H}$-matrix format for 2D systems in the numerical examples below.

\section{Numerical examples} \label{sec:numerical}

In this section we present numerical results for 1D and 2D systems using the methods we introduced above. As before, we will only focus on long-range pairwise interactions of the form (\ref{eqn:pairwise}).

\subsection{Quasi 1D system}

We compare the numerical performance of the exponential fitting method and the modified ISVD described in Algorithm \ref{alg:mpo_mc_modified} on Coulomb interaction of the form
\begin{equation*}
    \hat{V} = \sum_{1\leq i<j\leq N} \frac{1}{|\mathbf{r}_i - \mathbf{r}_j|}\hat{n}_i \hat{n}_j,
\end{equation*}
where $\mathbf{r}_i=(x_i,y_i)$. We always choose $x_i=i$, $|y_i|<1$ so that this mimics a quasi-1D system. 
Define the error matrix as $\mathbf{E}(i,j)=\tilde{\mathbf{V}}(i,j)-\mathbf{V}(i,j)$, and we measure the error using the 2-norm of the error matrix.

\begin{figure}[h]
    \centering
    \subfloat[Error versus rank]{\includegraphics[width=0.45\textwidth]{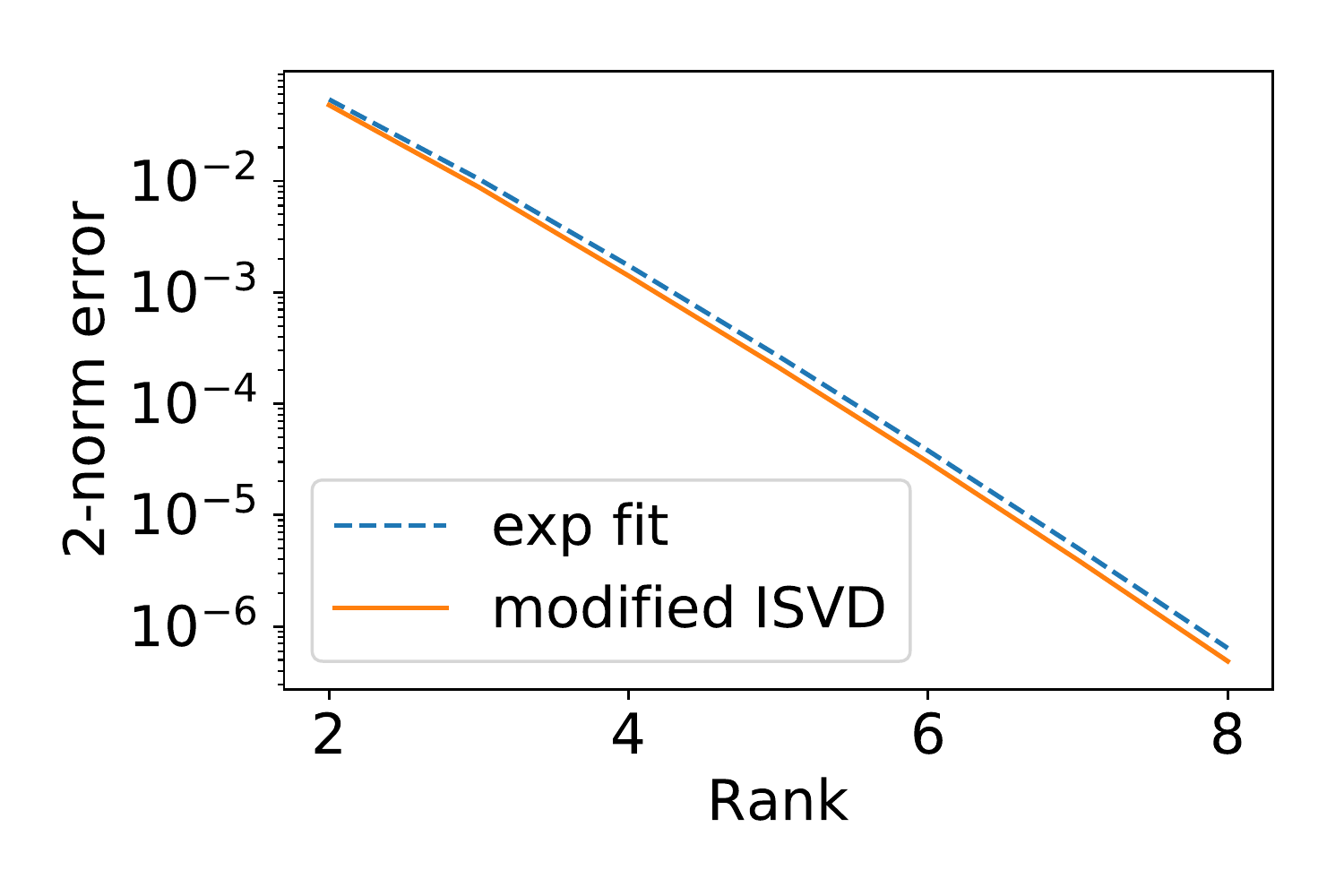}}
    \subfloat[Error versus system size]{\includegraphics[width=0.45\textwidth]{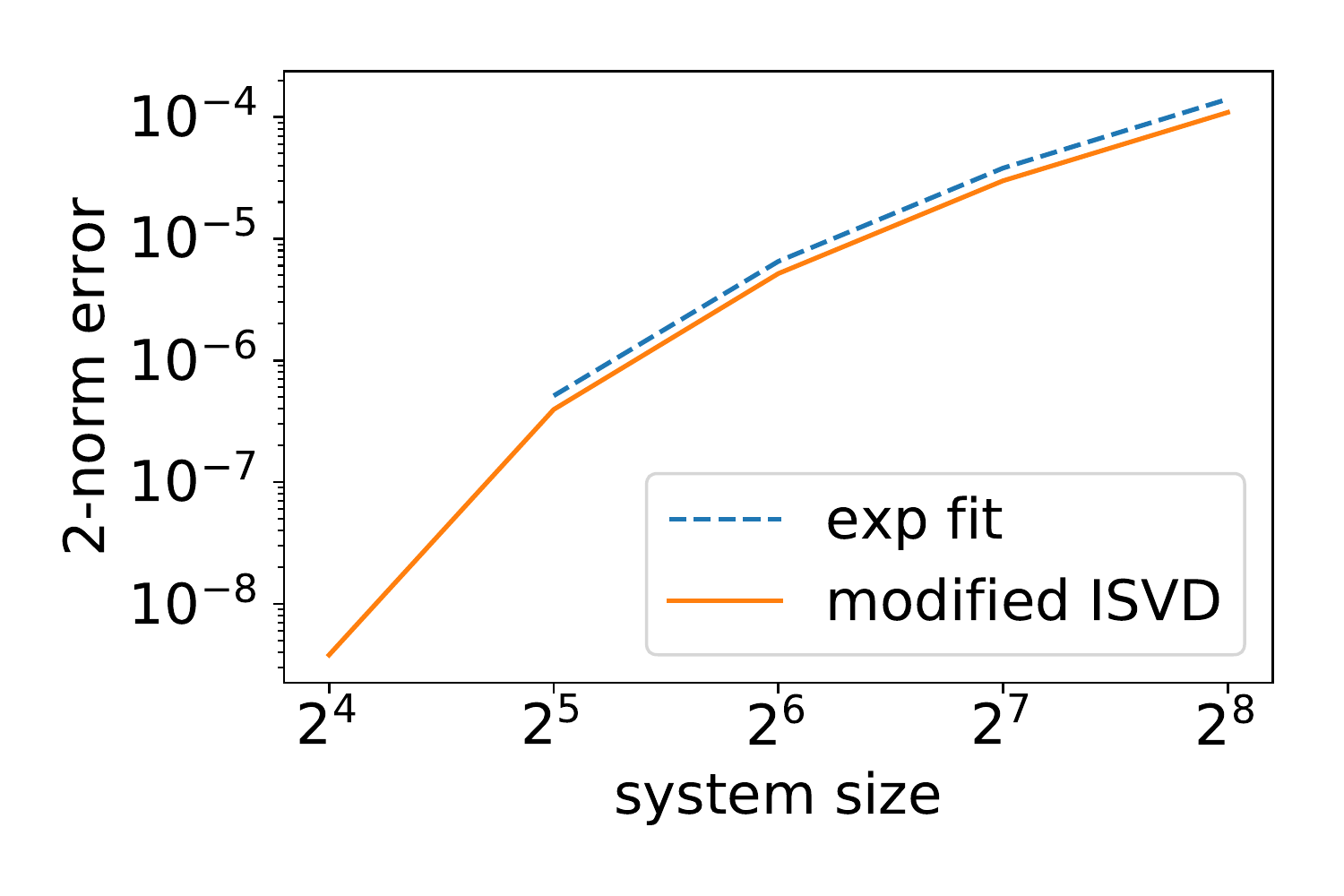}}
    \\
    \subfloat[Error distribution: exponential fitting]{\includegraphics[width=0.45\textwidth]{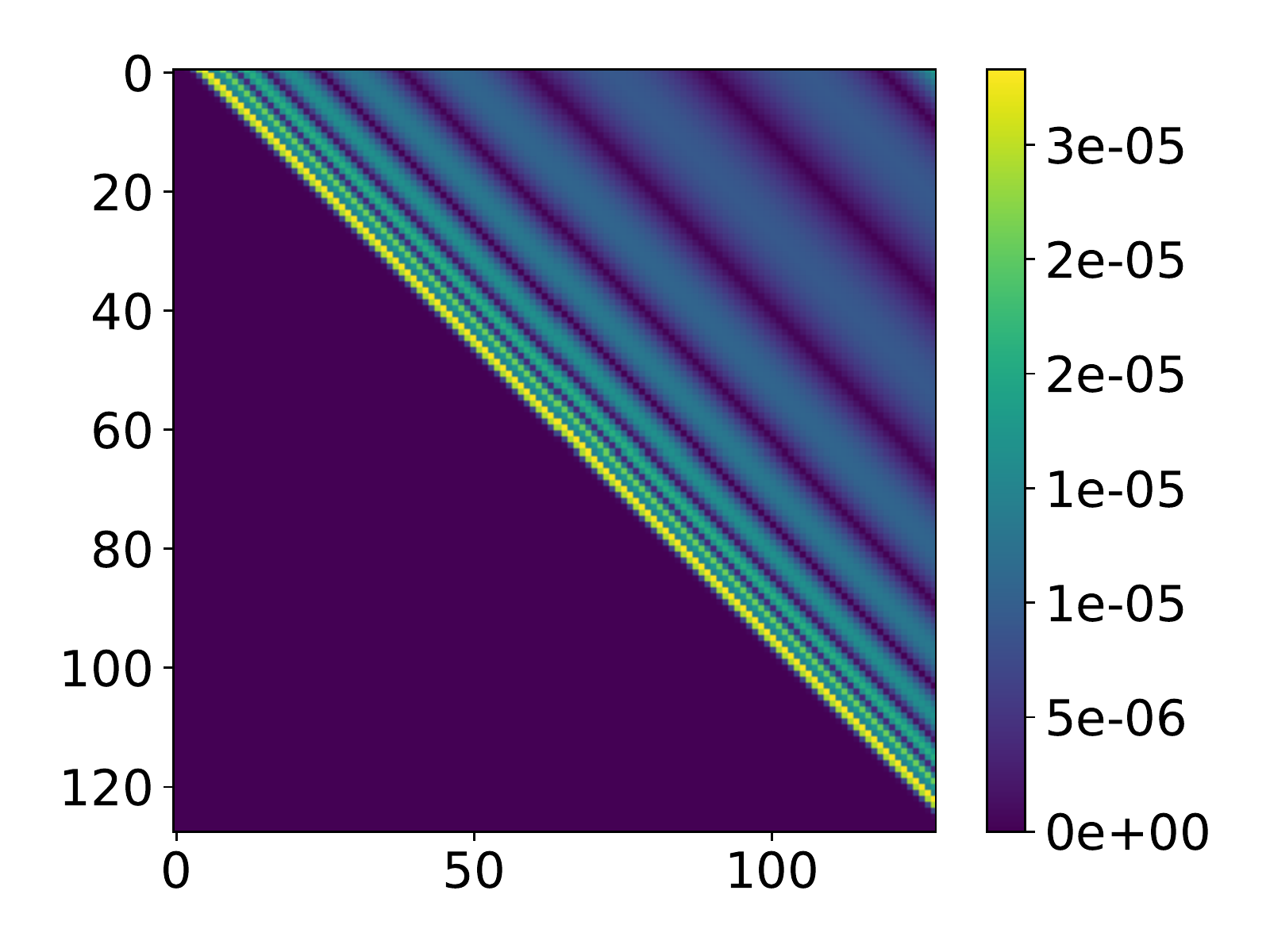}}
    \subfloat[Error distribution: modified ISVD]{\includegraphics[width=0.45\textwidth]{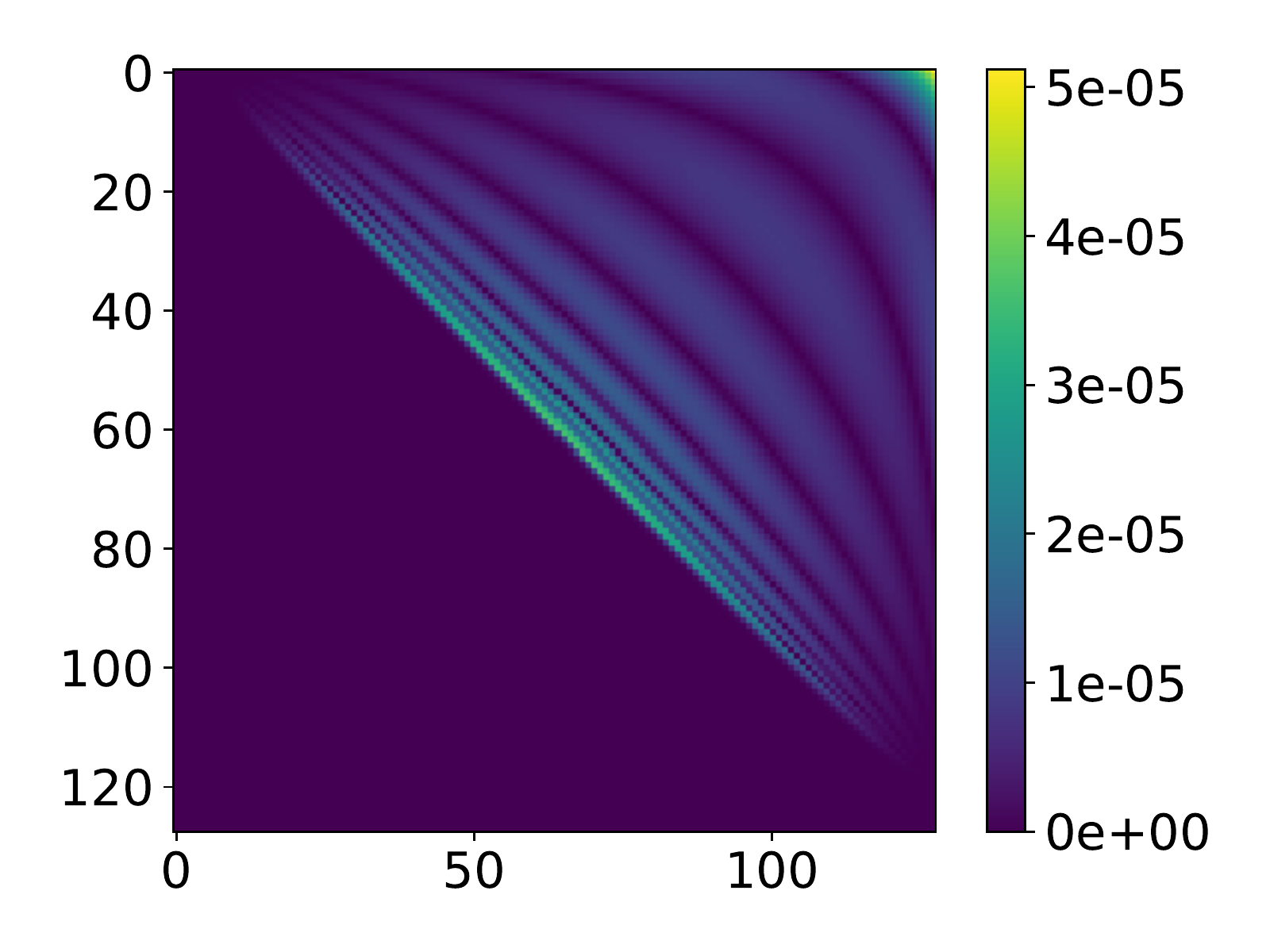}}
    \caption{2-norm error with exponential fitting and modified ISVD in Algorithm \ref{alg:mpo_mc_modified} for (a) system size 128 and rank from 2 to 8, and (b) system size from 16 to 256 and rank 6. (c) (d) are error distributions for the exponential fitting and modified ISVD respectively.}
    \label{fig:err_isvd_exp_fit}
\end{figure}

In the first example we let all $y_i=0$. The resulting system is translation invariant, and we can apply the exponential fitting method as well as modified ISVD. The numerical results are shown in Fig.~\ref{fig:err_isvd_exp_fit}.

Fig.~\ref{fig:err_isvd_exp_fit} shows that the accuracy of the two methods is comparable, and the modified ISVD outperforms exponential fitting by a small margin in terms of 2-norm error. However, the distribution of the error of the two methods is very different. The error of the modified ISVD is concentrated at the upper-right corner of the matrix, which is the interaction between the two ends of the system.  This can be expected from Eq.~(\ref{eq:isvd_approx}), where the error accumulation becomes more significant as more matrices $\tilde{\mathbf{X}}_k$ are multiplied together.  On the other hand, the error of exponential fitting is concentrated some distance away from the diagonal.

\begin{figure}
    \centering
    \subfloat[Error versus rank]{\includegraphics[width=0.45\textwidth]{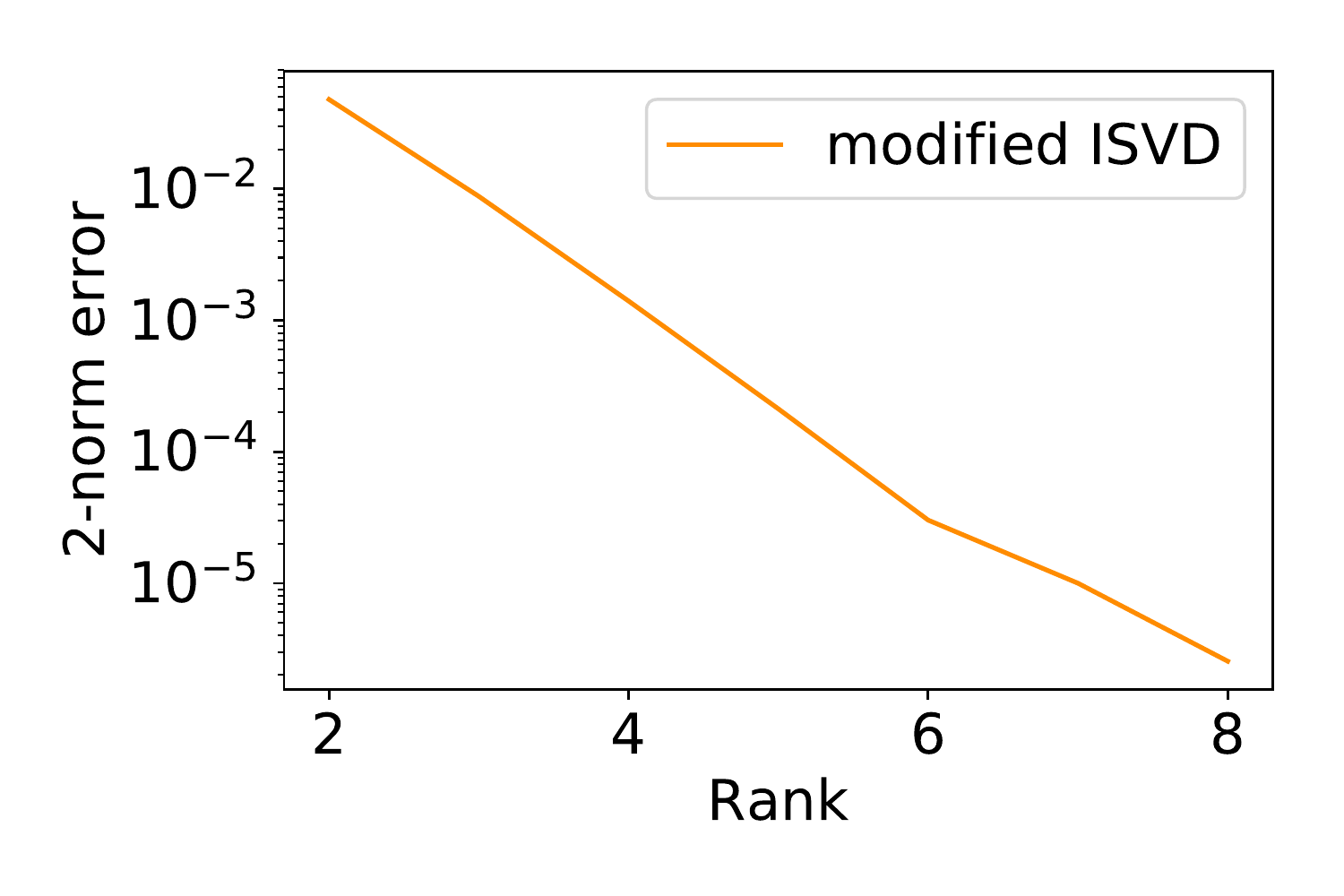}}
    \subfloat[Error distribution]{\includegraphics[width=0.45\textwidth]{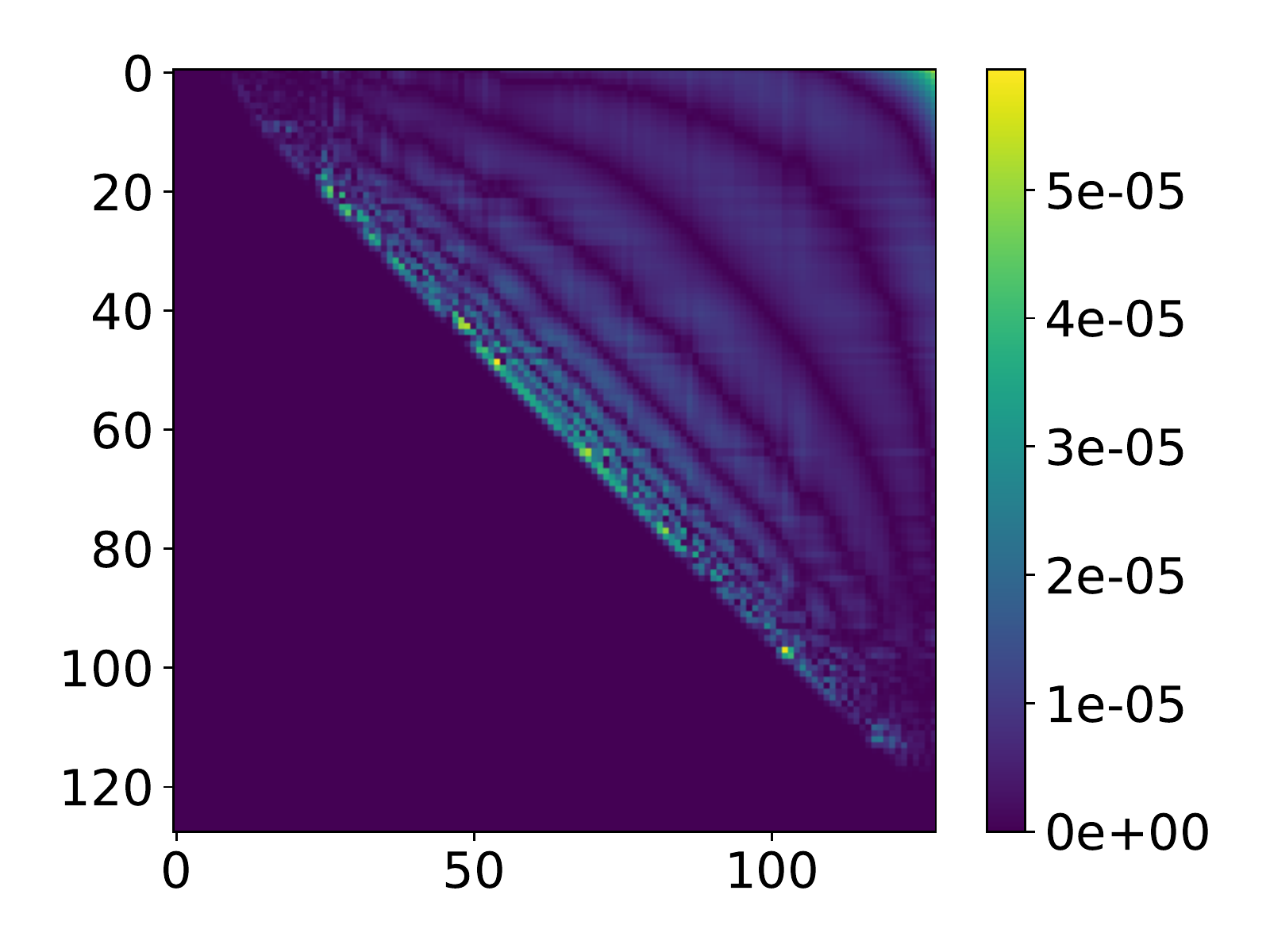}}
    \caption{Relative 2-norm errors versus rank $M$ and error distribution of modified ISVD for non-translation-invariant interactions.}
    \label{fig:err_nti}
\end{figure}

In the second example we set $y_i$ to be random numbers drawn uniformly from $[-0.2,0.2]$. system is then no longer translation-invariant, and the exponential fitting method is not applicable. However, the modified ISVD still performs well. In Fig.~\ref{fig:err_nti} we can see that the convergence of error with respect to $M$ and error distribution are both very similar to the translation-invariant case.

\begin{figure}
    \centering
    \subfloat[HODLR error]{\includegraphics[width=0.45\textwidth]{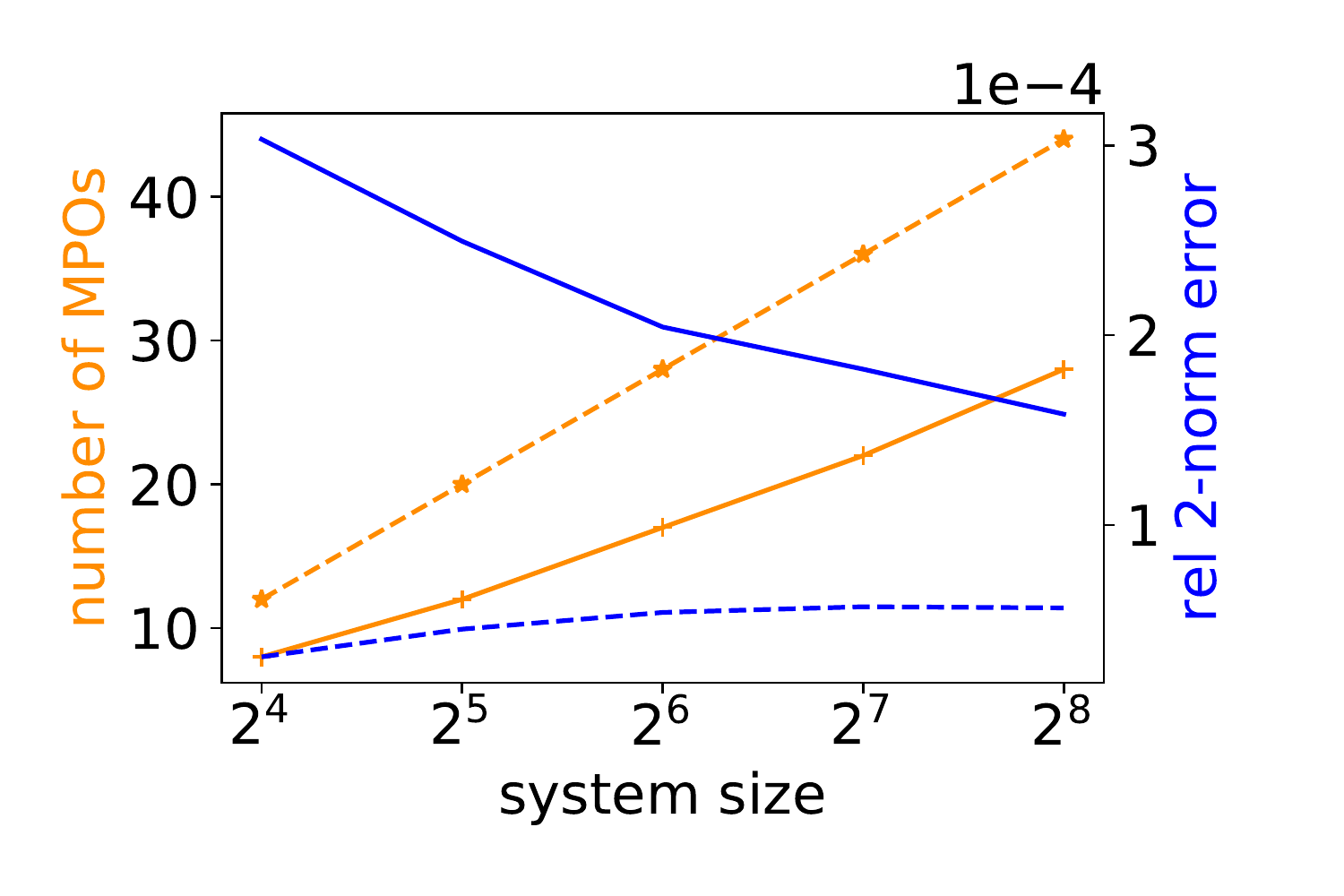}}
    \subfloat[HODLR]{\includegraphics[width=0.4\textwidth]{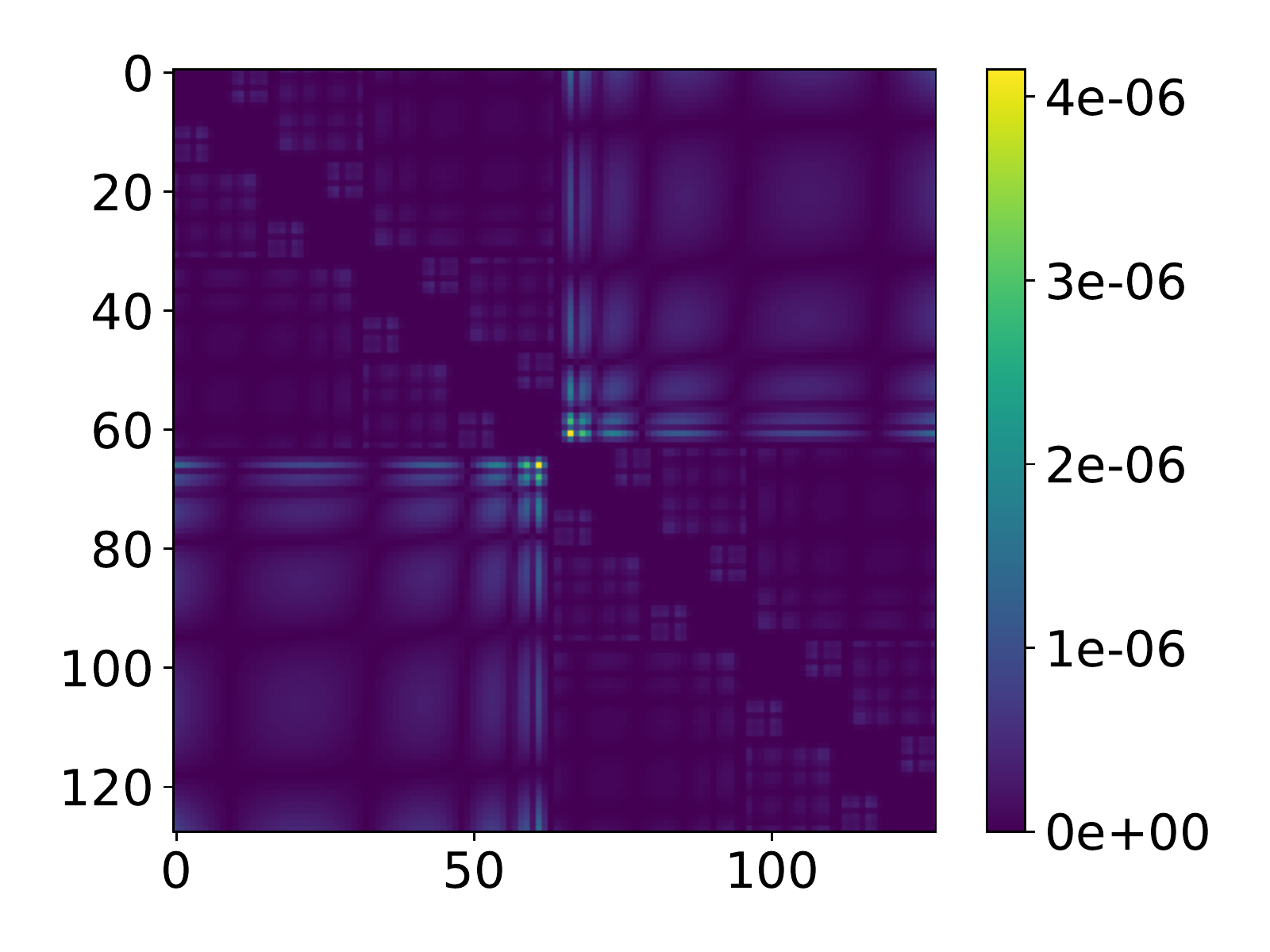}}
    \\
    \subfloat[$\mathcal{H}$-matrix 1D]{\includegraphics[width=0.4\textwidth]{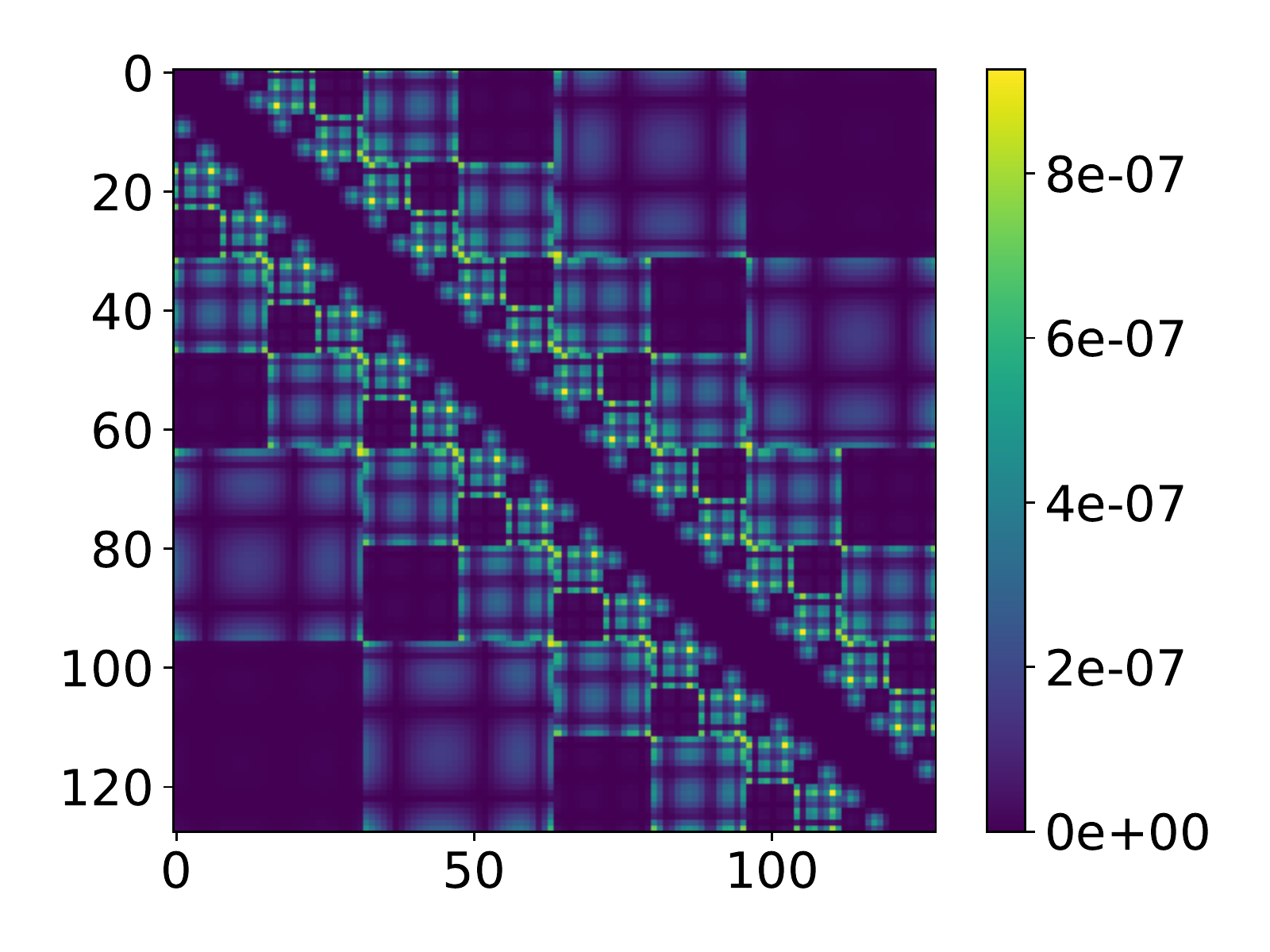}}
    \subfloat[$\mathcal{H}$-matrix 2D]{\includegraphics[width=0.4\textwidth]{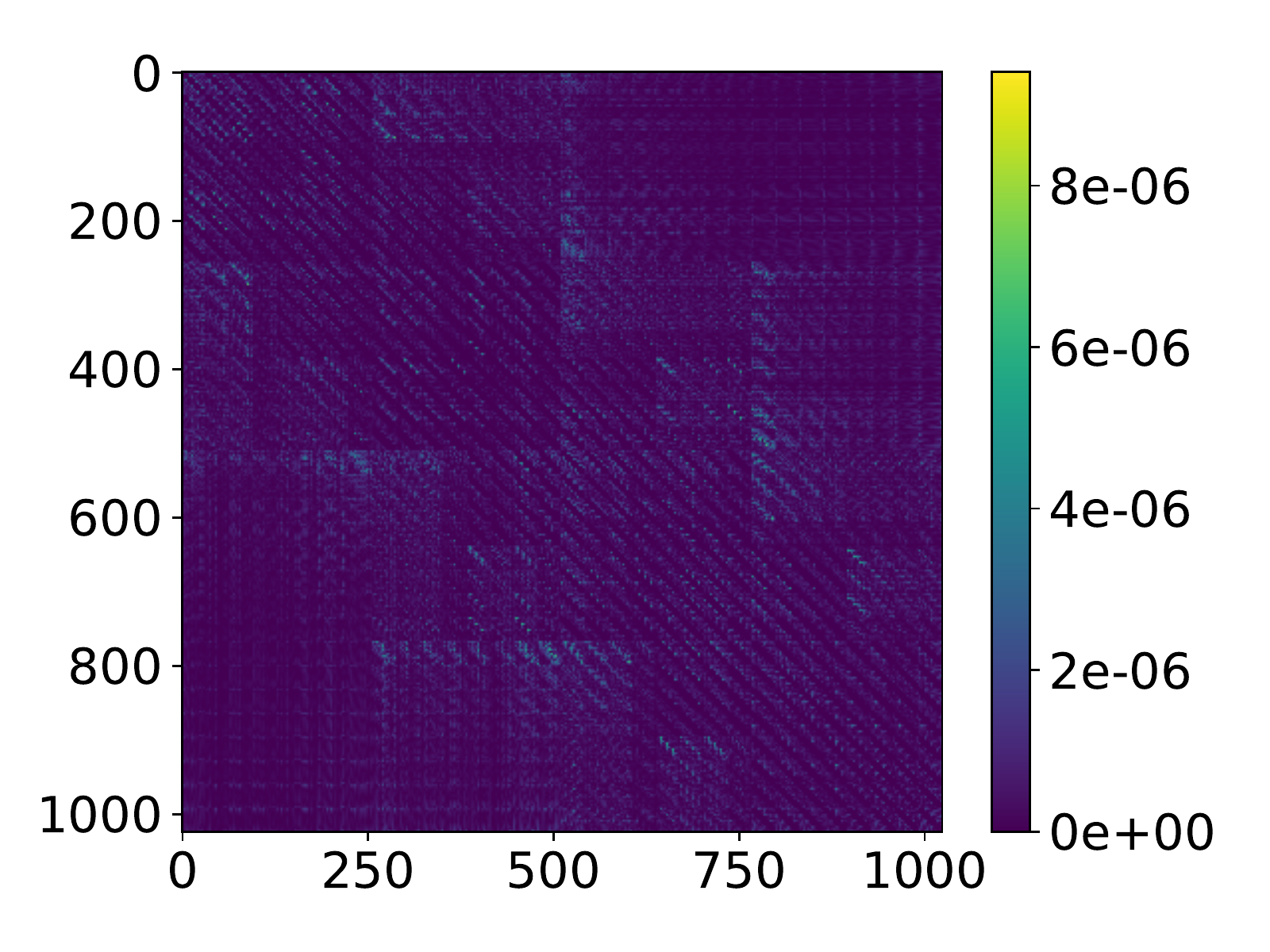}}
    \caption{(a) Number of MPOs needed (orange) and relative 2-norm errors (blue) for HODLR format (solid lines) and $\mathcal{H}$-matrix format (dashed lines) in 1D with threshold $10^{-3}$. (b) Error distributions of HODLR format in 1D. (c) $\mathcal{H}$-matrix in 1D. (d) $\mathcal{H}$-matrix in 2D. System size for (b) and (c) is 128, and system side length is $32$ for (d). Threshold is $\epsilon'=10^{-5}$ for (b), (c), (d).}
    \label{fig:err_dist_hodlr_and_hmat}
\end{figure}

We also test the performance of HODLR and $\mathcal{H}$-matrix based MPO construction on this model with $y_i=0$. The number of MPOs needed with fixed singular value threshold $\epsilon'=10^{-3}$ and the relative 2-norm error is plotted in Fig.~\ref{fig:err_dist_hodlr_and_hmat} (a). The $\mathcal{H}$-matrix format requires a large number of MPOs than the HODLR format. 
The error distribution is plotted for system size $N=128$, in Fig.~\ref{fig:err_dist_hodlr_and_hmat} (b) (c).

\begin{table}[]
    \centering
    \begin{tabular}{c|cc}
        \hline
        system size & number of PEPOs & maximal rank  \\
        \hline
        $16\times 16$ & 328 & 13 \\
        $32\times 32$ & 756 & 14 \\
        $64\times 64$ & 1221 & 15 \\
        $128\times 128$ & 1696 & 15 \\
        \hline
    \end{tabular}
    \caption{Performance of the $\mathcal{H}$-matrix format for the 2D system. The truncation threshold $\epsilon'=10^{-5}$.}
    \label{tab:num_pepo}
\end{table}

\subsection{2D system}

We then consider a 2D system with sites on a uniform $N\times N$ grid.
For the 2D system we only show the numerical results for the $\mathcal{H}$-matrix format, since the Coulomb interaction in 2D only satisfies the strong admissibility condition. In other words, the growth of the rank $M$ becomes unfavorable using the HODLR format. The error distribution is plotted in 
Fig.~\ref{fig:err_dist_hodlr_and_hmat} (d). 
Table~\ref{tab:num_pepo} shows the number of PEPOs, as well as the maximal PEPO rank for different system sizes in order to achieve the same truncation criterion $\epsilon'$ (we truncate singular values for block $\mathbf{V}(\mathcal{I},\mathcal{J})$) at threshold $\epsilon'\|\mathbf{V}(\mathcal{I},\mathcal{J})\|_2$). We observe that the maximal rank $M$ grows very mildly with respect to the system size. The number of PEPOs increases more noticeably when the system size is small, and the growth becomes slower as the side length $N$ increases from $64$ to $128$. The number of PEPOs should eventually become only logarithmic with respect to $N$.

\section{Conclusion} \label{sec:conclusion}

We introduced two methods for constructing efficient MPO and PEPO representation of tensors with pairwise long-range interactions, without assuming that the system is translation-invariant.  Using the upper-triangular low-rank (UTLR) structure, the construction of the MPO representation can be solved as a matrix completion problem. We developed a modified ISVD algorithm to solve the matrix completion problem with in a numerically stable fashion.  
Although we do not yet have a theoretical error bound for the accuracy of the modified ISVD algorithm, numerical results indicate that the error of the modified ISVD method is comparable to that of the exponential fitting method, which indicates that for Coulomb interaction, the MPO rank from the modified ISVD method can be bounded by $\log(N/\epsilon))$. 
The hierarchical low-rank format can be used to construct both MPOs and PEPOs. For the Coulomb interaction, the tensor can be approximated by the linear combination of $\Or(\log(N)\log(N/\epsilon))$ MPOs/PEPOs, and the rank of each MPO/PEPO is bounded by a constant.  The hierarchical low-rank format leads to an MPO representation with a higher MPO rank than that from the modified ISVD method. However, the linear combination representation 
may naturally facilitate parallel computation, and this is particularly the case for the PEPO representation where a relatively large number of PEPOs can be needed.
On the other hand, the MPOs/PEPOs from our hierarchical low-rank format have many zero entries. It would be of practical interest to develop alternative algorithms to reduce the number of zeros and to obtain a more compact representation with reduced preconstants.

\subsection*{Acknowledgments} 

This work was partially supported by the Department of Energy under Grant No. DE-SC0017867, No. DE-AC02-05CH11231 (L. L.), by the Air Force Office of Scientific Research under award number FA9550-18-1-0095 (L.L. and Y. T.). We thank Garnet Chan, Matthew O'Rourke and Miles Stoudenmire for helpful discussions.

\appendix

\section{FSM rules for sum of non-overlapping MPOs}
\label{sec:appendix_mpo}

For a 1D system we divide into intervals $\{\mathcal{I}_\alpha\}$ such as done in our construction of the hierarchical low-rank matrix. First we clarify some notations. For an operator defined in an interval we use the notation $\hat{O}$. For example, the identity operator defined in an interval $\{\mathcal{I}_\alpha\}$ is $\hat{I}_\alpha$. Tensors and operators defined on a single site are still denoted as $\mathbf{T}$, $\mathbf{O}$, etc. We want to represent an operator of the form 
\begin{equation*}
    \hat{H} = \sum_{\alpha} \left(\bigotimes_{\beta<\alpha} \hat{I}_{\beta}\right) \otimes\hat{H}_\alpha\otimes \left( \bigotimes_{\beta>\alpha} \hat{I}_{\beta}\right),
\end{equation*}
using an MPO, where we already have an MPO representation of each operator $\hat{H}_\alpha$ defined on interval $\mathcal{I}_\alpha$.

Using the finite state machine (FSM) description of MPOs \cite{crosswhite2008finite,frowis2010tensor}, we assume there are the following general finite state machine rules for each $\hat{H}_\alpha$ shown in Table~\ref{tab:mpo_rule_original}. It is equivalent to the equation
\begin{equation*}
    \hat{H}_\alpha = \sum_{r_1,r_2,\cdots,r_{\|\mathcal{I}_\alpha\|-1}}\bigotimes_{i=1}^{|\mathcal{I}_\alpha|}\mathbf{H}_i(r_{i-1},r_i),
\end{equation*}
where $r_0=r_{\|\mathcal{I}_\alpha\|}=1$, and $\mathbf{H}(l,r)$ is an operator for each $l,r$.
The MPO rules for $\hat{H}$ are shown in Table~\ref{tab:mpo_rule_sum}.

\begin{table}[ht!]
	\centering
	\begin{tabular}{c|c|ccc}
		\hline
		 position & rule num & input state & operator & output state \\
		\hline
		 ALL & 1 & $l$ & $\mathbf{H}_i(l,r)$ & $r$  \\
		\hline
	\end{tabular}
	\caption{MPO rules for the $\hat{H}_\alpha$ on interval $\mathcal{I}_\alpha$.}
	\label{tab:mpo_rule_original}
\end{table}

\begin{table}[ht!]
	\centering
	\begin{tabular}{c|c|ccc}
		\hline
		 & rule num & input state & operator & output state \\
		\hline
		\multirow{3}{*}{HEAD} & 1 & $\bigtriangleup$ & $\mathbf{I}_i$ & $\bigtriangleup$ \\
		 & 2 & $\bigtriangledown$ & $\mathbf{I}_i$ & $\bigtriangledown$ \\
		 & 3 & $\bigtriangleup$ & $\mathbf{H}_i(1,r)$ & $r$  \\
		\hline
		\multirow{3}{*}{BODY} & 4 & $\bigtriangleup$ & $\mathbf{I}_i$ & $\bigtriangleup$ \\
		 & 5 & $\bigtriangledown$ & $\mathbf{I}_i$ & $\bigtriangledown$ \\
		 & 6 & $l$ & $\mathbf{H}_i(l,r)$ & $r$  \\
		\hline
		\multirow{3}{*}{TAIL} & 7 & $\bigtriangleup$ & $\mathbf{I}_i$ & $\bigtriangleup$ \\
		 & 8 & $\bigtriangledown$ & $\mathbf{I}_i$ & $\bigtriangledown$ \\
		 & 9 & $l$ & $\mathbf{H}_i(l,1)$ & $\bigtriangledown$  \\
		\hline
	\end{tabular}
	\caption{MPO rules for the sum of operators $\left(\bigotimes_{\beta<\alpha} \hat{I}_{\beta}\right) \otimes\hat{H}_\alpha\otimes \left( \bigotimes_{\beta>\alpha} \hat{I}_{\beta}\right)$.}
	\label{tab:mpo_rule_sum}
\end{table}

For the FSM rules of the operators $\hat{H}_\alpha$, we assume the FSM is in state 1 at the very beginning and will be in state 1 at the end. This means on the first site in interval $\mathcal{I}_\alpha$ an operator $\mathbf{H}_i(1,r)$ is activated and then the FSM goes to state $r$, and on the last site an operator $\mathbf{H}_i(l,1)$ is activated, and the FSM goes into state 1. The first site in the interval is called the head and the last site is called the tail. Note that the FSM rules on the head and tail sites are different from on other sites inside the interval.

Now we construct the FSM rules for $\hat{H}$. We add two states $\tup$ and $\tdown$ into the FSM. $\tup$ is the default state meaning a non-trivial operator, i.e. $\hat{H}_\alpha$, has not been activated, and $\tdown$ means such an operator has been activated, and only identity operators should be picked afterwards. The FSM is in state $\tup$ both at the very beginning and at the end.

At the first site, i.e. the head, of each interval $\mathcal{I}_\alpha$, the input signal from the left is either $\tup$ or $\tdown$. If the signal is $\tup$, then we choose either to not activate $\hat{H}_\alpha$ and pass along the signal $\tup$, as in rule 1 of Table~\ref{tab:mpo_rule_sum}, or to activate $\hat{H}_\alpha$, as in rule 3 of the same table. If the signal is $\tdown$, then $\hat{H}_\alpha$ cannot be activated and a signal $\tdown$ is passed down. This is done in rule 2.

In the middle, or body, of each interval $\mathcal{I}_\alpha$, if $\hat{H}_\alpha$ has been activated at the first site, then we follow the FSM rules for the individual $\hat{H}_\alpha$ (rule 6). If we receive either $\tup$ or $\tdown$ then we pass the signal down (rules 4 and 5).

At the last site, or the tail, of each interval, if $\hat{H}_\alpha$ has been activated, then the output signal is $\tdown$ (rule 9). Otherwise the output signal is whatever this interval received from the beginning (rules 7 and 8).

An important fact is that we have only added two signals into the original set of signals. This means if the original MPOs $\{\hat{H}_\alpha\}$ have a maximum bond dimension $D$, then the new MPO for the sum has a bond dimension $D+2$. In the case of the sum of non-overlapping rank-one operators as discussed in \ref{sec:rank_one_mpos}, the bond dimension of the sum is 5.

\section{FSM rules for sum of non-overlapping PEPOs}
\label{sec:appendix_pepo}

\begin{figure}[ht!]
    \centering
    \includegraphics[width=0.7\textwidth]{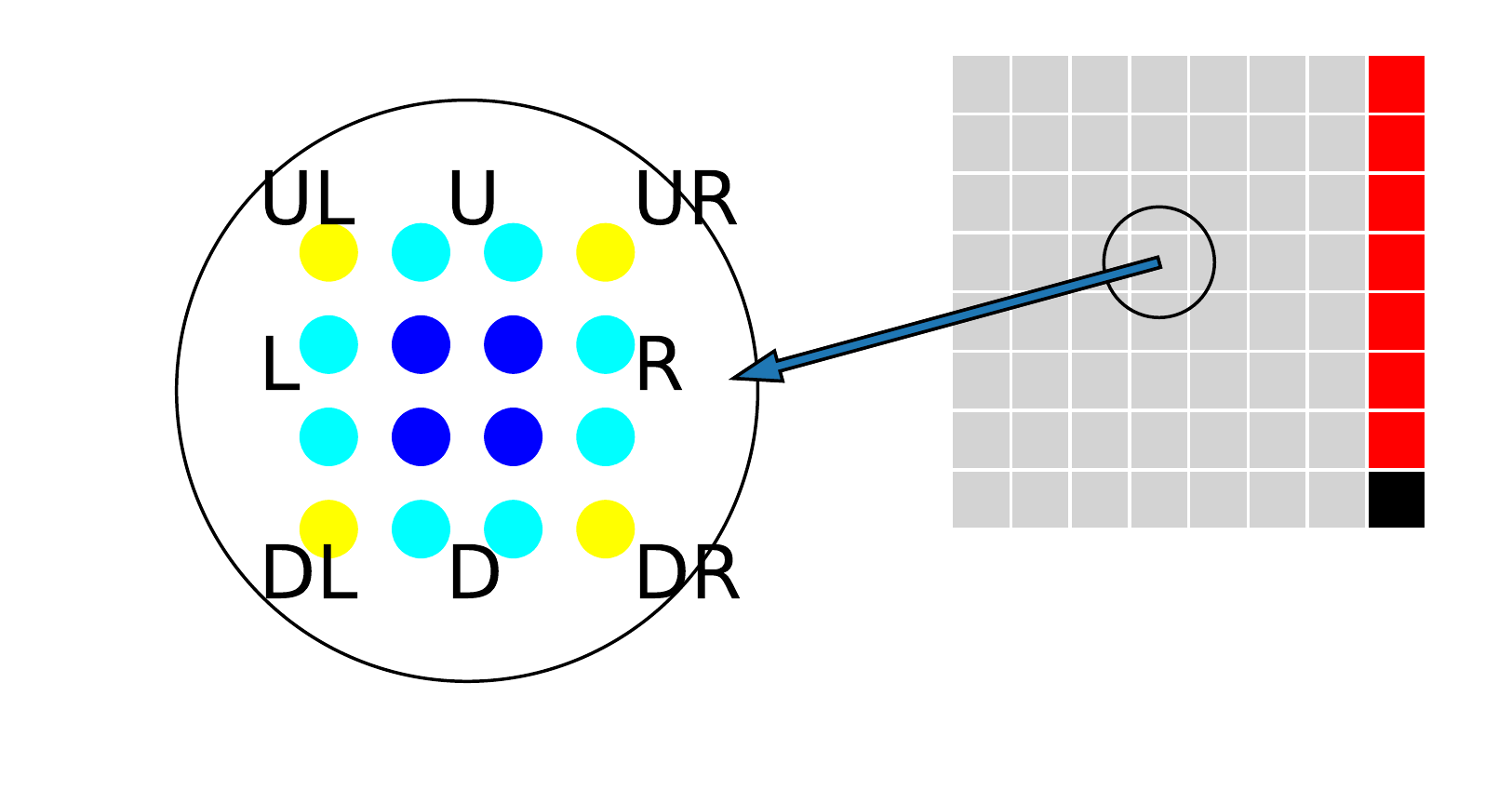}
    \caption{At level $\ell=3$ the 2D system is partitioned into $8\times 8$ boxes as shown on the right. Each box, such as the one in the small circle, has its sites labeled either body, UL, U, UR, R, DR, D, DL, or L as shown on the left. Different FSM rules apply in different regions in the system. FSM rules in Tables~\ref{tab:pepo_rule_body}, \ref{tab:pepo_rule_right}, and \ref{tab:pepo_rule_lower_right} apply in the grey, red, and black boxes respectively.}
    \label{fig:partition_pepo_rule}
\end{figure}

The FSM rules for PEPOs are more complex.  To describe our FSM rules for PEPOs, we first divide the sites inside a given box into different parts, as shown in the right part of Fig.~\ref{fig:partition_pepo_rule}. Different FSM rules should be applied to different parts. Sites in a box are labeled based on their position in the box, as shown in the left part of Fig.~\ref{fig:partition_pepo_rule}. For example, the box at the upper-right corner of a box is labeled UR. 

We assume for each individual PEPO $\hat{H}_{\bm{\alpha}}$, we have the FSM rules specified in Table~\ref{tab:pepo_original}. Note that not all PEPO tensors have 4 internal indices and 2 physical indices, as can be seen from Fig.~\ref{fig:peps_pepo} (b). For example, the tensors on the edges have 3 internal indices. In this case,  we simply add bonds with bond dimension 1 to make them have 4 internal indices, similar to what we have done in the snake-shaped PEPO construction in \ref{sec:rank_one_mpos}. These extra indices all take the value 1.

Now we design a set of FSM rules to represent
\begin{equation*}
    \hat{H} = \sum_{\bm{\alpha}} \left(\bigotimes_{\bm{\beta}\prec\bm{\alpha}}\hat{I}_{\bm{\beta}}\right)\otimes\hat{H}_{\bm{\alpha}}
    \otimes\left(\bigotimes_{\mathbf{\bm{\beta}} \succ \bm{\alpha}}\hat{I}_{\bm{\beta}}\right),
\end{equation*}
where $\prec$ gives the lexicographical order defined in Section \ref{sec:intro}. One key difference between 2D and 1D is that in 2D we no longer have a fixed direction for the signal of FSM to be passed along. Therefore we require that signals between boxes (there are signals passing within boxes according to the original FSM rules in Table~\ref{tab:pepo_original}, which we leave unchanged) only pass rightward or downward, making it easier to design FSM rules.

\begin{table}[ht!]
    \centering
    \begin{tabular}{c|c|cc}
        \hline
         position & rule num & $(l,u,d,r)$ & operator  \\
        \hline
         ALL & 1 & $(l,u,d,r)$ & $\mathbf{H}_i(l,u,d,r)$ \\
        \hline
    \end{tabular}
    \caption{PEPO rules for $\hat{H}_{\bm{\alpha}}$ in the box $\mathcal{I}_{\bm{\alpha}}$}
    \label{tab:pepo_original}
\end{table}

We first design the FSM rules for all parts of the system except for the upper-right corner, the right edge, and the lower-right corner (the colored parts in Fig.~\ref{fig:partition_pepo_rule} (b)). The rules are shown in Table~\ref{tab:pepo_rule_body}. Just like in the 1D case, we introduce two signals $\tup$ and $\tdown$. $\tup$ is the default signal. We change the value of the extra indices we added at the edges and corners from 1 to $\tup$.

At the UL site of each box $\mathcal{I}_{\bm{\alpha}}$, there are two input signals from left and up. When both are $\tup$, then it means $\hat{H}_{\bm{\alpha}}$ can be activated in this box (rule 3), but we can also choose not to activate it and pass along the signal (rule 1). If the site receives a signal $\tdown$ from the left, and $\tup$ from above, then it means an $\hat{H}_{\bm{\alpha}'}$ has already been activated for some $\bm{\alpha}'$ to the left, and the only thing that can be done now is to pass down the signal $\tdown$ to the right until the right edge (rule 2), where other FSM rules will process it.

\begin{table}
	\centering
	\begin{tabular}{c|c|cc}
	\hline
	 & rule num & $(l,u,d,r)$ & operator \\
	\hline
	\multirow{3}{0.13\textwidth}{UL} & 1 & $(\tup,\tup,\tup,\tup)$ & $\mathbf{I}_i$ \\
	 										 & 2 & $(\tdown,\tup,\tup,\tdown)$ & $\mathbf{I}_i$ \\
	 										 & 3 & $(\tup,\tup,d,r)$ & $\mathbf{H}_i(1,1,d,r)$ \\
	\hline
	\multirow{3}{0.13\textwidth}{U}  & 4 & $(\tup,\tup,\tup,\tup)$ & $\mathbf{I}_i$ \\
	 										 & 5 & $(\tdown,\tup,\tup,\tdown)$ & $\mathbf{I}_i$ \\
	 										 & 6 & $(l,\tup,d,r)$ & $\mathbf{H}_i(l,1,d,r)$ \\
	\hline
	\multirow{3}{0.13\textwidth}{UR} & 7 & $(\tup,\tup,\tup,\tup)$ & $\mathbf{I}_i$ \\
	 										 & 8 & $(\tdown,\tup,\tup,\tdown)$ & $\mathbf{I}_i$ \\
	 										 & 9 & $(l,\tup,d,\tdown)$ & $\mathbf{H}_i(l,1,d,1)$ \\
	\hline
	\multirow{2}{0.13\textwidth}{R }  & 10 & $(\tup,\tup,\tup,\tup)$ & $\mathbf{I}_i$ \\
	 										 & 11 & $(l,u,d,\tup)$ & $\mathbf{H}_i(l,u,d,1)$ \\
	\hline
	\multirow{2}{0.13\textwidth}{DR} & 12 & $(\tup,\tup,\tup,\tup)$ & $\mathbf{I}_i$ \\
	 										 & 13 & $(l,u,\tup,\tup)$ & $\mathbf{H}_i(l,u,1,1)$ \\
	\hline
	\multirow{2}{0.13\textwidth}{D}  & 14 & $(\tup,\tup,\tup,\tup)$ & $\mathbf{I}_i$ \\
	 										 & 15 & $(l,u,\tup,r)$ & $\mathbf{H}_i(l,u,1,r)$ \\
	\hline
	\multirow{2}{0.13\textwidth}{DL} & 16 & $(\tup,\tup,\tup,\tup)$ & $\mathbf{I}_i$ \\
	 										 & 17 & $(\tup,u,\tup,r)$ & $\mathbf{H}_i(1,u,1,r)$ \\
	\hline
	\multirow{2}{0.13\textwidth}{L}  & 18 & $(\tup,\tup,\tup,\tup)$ & $\mathbf{I}_i$ \\
	 										 & 19 & $(\tup,u,d,r)$ & $\mathbf{H}_i(1,u,d,r)$ \\
	\hline
	\multirow{2}{0.13\textwidth}{BODY} & 20 & $(\tup,\tup,\tup,\tup)$ & $\mathbf{I}_i$ \\
	 										   & 21 & $(l,u,d,r)$ & $\mathbf{H}_i(l,u,d,r)$ \\
	\hline
	\end{tabular}
	\caption{PEPO rules for the sum of operators $\left(\bigotimes_{\bm{\beta}\prec\bm{\alpha}}\hat{I}_{\bm{\beta}}\right)\otimes\hat{H}_{\bm{\alpha}}
    \otimes\left(\bigotimes_{\mathbf{\bm{\beta}} \succ \bm{\alpha}}\hat{I}_{\bm{\beta}}\right)$ in the whole system except for the upper-right corner, the right edge, and the lower-right corner.}
	\label{tab:pepo_rule_body}
\end{table}

Other rules in Table~\ref{tab:pepo_rule_body} are either identical to the FSM rules for $\hat{H}_{\bm{\alpha}}$, or to pass along signals $\tup$ and $\tdown$. Note that the signal $\tdown$ is only passed along the upper edge (UL, U, and UR in Fig.~\ref{fig:partition_pepo_rule} (a)) as specified in rules 2, 5, 8. If $\hat{H}_{\bm{\alpha}}$ has been activated then a signal $\tdown$ will be passed to the box to the right from the UR site (rule 9).

\begin{table}[ht!]
	\centering
	\begin{tabular}{c|c|cc}
	\hline
	 & rule num & $(l,u,d,r)$ & operator \\
	\hline
	\multirow{4}{0.13\textwidth}{UL} & 1 & $(\tup,\tup,\tup,\tup)$ & $\mathbf{I}_i$ \\
	 									  & 2 & $(\tdown,\tup,\tdown,\tup)$ & $\mathbf{I}_i$ \\
	 									  & 3 & $(\tup,\tdown,\tdown,\tup)$ & $\mathbf{I}_i$ \\
	 									  & 4 & $(\tup,\tup,d,r)$ & $\mathbf{H}_i(1,1,d,r)$ \\
	\hline
	\multirow{2}{0.13\textwidth}{U}  & 5 & $(\tup,\tup,\tup,\tup)$ & $\mathbf{I}_i$ \\
	 									  & 6 & $(l,\tup,d,r)$ & $\mathbf{H}_i(l,1,d,r)$ \\
	\hline
	\multirow{2}{0.13\textwidth}{UR} & 7 & $(\tup,\tup,\tup,\tup)$ & $\mathbf{I}_i$ \\
	 									  & 8 & $(l,\tup,d,\tup)$ & $\mathbf{H}_i(l,1,d,1)$ \\
	\hline
	\multirow{2}{0.13\textwidth}{R}  & 9 & $(\tup,\tup,\tup,\tup)$ & $\mathbf{I}_i$ \\
	 									  & 10 & $(l,u,d,\tup)$ & $\mathbf{H}_i(l,u,d,1)$ \\
	\hline
	\multirow{2}{0.13\textwidth}{DR} & 11 & $(\tup,\tup,\tup,\tup)$ & $\mathbf{I}_i$ \\
	 								      & 12 & $(l,u,\tup,\tup)$ & $\mathbf{H}_i(l,u,1,1)$ \\
	\hline
	\multirow{2}{0.13\textwidth}{D}  & 13 & $(\tup,\tup,\tup,\tup)$ & $\mathbf{I}_i$ \\
	 									  & 14 & $(l,u,\tup,r)$ & $\mathbf{H}_i(l,u,1,r)$ \\
	\hline
	\multirow{3}{0.13\textwidth}{DL} & 15 & $(\tup,\tup,\tup,\tup)$ & $\mathbf{I}_i$ \\
										  & 16 & $(\tup,\tdown,\tdown,\tup)$ & $\mathbf{I}_i$ \\
	 									  & 17 & $(\tup,u,\tdown,r)$ & $\mathbf{H}_i(1,u,1,r)$ \\
	\hline
	\multirow{3}{0.13\textwidth}{L}  & 18 & $(\tup,\tup,\tup,\tup)$ & $\mathbf{I}_i$ \\
										  & 19 & $(\tup,\tdown,\tdown,\tup)$ & $\mathbf{I}_i$ \\
	 									  & 20 & $(\tup,u,d,r)$ & $\mathbf{H}_i(1,u,d,r)$ \\
	\hline
	\multirow{2}{0.13\textwidth}{BODY} & 21 & $(\tup,\tup,\tup,\tup)$ & $\mathbf{I}_i$ \\
	 										& 22 & $(l,u,d,r)$ & $\mathbf{H}_i(l,u,d,r)$ \\
	\hline
	\end{tabular}
	\caption{PEPO rules for the upper-right corner and the right edge. Main differences are in the rules for UL, UR, L, DL, U.}
	\label{tab:pepo_rule_right}
\end{table}

We then describe the FSM rules for the upper-right corner and the right edge of the system as shown in Table~\ref{tab:pepo_rule_right}. Together with the FSM rules for the lower-right corner in Table~\ref{tab:pepo_rule_lower_right}, these FSM rules are to ensure that either only one signal $\tdown$ is passed down from the left, or exactly one $\hat{H}_{\bm{\alpha}}$ is activated in this part of the system.

Here we explain the rules in Table~\ref{tab:pepo_rule_right}. For the UL site of each box $\mathcal{I}_{\bm{\alpha}}$, there are four possible incoming signal combinations it can receive. The first possibility is that it receives two $\tup$ signals. In this case it means no $\hat{H}_{{\bm{\alpha}}'}$ has been activated to the left or from above, we can choose whether to activate $\hat{H}_{\bm{\alpha}}$ or not (rules 1 and 4). If only one of the signals is $\tdown$ then it means an $\hat{H}_{{\bm{\alpha}}'}$ has been activated elsewhere, and it is only allowed to pass the signal $\tdown$ downward (instead of rightward as in Table~\ref{tab:pepo_rule_body}). This is specified in rule 2. Other rules exist to pass along the signals or are the same as those for the individual $\hat{H}_{\bm{\alpha}}$. At the DL site, if $\hat{H}_{\bm{\alpha}}$ has been activated, then a signal $\tdown$ is passed to the next box below (rule 16). The signal $\tdown$ is only passed along the left boundary, as opposed to along the upper boundary in Table~\ref{tab:pepo_rule_body}.

\begin{table}
	\centering
	\begin{tabular}{c|c|cc}
	\hline
	 & rule num & $(l,u,d,r)$ & operator \\
	\hline
	\multirow{4}{0.13\textwidth}{UL of R} & 1 & $(\tdown,\tup,\tup,\tup)$ & $\mathbf{I}_i$ \\
	 									  & 2 & $(\tup,\tdown,\tup,\tup)$ & $\mathbf{I}_i$ \\
	 									  & 3 & $(\tup,\tup,d,r)$ & $\mathbf{H}_i(1,1,d,r)$ \\
	\hline
	\multirow{2}{0.13\textwidth}{U of R}  & 4 & $(\tup,\tup,\tup,\tup)$ & $\mathbf{I}_i$ \\
	 									  & 5 & $(l,\tup,d,r)$ & $\mathbf{H}_i(l,1,d,r)$ \\
	\hline
	\multirow{2}{0.13\textwidth}{UR of R} & 6 & $(\tup,\tup,\tup,\tup)$ & $\mathbf{I}_i$ \\
	 									  & 7 & $(l,\tup,d,\tup)$ & $\mathbf{H}_i(l,1,d,1)$ \\
	\hline
	\multirow{2}{0.13\textwidth}{R of R}  & 8 & $(\tup,\tup,\tup,\tup)$ & $\mathbf{I}_i$ \\
	 									  & 9 & $(l,u,d,\tup)$ & $\mathbf{H}_i(l,u,d,1)$ \\
	\hline
	\multirow{2}{0.13\textwidth}{DR of R} & 10 & $(\tup,\tup,\tup,\tup)$ & $\mathbf{I}_i$ \\
	 								      & 11 & $(l,u,\tup,\tup)$ & $\mathbf{H}_i(l,u,1,1)$ \\
	\hline
	\multirow{2}{0.13\textwidth}{D of R}  & 12 & $(\tup,\tup,\tup,\tup)$ & $\mathbf{I}_i$ \\
	 									  & 13 & $(l,u,\tup,r)$ & $\mathbf{H}_i(l,u,1,r)$ \\
	\hline
	\multirow{3}{0.13\textwidth}{DL of R} & 14 & $(\tup,\tup,\tup,\tup)$ & $\mathbf{I}_i$ \\
	 									  & 15 & $(\tup,u,\tup,r)$ & $\mathbf{H}_i(1,u,1,r)$ \\
	\hline
	\multirow{3}{0.13\textwidth}{L of R}  & 16 & $(\tup,\tup,\tup,\tup)$ & $\mathbf{I}_i$ \\
	 									  & 17 & $(\tup,u,d,r)$ & $\mathbf{H}_i(1,u,d,r)$ \\
	\hline
	\multirow{2}{0.13\textwidth}{BODY of R} & 18 & $(\tup,\tup,\tup,\tup)$ & $\mathbf{I}_i$ \\
	 										& 19 & $(l,u,d,r)$ & $\mathbf{H}_i(l,u,d,r)$ \\
	\hline
	\end{tabular}
	\caption{PEPO rules for the lower-right corner of the system.}
	\label{tab:pepo_rule_lower_right}
\end{table}

Now all that is left is to formulate the FSM rules for the lower-right corner of the system. The rules are shown in Table~\ref{tab:pepo_rule_lower_right}. If the UL site receives exactly one signal $\tup$ and exactly one signal $\tdown$ form left and above respectively, then this means everything is good and we only need to pass the signal $\tup$ to the boundaries without activating the operator (rules 1 and 2). If it receives two signals $\tup$ then it means no $\hat{H}_{{\bm{\alpha}}'}$ has been activated in any other box, and therefore $\hat{H}_{\bm{\alpha}}$ has to be activated in the current box (rule 3). No signal $\tdown$ will be passed down beyond this point.

The rules in Tables~\ref{tab:pepo_rule_body}, \ref{tab:pepo_rule_right}, \ref{tab:pepo_rule_lower_right} ensure that exactly one operator $\hat{H}_{\bm{\alpha}}$ is going to be activated in the system. Otherwise some box on the right boundary (upper-right corner, right edge, and lower-right corner) will have two signals $\tdown$ incoming at the UL site, which is not allowed in the FSM rules and therefore will return a zero operator, making the whole tensor product zero.

Similar to the 1D case, the bond dimension of each PEPO $\hat{H}$ we constructed in this way is the bond dimension of the original PEPO $\hat{H}_{\bm{\alpha}}$ plus 2. In the case of rank-one operators the sum PEPO has a bond dimension of 5.

\bibliography{ref}
\bibliographystyle{siam}

\end{document}